\documentclass[aps,prx,reprint, superscriptaddress]{revtex4-2}
\usepackage{blindtext}
\usepackage[colorlinks=true,allcolors=blue]{hyperref}

\usepackage{blindtext}
\usepackage{bm}

\def \beq{\begin{eqnarray}}
\def \eeq{\end{eqnarray}}
\def \NRCHs{\mbox{NRCH}^*}
\def \bq{\bm{q}}
\def \br{\bm{r}}
\def \bn{\bm{n}}

\def \del{\partial}

\usepackage{amsmath}
\usepackage{amssymb}
\usepackage{graphicx}
\usepackage{graphicx,xcolor}

\graphicspath{{figs}}
\usepackage[capitalize]{cleveref}
\crefname{equation}{Eq.}{}
\crefrangeformat{equation}{Eqs.~(#3#1#4)-(#5#2#6)}

\begin{document}


\title{Spinning mixtures: nonreciprocity transfers chirality across scales in scalar densities}

\author{Giulia Pisegna}
\email{giulia.pisegna@ds.mpg.de}
\affiliation{Max Planck Institute for Dynamics and Self-Organization (MPI-DS), D-37077 Göttingen, Germany}
\author{Suropriya Saha}
\email{suropriya.saha@ds.mpg.de}
\affiliation{Max Planck Institute for Dynamics and Self-Organization (MPI-DS), D-37077 Göttingen, Germany}

\begin{abstract}

Mixtures of chiral active particles provide a minimal setting where nonreciprocity and chirality intertwine to shape collective dynamics. Here we show that nonreciprocal interactions can generate and transfer chirality across scales even in theories formulated solely in terms of scalar densities. We introduce a minimal two-species continuum model in which nonreciprocity and chirality enter through local and nonlocal parts of the chemical potential, both contributions equally relevant in terms of the number of spatial gradients involved. At the coarse-grained level, chirality manifests as localized domains (called charges) with a finite curl of the interspecies nonreciprocal current. These chiral charges actively redistribute particles, strongly modifying the underlying nonreciprocal dynamics. Their interplay gives rise to distinct nonequilibrium steady states, including phase separation with persistent edge currents and a spatiotemporally disordered phase combining chiral and nonreciprocal signatures. Our results establish nonreciprocity as a generic mechanism for generating chiral dynamics in scalar mixtures.  
\end{abstract}

\maketitle

\section{INTRODUCTION}

Chirality, the absence of mirror symmetry, is ubiquitous in both non-living and living matter~\cite{Harris_Kamien_Lubensky_RevModPhys.71.1745}. At the molecular scale, it manifests in the homochirality of life: nucleic acids such as DNA and RNA are built from D-sugars, while proteins are composed of L-amino acids. Why life selected specific handednesses, and how these preferences became universal, remains an open problem at the interface of physics, chemistry, and biology~\cite{tverdislov2013chirality}. Beyond the molecular scale, chirality reappears at higher levels of organization, from microtubule gliding assays~\cite{athani2025gliding} and the helical beating of bacterial flagella~\cite{berg2004coli, di2011swimming} to collective chiral flows in tissues~\cite{inaki2016cell, chen2025chirality}. These diverse examples raise a central question: how can chiral twist be transferred to macroscopic scales, and under what conditions is this process robust to fluctuations? From the perspective of physics, such a transfer requires mechanisms mediated by interactions capable of amplifying microscopic handedness to produce collective chiral effects.

\begin{figure}[t]
    \centering
    \includegraphics[width=0.7\linewidth]{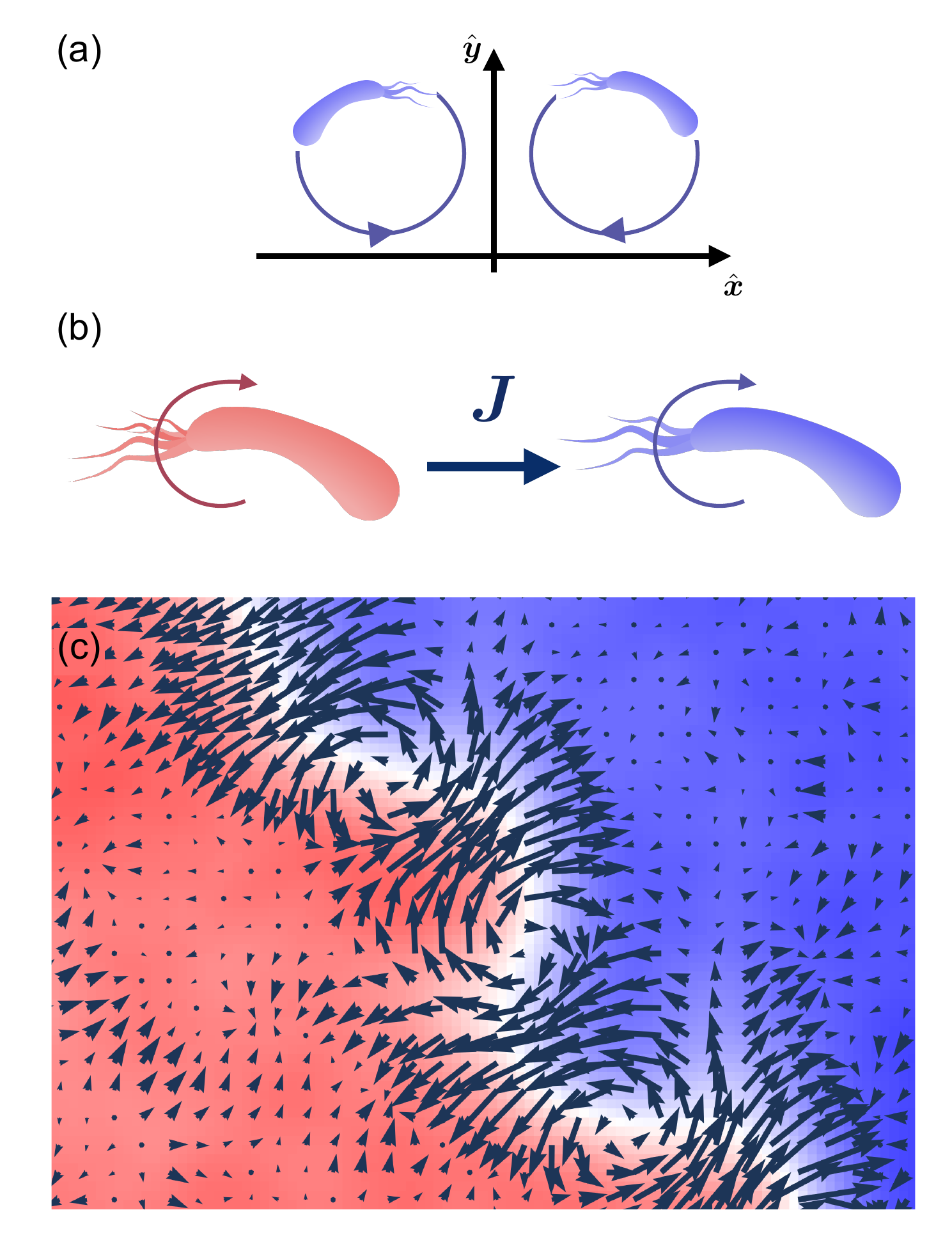}
    \caption{ \textbf{Chirality and nonreciprocity across scales.} (a) A broken mirror symmetry in flagellar rotation causes an opposite rotation of the bacterial cell body \cite{beppu2021edge}. As a result, the motion of the cell is chiral, meaning it cannot be mapped onto its mirror image by any rotation or translation.
    (b) Two species of chiral bacteria: nonreciprocity induces an interspecies chasing dynamics with directed motion along a polar order parameter $\bm J$. (c) The continuum description of the model, valid at large length scales, provides an opportunity for chirality and nonreciprocity to waltz together. The snapshot of $\bm{J}$, superimposed on the density fields, shows the organization into separated bulk phases with edge currents and vortices of $\bm{J}$. 
}
    \label{fig:schematic}
\end{figure}

At microscopic scales, many biological and non-equilibrium systems exhibit intrinsic chiral dynamics at the level of individual agents. However, microscopic chirality does not necessarily propagate to the collective scale. Dense assemblies of chiral particles, such as bacterial suspensions, can instead develop turbulent bulk dynamics that effectively wash out chirality, resulting in macroscopically achiral motion \cite{dunkel2013fluid,wolgemuth2008collective}. In such cases, chiral signatures typically survive because of single particles interactions with boundaries or emerge through collective amplification of weak microscopic biases~\cite{li2024robust,beppu2021edge}. These observations indicate that chirality alone is often insufficient to ensure macroscopic chiral order.

Understanding the large scale dynamics of these systems is challenging, especially when active agents of different chirality coexist and interact. Several theoretical models for interacting chiral active particles have been considered \cite{digregorio2025phaseseparationchiralactive,Caprini2025,liebchen2017collective, sese2022microscopic, caporusso2024phase}, with chiral edge properties unveiled in experiments with spinners \cite{soni2019odd, li2024robust,chen2025self}. Additionally, theoretical studies modelled
mixtures of particles spinning at different rates but interacting symmetrically
\cite{ChiralMixtures_PhysRevE.100.012406, Levis_Pagonabarraga_Liebchen_PhysRevResearch.1.023026}. Importantly, heterogenous chiral mixtures with asymmetric microstructures are naturally accompanied by effectively non-conservative and nonreciprocal interactions \cite{nguyen2014emergent, yeo2015collective}.


Nonreciprocity, namely an effective violation of action-reaction symmetry, characterizes interspecies interactions that cannot be derived from a potential and therefore intrinsically break detailed balance and time-reversal symmetry. In minimal settings, nonreciprocity alone can generate non-equilibrium self-organized states that are forbidden in equilibrium systems, including propelled clusters \cite{soto2014self, agudo2019active, ouazan2023self}, propagating density waves \cite{saha2020scalar, you2020nonreciprocity, pisegna2024emergent}, and spontaneous pattern formation \cite{rana2023defect}. In this sense, nonreciprocity acts as a fundamental promoter and driver of collective organization. This naturally raises the question of whether nonreciprocal interactions can play an analogous role in chiral systems, enabling microscopic chirality to propagate across scales and stabilize macroscopic chiral dynamics.

In this paper, we formulate a phenomenological approach to the problem by considering the coupled dynamics of two density fields interacting via minimal nonreciprocal and chiral couplings. Chirality, which corresponds to broken left-right and up-down symmetry in two dimensions, see Fig.~\ref{fig:schematic}(a), cannot be associated with a point-like object, or, at the level of fields, with a scalar number density. To introduce chiral particle currents in a scalar mixture, one must consider a vectorial quantity, thereby bringing the nonreciprocal current $\bm{J}$ -- which quantifies the chasing dynamics of two density fields (Fig.~\ref{fig:schematic}(b)) -- to prominence. Chirality causes $\bm{J}$ to twist in either clockwise or counter-clockwise directions, thus fundamentally altering the macroscopic dynamics of the underlying scalar fields, see Fig.~\ref{fig:schematic}(c). Based on symmetry arguments, we propose a field theory, the chiral Nonreciprocal Cahn-Hilliard$^*$ model ($\NRCHs$), that includes both nonreciprocity and chirality in a binary mixture. The presence of two species allows for chiral interactions that lead to dynamics fundamentally different from the single-species case~\cite{kole2021layered, maitra2020chiral, Maitra_annurev}.


Chiral coupling alters the phenomenology associated with nonreciprocity: beyond a threshold, it suppresses the traveling-wave states characteristic of nonreciprocal mixtures and instead promotes spatial organization driven by localized domains, which we call chiral charges, with circulating currents. 
When charges of opposite circulation pairwise cancel, the system undergoes bulk phase separation accompanied by persistent edge currents confined to interfaces. By contrast, when such cancellation fails, the system remains disordered with small scale chiral currents. While these non-linear effects lie beyond a complete analytical description, we provide physical arguments for how the interplay between chirality and nonreciprocity leads to phase separation and explain why chirality is essential for the emergence of circulating edge currents at the interfaces. The active contributions to the dynamics allowed by symmetry are justified by coarse-graining different microscopic models of active chiral particles, demonstrating that the theoretical framework holds for a wide range of models with two key ingredients: chirality and nonreciprocity.

The manuscript is divided into three parts -  in the first part covered in Sections \ref{sec:model}, \ref{Sec:Microscopic}, and \ref{Sec:Phenomenology} we introduce $\NRCHs$ and establish its microscopic connections, and general phenomenology.
Specifically, in Sec.~\ref{Sec:Phenomenology}, we explore the rich dynamical phases of the model showing that in the absence of stochastic noise, chirality manifests at large scales only in the presence of nonreciprocal interactions, giving rise to two novel steady states: a phase-separated state with chiral edge currents, and a chiral disordered phase. 
In the second part in Sec.~\ref{Sec:LinearStability}, we show the manner in which fluctuations about a nonreciprocal traveling state grow in the presence of strong enough chirality to modify defining signatures of nonreciprocity. In the third part covered in Sec.~\ref{sec:PhaseSeparation} and \ref{sec:interfaces} we present insights from comparing some of the dynamical states to bulk-phase separation. 
We first derive an out-of-equilibrium analogue of pressure, which balances the effect of the chiral edge contribution. We then study the dynamics of these edge currents by linearizing the coupled interface fluctuations of the two fields (Sec.\ref{sec:interfaces}). This analysis reveals that edge currents correspond to propagating interface waves, whose amplitude is proportional to the total microscopic chirality. Altogether, our results demonstrate that mixtures containing particles of opposite or equal chirality can exhibit a unified, emergent large-scale chiral behavior mediated by nonreciprocal interactions.

\section{RESULTS}
\subsection{The model}
\label{sec:model}

To recall the notion of chirality, consider the bacterium illustrated in Fig.~\ref{fig:schematic}(a). Its flagella rotate in a fixed direction, inducing an opposite rotation of the cell body \cite{berg2004coli}. The resulting motion cannot be superimposed onto its mirror image, as both head-tail and left-right symmetries are broken simultaneously. Therefore, the simplest example of a chiral particle can be described through its orientational unit vector $\bm n$, which rotates within a two-dimensional plane spanned by unit vectors $\hat{\bm x}$ and $\hat{\bm y}$. In the absence of fluctuations, $\bm n$ undergoes a steady rotation with constant angular velocity $\bm \omega$ directed along the $\hat{\bm z}$ axis, perpendicular to the plane to which it is confined. To account for stochastic fluctuations, we include a noise term $\bm \xi$ of amplitude $D_r$ so that the orientational dynamics is governed by
\beq 
\label{eq:micron}
\frac{\mbox{d} \bm n}{ \mbox{d} t } = (\omega \hat{\bm{z}} + \bm \xi)\times \bm{n}.
\eeq 
This equation is not invariant under a parity transformation of one of the space axes, such as $\hat{\bm x} \to - \hat{\bm x}$, under which $\hat{\bm z} \to - \hat{\bm z}$. The dynamics is preserved only if the handedness of the particle is simultaneously reversed, i.e., $\omega \to - \omega$.

To investigate how this type of chirality is transferred from single particle to larger length-scales in multispecies systems, we consider a binary mixture of conserved scalar fields $\phi_a$ ($a=\{1,2\}$) with nonreciprocal interactions, and chiral currents that reflect the handedness of the constituent particles. The fields evolve as
\beq 
\partial_t \phi_a = \nabla^2 \mu_a - \bm{\nabla} \cdot \bm{j}^{c}_{a} + \bm{\nabla}\cdot \bm{\zeta}_a.
\label{eq:ChiralCon}
\eeq 
The chemical potential $\mu_a = \mu_{eq,a} + \alpha \epsilon_{ab} \phi_b$ receives two contributions. The first, $\mu_{eq,a}$, is the functional derivative of a free energy density that leads to bulk phase separation in the passive limit. The second term stems from inter-species nonreciprocal interactions, which we minimally include in $\mu_a$ introducing the strength of nonreciprocity, $\alpha$. $\epsilon_{ab}$ is the two dimensional Levi-Civita tensor acting on the internal space of densities \cite{saha2020scalar, you2020nonreciprocity} and we assume summation convention for the species index. The nonreciprocal term  does not have a variational form, and introduces frustration in the dynamics of the two species. A conserved white and Gaussian stochastic noise $\bm \zeta_a$ acts on the densities, as in standard Model B dynamics \cite{hohenberg1977theory}.

The starting point of our work is to propose a phenomenological form for the chiral currents $\bm j_a^{c}$ based on symmetry arguments. Consider the following form,
\beq \label{eq:ChiralCurrent}
\bm{j}^{c}_{a} = - \beta_a  \mathcal{G}' (\mathbb{A} \cdot \bm{\nabla}) \mathcal G,
\eeq 
where $\mathcal G,\mathcal G'$ are functionals of the fields and their gradients, and $\mathbb{A}$ is the fully anti-symmetric tensor acting on the two dimensional Cartesian subspace, $A_{xx} = A_{yy} = 0$, and $A_{xy}=-A_{yx} = 1$. The simplest term involving the fewest number of fields and gradients is obtained for $\mathcal{G} = \phi_1, \mathcal{G}' = \phi_2$ such that \footnote{Three different choices lead to the same term on the R.H.S. of Eq.~\eqref{eq:ChiralCon}. They are $\mathcal{G}=c_1, \, \mathcal{G}' = c_2 \phi_1 \phi_2$, $\mathcal{G}=c_1 \phi_1, \, \mathcal{G}' = c_2 \phi_2$, and $\mathcal{G}=c_1 \phi_2, \, \mathcal{G}' =c_2 \phi_1$, with the redefinition $\beta \equiv \beta c_1 c_2$},
\beq 
\label{eq:betaterm}
\partial_t \phi_a = \nabla^2 \mu_a - \beta_a \hat{\bm z} \cdot ( \bm \nabla \phi_1 \times \bm \nabla \phi_2) + \sqrt{2D}\bm{\nabla}\cdot \bm{\zeta}_a .
\eeq

To ascertain the nature of the second term on the R.H.S. of Eq.~\eqref{eq:betaterm}, consider how the equation transforms when the handedness of the coordinate-axes is changed, akin to our examination of Eq.~\eqref{eq:micron}. The first term on the R.H.S. of Eq.~\eqref{eq:betaterm}, being a scalar, is invariant under the transformation, while the second term within the parentheses reverses sign. The dynamics in Eq.~\eqref{eq:betaterm} is invariant under  parity, and thus well defined, only if the coefficients  $\beta_a$ are chiral in nature, i.e. they are pseudo-scalars which change sign as well. This situation arises, for instance, when $\beta_a$ are proportional to the intrinsic spin of particles rotating in the two dimensional plane, following the microscopic interpretation of Eq.~\eqref{eq:micron}. 
The fields themselves are number densities, i.e. scalars, such that these considerations hold for all choices of $\mathcal{G}, \mathcal{G}'$. At this point we introduce a useful quantity related to the divergence of the chiral current name the chiral vorticity $\omega_c$: $\bm{\nabla} \cdot \bm{j}_a^c = \beta_a \omega_c$.

The novelty of $\bm{j}^c_a$ is rationalized considering again the following      
\beq \label{eq:GG'}
\bm{\nabla} \cdot \bm{j}_a^c = \beta_a(\partial_x \mathcal{G} \partial_y \mathcal{G}' - \partial_x \mathcal{G}' \partial_y \mathcal{G}).
\eeq 
For a non-vanishing L.H.S. of Eq.~\eqref{eq:GG'}, we require $\mathcal{G}$ and $\mathcal{G}'$ to have different functional forms. 
For a single species, the simplest term at lowest order in gradients and powers of fields that we are allowed to construct is with $\mathcal{G} = \phi$ and $\mathcal{G}' = \nabla^2 \phi$, see also \cite{bureković2026bulkmicrophaseseparationchiral}, leading to
\beq \label{eq:singleSpecies}
{\bm{j}}^{c} =  - \beta \phi (\mathbb{A} \cdot \bm{\nabla}) \nabla^2 \phi,
\eeq 
which involves three gradients and is thus less relevant (in the hydrodynamic limit) than the term we include in Eq.~\eqref{eq:betaterm}.

To complete the model we need to specify the passive part of the $\mu_a$, which is written in terms of a Cahn-Hilliard free energy functional $\mu_{eq,a} = \delta \mathcal F/\delta \phi_a$. This includes general linear and non-linear couplings between species \cite{cahn1958free}. The associated free-energy density $F = K  \bm  \nabla \phi_a \cdot \bm \nabla \phi_a /2 + V(\phi_a)$ is composed by a surface-tension contribution, assumed equal for all the species \cite{frohoff2021suppression}, and a potential $V$. Here we focus on two different forms of $V$ distinguishing different symmetries: a Mexican-hat potential, 
\beq \label{eq:mexican-hat}
V_{m} (\phi_a) = - \frac 12 \phi_a \phi_a + \frac 14 (\phi_a \phi_a)^2
\eeq 
that is $SO(2)$ symmetric, and a double-well potential,
\beq \label{eq:double-well}
V_{d}(\phi_a) = \sum_a \left[ - \frac 12 \phi_a ^2 + \frac 14 \phi_a^4 \right]
\eeq 
that is an example of discrete symmetry. The impact of these two alternatives on nonreciprocal phase separation has been studied in~\cite{johnsrud2025state}. Here, we demonstrate that the novel phenomenology arising from chiral interactions is independent of the specific choice of the potential. However, one of the two forms allows for more tractable analytical calculations in different contexts we explore; we use Eq.~\eqref{eq:mexican-hat} in Sec.~\ref{Sec:LinearStability} and Eq.~\eqref{eq:double-well} in Sec.~\ref{sec:phaseComposition}. The resulting analytical predictions are always validated by numerical simulations performed with the alternative potential, thus testing the general application of our results. 

\begin{figure}
    \centering
    \includegraphics[width=1\linewidth]{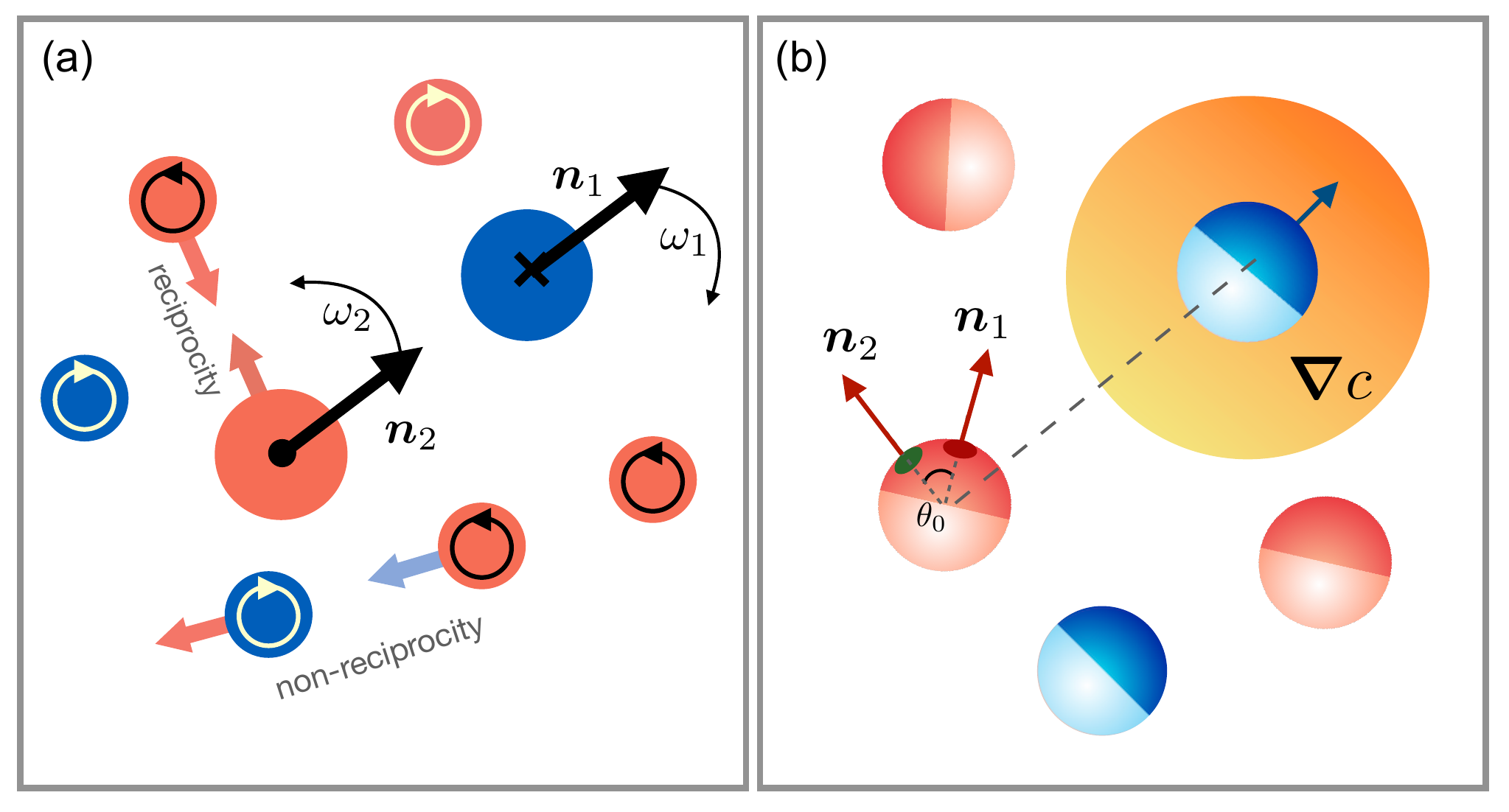}
    \caption{\textbf{Agent-based models of chiral particles with nonreciprocity}. (a) Two species of active particles with an intrinsic angular velocity: activity arises also through effective nonreciprocal interactions between species, which couple with the particles’ internal chirality. (b) Binary mixtures of chemically active colloids with chiral chemotaxis: a non-axisymmetric particle (for example with two mobility patches separated by a fixed angle $\theta_0$) responds to the chemical fields produced by other colloids by drifting to the left or the right with respect to the local chemical gradients.}
    \label{fig:micro_only}
\end{figure}

\begin{figure*}[t]
    \centering
\includegraphics[width=\linewidth]{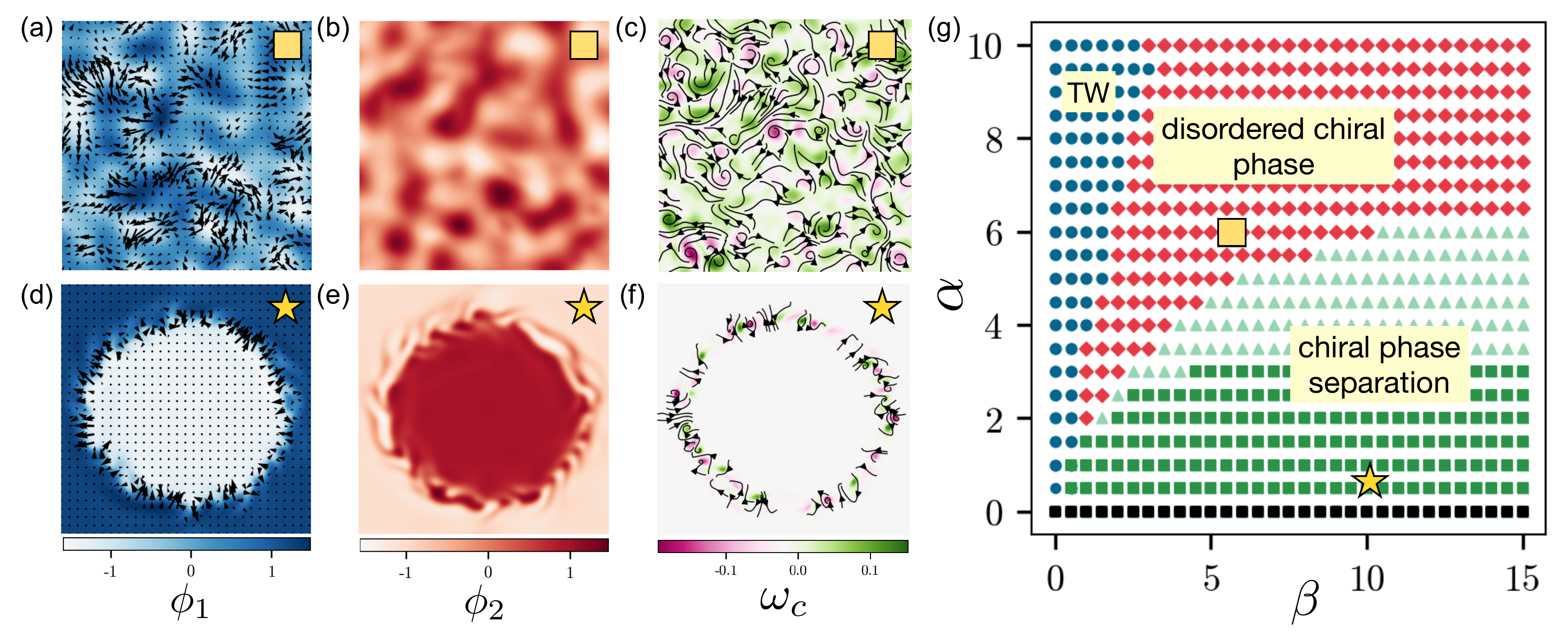}  
\caption{ \textbf{Dynamics in the steady state.} (a-c) Chiral disordered state: when chirality and nonreciprocity compete, the system remains disordered in the steady state; the fields lack any definite pattern and fluctuate across a broad spectrum of length and time scales. The polar order (depicted with streamlines) and the chiral vorticity (the magnitude of the circulation of $\bm{J}$) also appear featureless. (d-f) Phase separation with edge currents:  increasing $\beta$ drives the system into a macroscopic bulk phase separated state, with circulating currents localized at the interface. Both forms of activity produce effective repulsive interactions between the two fields for the chosen parameters, as seen in panels (d, e). In this phase, $\omega_c = (\bm \nabla \times \bm J)_{\hat{\bm z}}$ behaves strikingly differently: charges of opposite sign form along the interface, around which $\bm J$ circulates with alternating handedness. (g) The phase diagram shows the existence of four distinct phases at the steady state. Black markers: in the absence of stochastic terms, for $\alpha=0$ and $\beta>0$, we observe only bulk phase separation in the steady state. Blue markers: when $\alpha$ dominates over $\beta$, chiral effects are undetectable in the stable traveling waves (TW). 
Red markers: fully disordered chiral phase where we observe spatiotemporal chaos. Green markers: bulk phase separation with edge waves. Light-green triangles correspond to simulations initialized from an already phase-separated state obtained at same $\beta$ with $\alpha=0$, to avoid the prohibitively long coarsening times. Parameters for panels (a-c): $ \alpha=6, \beta=6 ,K=1, dt=0.05, T=3\times 10^4$. Parameters for bottom panels (d-f): $\alpha=1, \beta=10,K=1, dt=0.05, T=3\times 10^4$. }
\label{fig:phase_diagram}
\end{figure*}

\subsection{Microscopic derivation of the chiral current.}\label{Sec:Microscopic}

We have motivated the theoretical model in Eq.~\eqref{eq:betaterm}  through a phenomenological approach based on symmetry considerations under coordinates transformation and number conservation laws. Nevertheless, the chiral current $\bm j_a^c$ has a robust and generic microscopic interpretation. In Methods~\ref{MethodsA} we report three distinct examples. 

We consider a multi-species generalisation of the simple model of Eq.~\eqref{eq:micron}. Particles of species $a$ self-propel with speed $v_a$ and rotate with frequency $\omega_a$ (Fig.~\ref{fig:micro_only}(a)). When the propulsion speed depends on the local particles density  $v_a(\{\rho \})$, the system exhibits quorum sensing interactions, a mechanism known to drive motility induced phase separation \cite{cates2015motility} and nonreciprocal dynamical states \cite{dinelli2022self, duan2023dynamical}.  In Methods \ref{sec:ChiralBacteris} we present a coarse-graining of this microscopic model, showing that the mesoscopic density fields evolve according to $\partial_t \rho_a + \bm \nabla \cdot ( v_a \bm p_a + \mbox{h.o.t.})=0 $, receiving contributions  from the average polarisation $\bm{p}_a$. In the regime where the orientation relaxes faster than the conserved density field, we find
\beq \label{eq:Polarisation}
( 2 D_{ra} \mathbb I + \omega_a \mathbb A) \cdot \bm p_a = - \frac 12 \bm \nabla ( v_a \rho_a) \ ,
\eeq 
where $\mathbb{I}$ is the two-dimensional identity tensor, $I_{xx} = I_{yy} = 1$, and $I_{xy}=I_{yx} = 0$. For $\omega_a = 0$ the active contribution to the effective pressure drives an instability promoting particles aggregation \cite{cates2015motility}. In contrast, chirality introduces an odd rotational diffusivity, giving rise to chiral currents in the evolution of $\rho_a$. Comparing Eq.~\eqref{eq:Polarisation} and \eqref{eq:ChiralCurrent}, we identify $\mathcal G = v_a(\{\rho \}) \rho_a$ and $\mathcal G' \simeq \omega_a \rho_a$.

A second example consists of chemically active particles, responding to gradients of solute concentration $c$ produced by other colloids with surface activity rate $\sigma_a$. As described in Methods, 
there are multiple ways in which chirality manifests. For instance when the chemotactic drift of the particle is not along the gradient of the chemical field but has a component perpendicular to it (Fig.~\ref{fig:micro_only}(b)) \cite{saha2019pairing}. This lateral drift happens in two cases - for spinning chemical colloids, and also when the orientational response to the gradient is non-linear creating the possibility that it aligns to solute gradients at a finite angle determined by its internal parameters. We refer to this mechanism  shown in Fig.~\ref{fig:micro_only}(b) as chiral chemotaxis. The orientational degree of freedom of chemical colloids with a non-axisymmetric mobility surface distribution reads,
\beq
\bm p_a = \frac{1}{2 D_{ra}} [- v_a \bm \nabla \rho_a + ( \Omega_a \mathbb I - \left (\bar \Omega_a/3 \right)  \mathbb A ) \cdot \bm \nabla c \rho_a ]
\eeq 
with $\Omega_a, \bar{\Omega}_a$ describing the alignment of the orientation to $\bm \nabla c$ coming from phoretic chemotactic interactions (see Methods \ref{sec:ChiralChemotaxis} for a detailed derivation). Because of $c \sim \sigma_b \rho_b$, the last term produces a chiral current with $\mathcal G \simeq \bar{\Omega}_a \sigma_b \rho_b $ and $\mathcal G' = \rho_a$.
These two examples illustrate that, regardless of the source of activity in a multispecies system, chiral currents inevitably emerge at quadratic order in gradients, as stated in Eq.~\eqref{eq:ChiralCurrent}. The chiral terms are generated by enslaving the orientation to the density, assuming that sufficiently high rotational diffusivity prevents long-range orientational order, and thus the development of net chirality. As we describe in the next section \ref{Sec:Phenomenology}, the rich phenomenology of $\NRCHs$ is dominated by localized chiral domains within a globally achiral system.

Additionally, the calculation shows that microscopic parameters can renormalize coefficients in equilibrium potentials. We do not discuss this aspect further here, since the coarse-graining approach presented in the Methods primarily aims to highlight the emergence of the chiral current in different contexts. Nevertheless, this effect should be taken into account when inferring active dynamics and continuum models from experimental or numerical data \cite{boffi2024deep, supekar2023learning}. 

\begin{figure}
    \centering
    \includegraphics[width=1\linewidth]{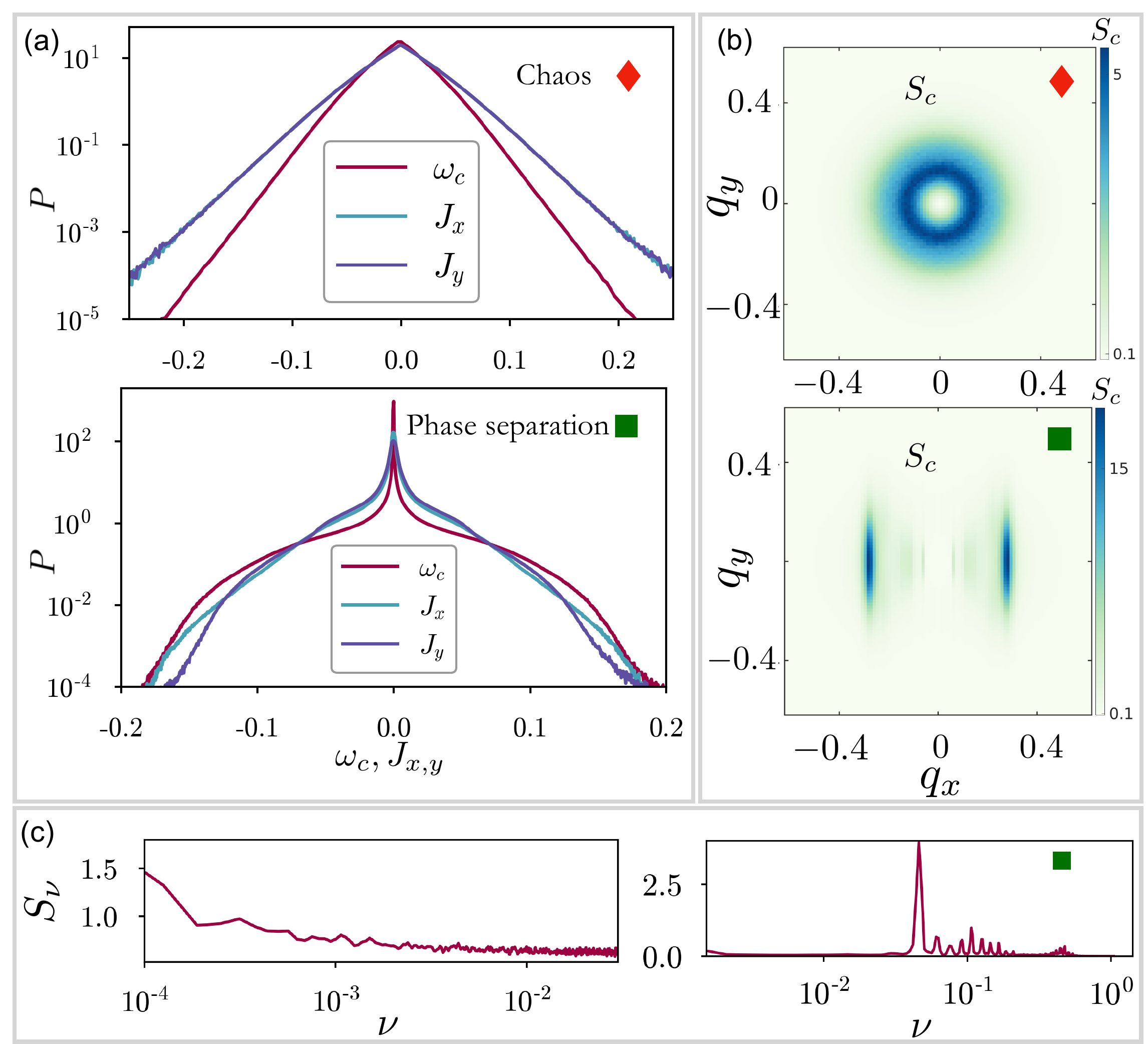}
    \caption{ \textbf{Characterising the dynamical steady states.} We compare and contrast the properties of the chiral steady states presented in Fig.~\ref{fig:phase_diagram}, evaluated in two distinct regimes in the state diagram: chaotic disorder ($\beta = 5, \, \alpha = 5$ red rhombus) and phase separation ($\beta = 5, \, \alpha = 3.25$ green square). (a) The probability distribution functions $P$ of the chiral vorticity $\omega_c$ and the two components $J_{x,y}$ are shown. In both states, $\omega_c, J_{x,y}$, $P$ are symmetric about zero, meaning that there is no coherent traveling state. The tails assume a clear exponential form in the disordered state, while in the phase separated state, it is less pronounced. (b) The static correlation function $S_c$ [see Eq.~\eqref{eq:StructureChirality}]  is isotropic in the disordered phase but anisotropic in the phase separated state (here a stripe, as in Fig.7(a)), reflecting that the directed edge currents still involve multiple length scales. (c) Power spectra of the densities in the disordered state reveal broad frequency fluctuations; in phase separation, edge currents produce multiple characteristic peaks.  }
    \label{fig:Sc}
\end{figure}

\subsection{Large-scale dynamics: interplay of chirality and nonreciprocity} \label{Sec:Phenomenology}
To underline the connection between nonreciprocity and the chiral term in Eq.~\eqref{eq:betaterm}, we notice that $ \omega_c = \hat{\bm{z}} \cdot (\bm \nabla \phi_1 \times \bm \nabla \phi_2) = \hat{\bm{z}} \cdot (\bm \nabla \times \bm J)/2$, where $\bm{J} = \phi_1\bm{\nabla}\phi_2-\phi_2\bm{\nabla}\phi_1$~\cite{saha2020scalar,pisegna2024emergent}.  The vector $\bm{J}$, which  vanishes when the two densities are either exactly in phase or exactly out of phase, captures the chasing dynamics between the two densities induced by nonreciprocal interactions. It serves as an order parameter~\cite{pisegna2024emergent, pisegna2025non} signalling the transition to a state with broken parity and time-reversal symmetry. In absence of other non-equilibrium interactions, it changes to nonzero value  when nonreciprocity is sufficiently large and traveling waves are formed \cite{rana2023defect}. 

We can rewrite Eq.~\eqref{eq:betaterm} purely as gradient dynamics in the form $\partial_t \phi_a = \nabla^2 M_a$ by introducing a non-local chemical potential $\mu_a^c$, i.e. $M_a = \mu_a + \mu_a^c$. The chiral contribution $\mu_a^c$ satisfies a Poisson equation sourced by the chiral vorticity $\omega_c$
\beq \label{eq:chiralMu}
&& \nabla^2 \mu_a^c = - \beta_a \omega_c, 
\eeq 
subject to the constraint
\beq \label{eq:MuConserved}
&& \int_A d^2 r \omega_c = \int_S \bm{J} \cdot \mbox{d} \bm{l}  =  0.
\eeq 
As a consequence of Eq.~\eqref{eq:chiralMu}, we have $\beta_2 \mu_1^c = -\beta_1 \mu_2^c$. From Eq.~\eqref{eq:MuConserved}, it is clear that the constraint on the area integral of $\omega_c$ is determined by the topology of the domain considered. Indeed the relation in Eq.~\eqref{eq:MuConserved} holds for periodic boundary conditions or for dynamics on a closed surface. Eq.~\eqref{eq:MuConserved} thus forces the system to be achiral on an average -  a direct consequence of number conservation. 

\begin{figure*}
    \centering
    \includegraphics[width=0.95\linewidth]{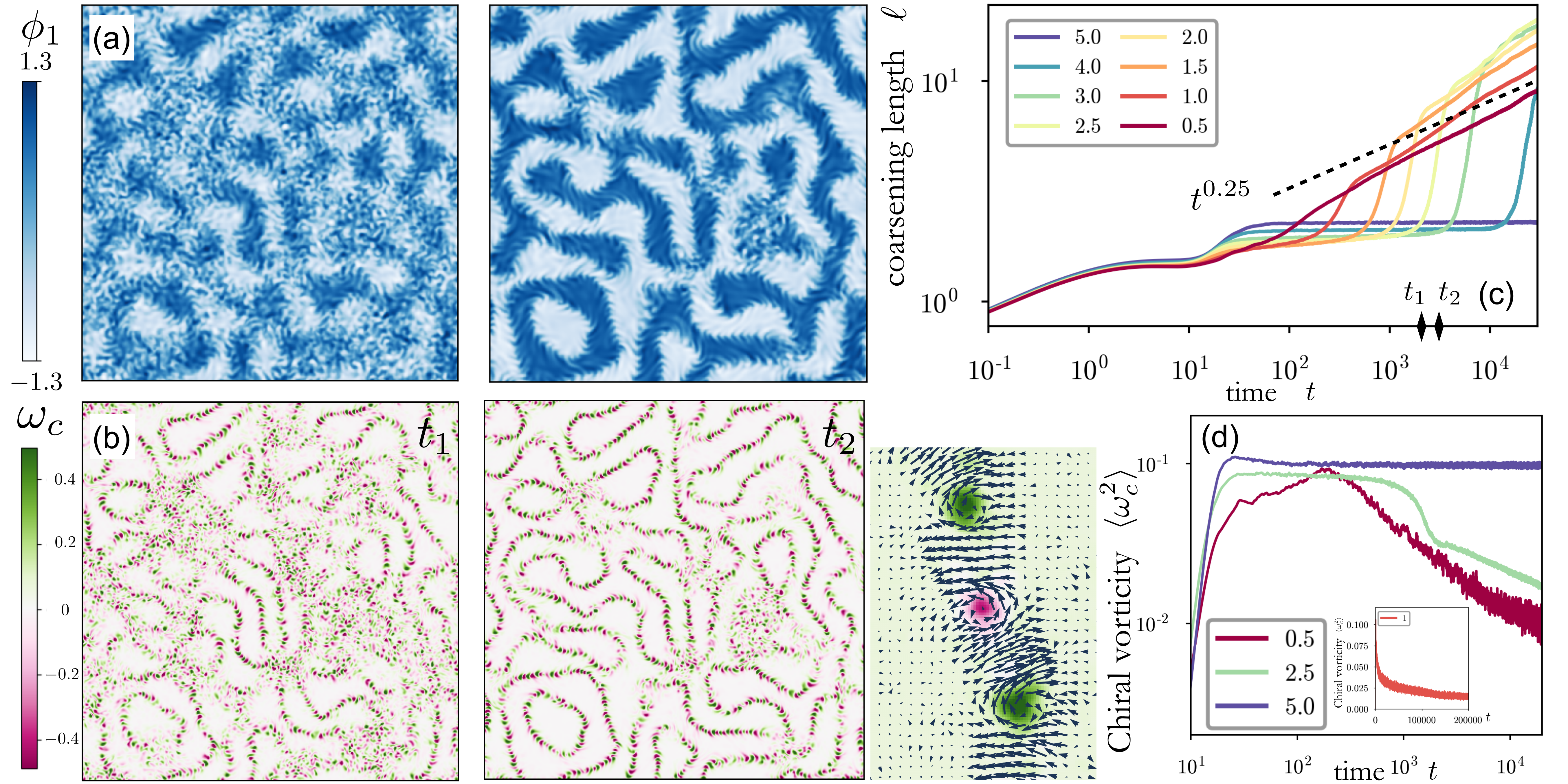}
    \caption{\textbf{Chiral vorticity and coarsening dynamics}. (a) The coarsening dynamics does not show self-similarity as clear from the snapshots of the density fields. (b) The differences are even more obvious when we look at patterns of $\omega_c$. At earlier times, the $\omega_c$ is disordered ($t_1 \approx 2000$). At later times, the sources of opposite signs cancel one another to produce linear arrangements of charges along the interface at $t_2 \approx 3000$. As is clear from the zoomed-in snapshot at the interface, $\bm{J}$ twists clockwise (anticlockwise) around local minima (maxima) of $\omega_c$. (c) The different coarsening regimes are evident in the time evolution of the length-scale associated with the domains of density for $\beta =5$, and for values of $\alpha$ indicated in the legend. (d) The disordered state is associated with a large, nearly constant value of $\langle \omega_c^2 \rangle$, as seen in the curve for $\alpha = 5.0$. For values of $\alpha$ where phase separation occurs, the quantity reaches a peak before decreasing to a constant value in the steady state (see inset for $\beta = 5$ and $\alpha = 1$), accompanied by a regime similar to Ostwald ripening in passive phase separation where the coarsening length scale increases with time with an exponent of $0.25$. Parameters for (a,b): $\beta = 5$, $\alpha = 1$.}
    \label{fig:coarsening}
\end{figure*}

From the discussion above, it follows that if nonreciprocity gives rise to local polar order in the form of a nonzero $\bm{J}$, then chirality emerges through the vorticity of this vector field. Consequently, nonreciprocal interactions are required to observe the effects of the chiral currents. 
This insight motivates a numerical exploration of the parameter space defined by $(\beta, \alpha)$, where we set $\beta = \beta_1 = \beta_2$.
We perform numerical simulations of Eq.~\eqref{eq:betaterm} without stochastic noise, using the double-well potential $V_d$ in Eq.~\eqref{eq:double-well}, at a fixed system size, and varying both the nonreciprocal and chiral strengths. The results are reported in the phase diagram of Fig.~\ref{fig:phase_diagram}(g).  First, we find that for $\alpha = 0$, and $\beta \neq 0$ the system evolves to a standard bulk phase separated state.
For very large $\alpha$ and small $\beta$, the chasing dynamics takes over and the system self-organizes in traveling patterns, which have been already extensively studied \cite{pisegna2024emergent}. Increasing the level of chirality, we  reveal two novel dynamical steady states that arise due to the competition of these non-equilibrium effects.

For $\beta \gg \alpha$, we find that the mixture coarsens to phase separated structures that are clearly out of equilibrium due to persistent currents that circulate at the edges (Fig.~\ref{fig:phase_diagram}(d-f)). A stable droplet with constant values of fields $\phi_{1,2}$ within and outside the interface exists in the steady state with transport at its edge in a direction determined by the sign of $\beta$ (see also movie M2). The polar order parameter $\bm{J}$ is null in the bulk, and it acquires non-zero values at the interfaces. From Fig.~\ref{fig:phase_diagram}(f) we observe that the edge currents correspond to local domains of pronounced chiral vorticity $\omega_c$, which we call charges, that alternate between positive and negative values so that their sum is zero, respecting the conservation law in Eq.~\eqref{eq:MuConserved}. The non-local chemical potential defined in Eq.~\eqref{eq:chiralMu} is thus sourced by charges of alternating signs that are located at the interface. The appearance of charges of alternating signs effectively screens the long ranged nature of $\mu_a^c$ leading to bulk-phase separation (see movie M5) details of which are presented in Section~\ref{sec:PhaseSeparation}.
Furthermore, the droplet configuration shown in Fig.~\ref{fig:phase_diagram} is not the only pattern observed in the phase-separated regime. The system can also spontaneously form stripes, as illustrated in Fig.~\ref{fig:PressureBalance}, revealing multistability in steady-state patterns selected by non-linearities. The relative probability of obtaining either morphology depends on several factors, including the symmetry of the equilibrium potential, the ratio between $\alpha$ and $\beta$, and the strength of stochastic noise when present (see Methods \ref{Sec:methodsNumerics} for further discussion). 

An interplay between comparable values of $\alpha$ and $\beta$ destroys both traveling waves and phase separation, and instead produces a spatio-temporally disordered state (see Fig.~\ref{fig:phase_diagram}(a-c) and movie M1). As we discuss later in the paper, the coarsening for this system is a slow process, therefore to distinguish better the transition between phase separated and disordered state, we start simulations from passive bulk separated initial conditions.

Even in the disordered state, the chiral current remains non-zero, with local sources of vorticity distributed throughout the system. To characterize its fluctuations quantitatively we calculate the histogram of values assumed by $\omega_c$ in the steady state as plotted in Fig.~\ref{fig:Sc}(a). In both the chiral disordered and the phase separated states, $P(\omega_c)$ peaks at zero and is symmetrical about this point enforcing the conservation law in Eq.~\eqref{eq:MuConserved}. Both $P(\omega_c)$ and $P(J_{x,y})$ have exponential tails in the disordered state indicating that locally the chiral fluctuations are strong as visible in Fig.~\ref{fig:phase_diagram}(c) (see also movie M1). In the phase separated state these tails are much less pronounced as expected from comparing the $\omega_c$ fields in Figs.~3(c, f). Next we compute the static correlation function of the chiral vorticity in Fourier space,
\beq \label{eq:StructureChirality}
S_c(\bq) = \frac{1}{T} \int_0^T \mbox{d}t|\omega_c(\bq,t)|^2.
\eeq 
where $T$ denotes the total simulation time. Fig.~\ref{fig:Sc}(b) shows $S_c$
 for both disordered and phase-separated configurations. In the disordered case, 
$S_c$ is isotropic in wave-vector space, while in the phase-separated state it becomes strongly anisotropic, reflecting the emergence of domains with well-defined vorticity charges. The wave-vector at which $S_c$
 reaches its maximum defines a characteristic length scale of vorticity fluctuations, (see Fig.~\ref{fig:Sc}(b)). These features are consistent with the power spectrum of the fluctuating densities - a broad range of frequencies are excited in the disordered state compared with the discrete frequencies associated with the interfacial edge currents, see Fig.~\ref{fig:Sc}(c). 

The spatial distribution of vorticity charges also plays an important role in the coarsening dynamics to form chiral phase separated droplets. On the route to phase separation, charges of opposite sign cancel one another to produce linear arrangements along the interfaces, which stop fusing only because they begin chasing one another without meeting, see Fig.~\ref{fig:coarsening}(a-b). The coarsening characteristic length scale $\ell(t)$ in Fig.~\ref{fig:coarsening}(c) calculated using the circularly averaged structure factor $S(q) = <|\phi_1(\bm q,t)|^2>$ for one of the two fields, such that $\ell(t)^{-1} = (\int d q \   S(q) q)/(\int d  q \  S(q))$, clearly illustrates the different regimes. The different effects of chirality and nonreciprocity compete resulting in a growth law that depends on both $\alpha$ and $\beta$. For comparable $\alpha$ and $\beta$, the coarsening arrests at a length-scale comparable to the surface tension. As nonreciprocity is decreased, chiral charges of opposite signs fuse, leading to a lower mean chiral vorticity in the system (Fig.~\ref{fig:coarsening}(d)). 
At the smallest values of $\alpha$, for which we still find phase separation with the transverse currents, the coarsening evolves with a power law of about $\ell \sim t^{1/4}$ before the saturation given by finite size effects. This exponent differs markedly from that of standard phase separation \cite{bray2003coarsening}, highlighting a distinct phenomenon and motivating further investigation in this direction. A possible explanation for the slower coarsening exponent is that phase separation is mediated by the annihilation of topological (vorticity) defects, a mechanism known to produce similarly slow coarsening dynamics in other physical systems \cite{harrison2000mechanisms}.

Our goal in the next three sections is to predict and understand analytically the new phases unveiled in $\NRCHs$. As the chiral terms are non-linear, a linear stability analysis around a state of uniform density does not bring further information. However, we will show that it is possible to study the stability of exact traveling-wave solutions, and show how chirality destabilizes them leading to the formation of different dynamical states. 

\subsection{Linear stability of the traveling wave}\label{Sec:LinearStability}

The numerical simulations summarised in Fig.~\ref{fig:phase_diagram}(g) depict a transition from traveling waves states to disordered chiral phase in the regime of large $\alpha$. To rationalize this part of the phase diagram we study the stability of the traveling wave solutions in Eq.~\eqref{eq:betaterm}. Specifically, we consider the system with a Mexican-hat potential $V_m$ in Eq.~\eqref{eq:mexican-hat} to describe free-energy-driven interactions. In this setting, a traveling wave is an exact solution, enabling a systematic study of fluctuations around it \cite{pisegna2024emergent,pisegna2025non}.

Defining the complex field $\phi = \phi_1 + i \phi_2$ 
and choosing $\beta_1 = \beta_2 = \beta$, the Eq.~\eqref{eq:betaterm} in terms of $\phi$ becomes,
\beq\label{eq:chiralComplexLG}
&& \partial_t \phi = \nabla^2 [-(1+i \alpha) \phi + |\phi|^2 \phi -K  
\nabla^2 \phi] \\ \nonumber
&& + \beta_c A_{ij} \partial_i \phi^* \partial_j \phi \ 
\eeq 
where $\beta_c = \beta(1-i)/2$ is a complex coefficient, and the dynamics resembles a conserved complex Ginzburg-Landau equation \cite{aranson2002world}. While the equilibrium and non-chiral active part of the dynamics are invariant  under a global phase transformation  $\phi \to e^{ i \varphi} \phi$, the chiral term $\beta_c$ is not. The lack of this invariance makes the linear stability analysis around this state quite non-trivial.  

\begin{figure}[h]
	\centering
	\includegraphics[width= \linewidth]{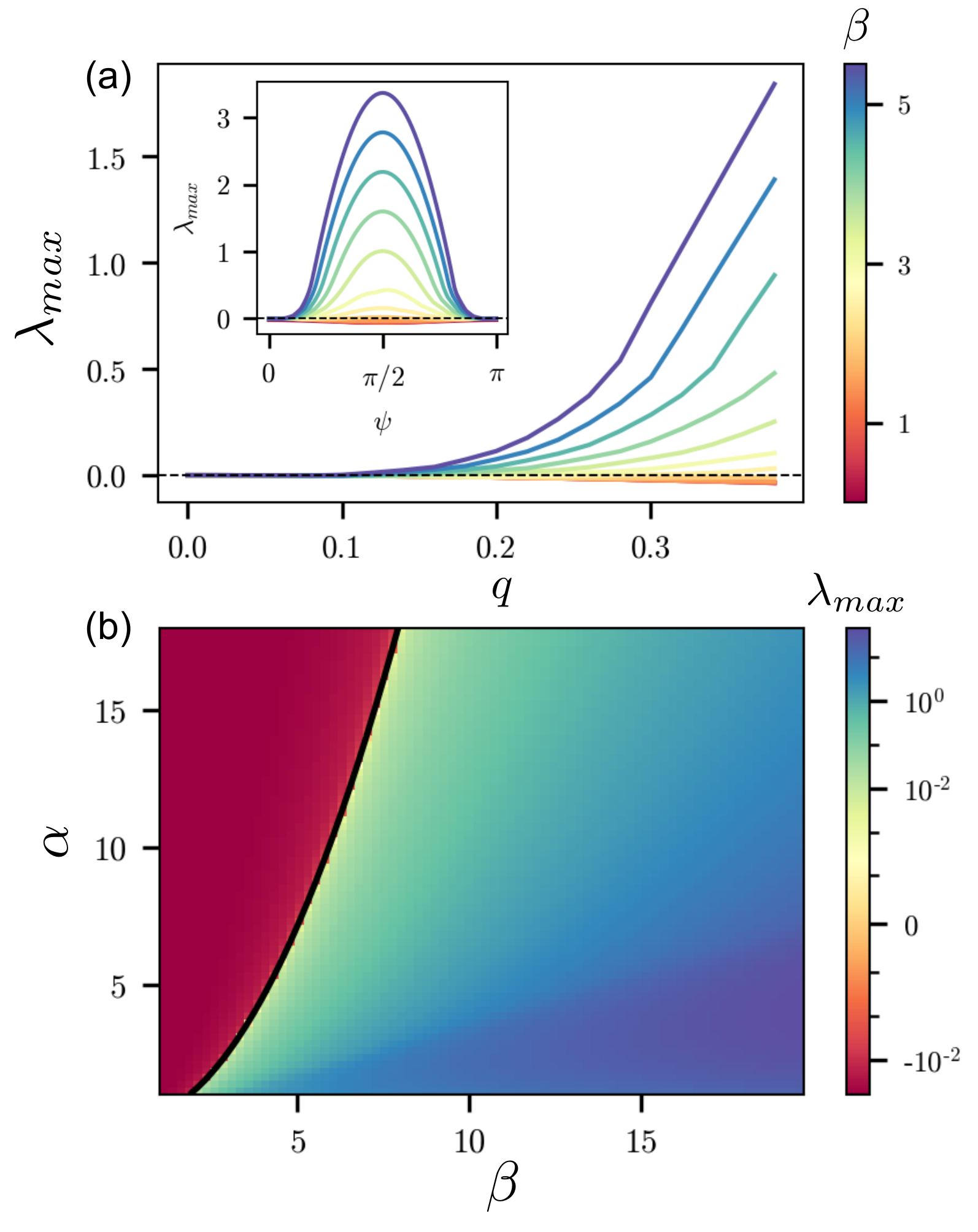}
\caption{\textbf{Linear stability of the traveling wave.} We solve for the largest real eigenvalue of the matrix with $n \in [-N,N]$ and $N=100$. Here $\psi$ is the angle of perturbation, such that $\bm q= q(\cos \psi, \sin \psi)$. (a) Largest eigenvalue at different $\beta$, at $q_0=0.5,\alpha=1$, $K=1$; inset: $q=0.5$, and varying angle $\psi$, $\pi/2$ is the direction of the most unstable mode; main panel, trend in $q$ varying $\beta$ at $\psi=\pi/2, q_0=0.5, \alpha=1$,$K=1$. (b) Largest eigenvalue as a function of $\alpha$ and $\beta$. Large nonreciprocity stabilizes a wave-like solution, while large $\beta$ makes the ordered pattern unstable (parameters: $q=0.5, \psi=\pi/2, q_0=0.5$.)}
	\label{fig:LinStabilityanalysis}
\end{figure}

We know that for $\beta_c = 0$ and large $\alpha$, the binary mixture organizes in traveling waves with wave number $q_0$ and frequency proportional to nonreciprocity \cite{rana2024defect, saha2020scalar}. Indeed, a plane wave $\phi(\bm x,t)= \rho_0 \exp[i \theta_0(\bm x,t)]$ with $\theta_0 = \bm q_0 \cdot \bm x - \omega_0 t$ is an exact solution of the model. This is valid also if the dynamics becomes chiral, preserving the dispersion relation $\rho_0^2 = 1 - K q_0^2$ and $\omega_0= -\alpha q_0^2$ for every $q_0<1$. We can thus investigate how chirality affects the stability of this pattern expanding in linear fluctuations around the traveling state. 

We expand the amplitude as $\rho_0 \to \rho_0+ \delta\rho$ and the phase of the complex field as  $\theta_0 \to \theta_0+ \delta \theta$. In Methods \ref{sec:travelingwaves} we show that the fluctuations in $\rho$ decay with a finite time scale, while those of the phase $\delta \theta$ correspond to the slow mode of the theory \cite{pisegna2024emergent}. We can thus express $ \delta \rho = - K \bm q_0 \cdot \bm \nabla \delta \theta/\rho_0$ and close the linear dynamics only in phase fluctuations:
\beq\label{eq:linTheta}
&& \partial_t \delta \theta = -2\alpha q_0 \partial_\parallel \delta \theta + D_p \partial_\perp^2 \delta \theta + D_l \partial_\parallel^2 \delta \theta \\ \nonumber
&&  + \hat{\beta} (\sin \theta_0 - \cos \theta_0 ) \partial_\perp \partial_\parallel \delta \theta
\eeq 
with $\hat{\beta}= \beta_c K q_0^2/\rho_0$, and diffusion coefficients $D_p=K q_0^2$, $D_l= \gamma K q_0^2$ , $\gamma =(1-3 q_0^2)/(1-q_0^2)$ in directions perpendicular and parallel to $\bm q_0$. If the wave propagates along the direction $\bm q_0 = q_0 \hat{\bm e}$, then the spatial derivatives read $\del_\parallel = \hat{\bm e} \cdot \bm \nabla$ and $\bm \nabla_\perp = \bm \nabla - \hat{\bm e} ( \hat{\bm e} \cdot \bm \nabla)$.  

\begin{figure*}
	\centering
	\includegraphics[width= 0.95\linewidth]{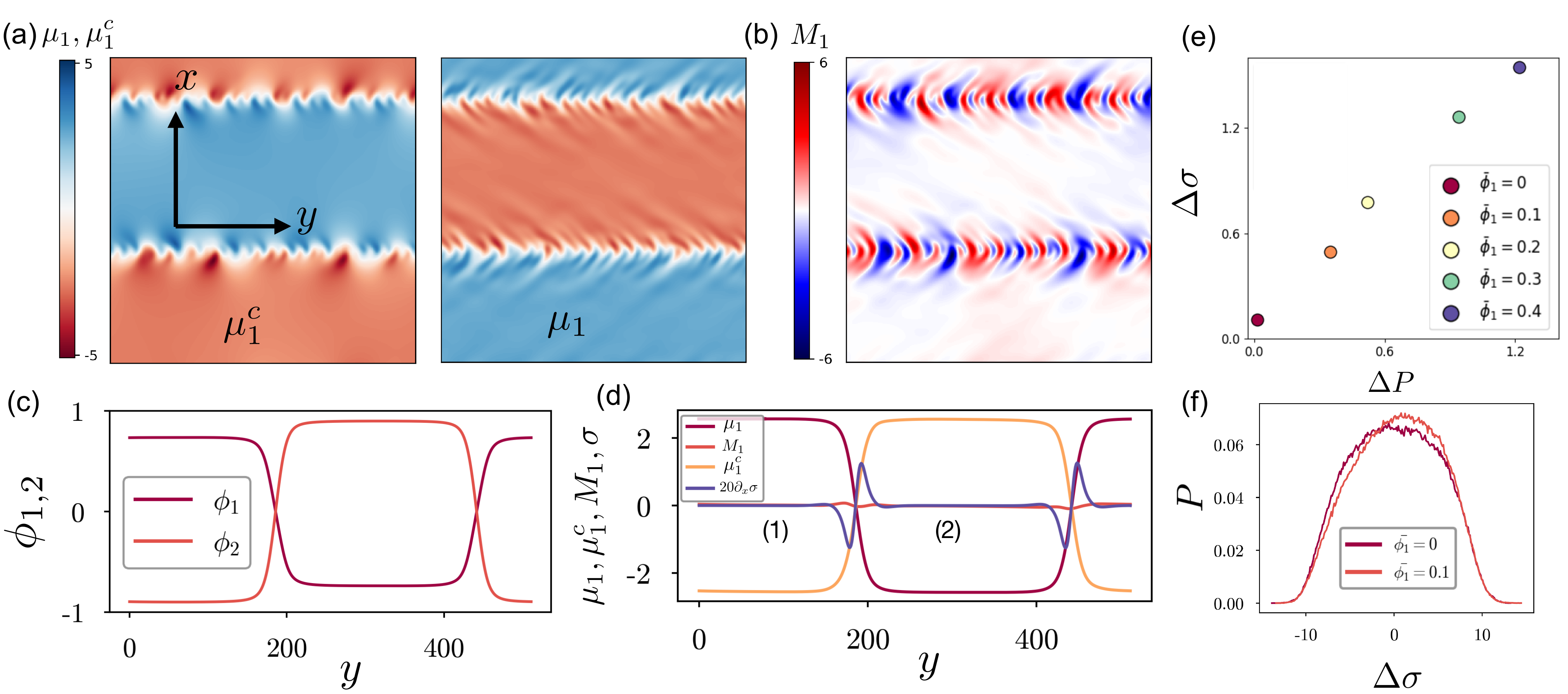}
\caption{\textbf{Chiral phase separation.} Phase separation occurs when the two contributions to the chemical potential sum to a nearly constant value in the bulk, with strong fluctuations at the interface. (a) The non-local and local part of $\mu$ are shown at a fixed instant of time. (b) $M_1$ fluctuates widely particularly at the interface. (c) Time averaged profiles of the fields at steady state. (d) Time averaged profiles of $\mu_{1}$ and $\mu_1^c$ and $M_1$. The time averaged pressure gradient $\partial_x \sigma$ plotted showing the excess pressure around the interface. (e) The average values of $\Delta P$ (difference in the nonrecprocal pressure in the two phases) and $\Delta \sigma$ (stress fluctuations due to chirality) are plotted against one another showing the balance required for phase separation. The calculations are done for $\alpha = 3.25$ and $\beta = 5$. (f) The probability distribution of fluctuations in stresses $\Delta \sigma$ along the interface for compositions shown in the legend demonstrate the large fluctuations.
}
\label{fig:PressureBalance}
\end{figure*}

For $\beta_c=0$ and $\gamma <0$ the planar wave is unstable and undergoes the well-known Eckhaus instability \cite{aranson2002world, cross2009pattern, pisegna2025non}. As we are interested primarily in instabilities appearing due to the chiral term, we assume $\gamma>0$. Using a standard Galilean transform $\bm r  \to \bm r - (\omega_0 t/q_0) \hat{\bm e}$ to change to a reference frame co-moving with the plane wave solution, the explicit time dependences in the chiral terms on the R.H.S. of Eq.\eqref{eq:linTheta} disappear. We can now expand $\delta \theta$ in the Fourier basis~\cite{risken1989fokker},
\beq 
&& \delta \theta( \bm r,t) = e^{\lambda t} \sum_{n=- \infty}^{+\infty} c_n(\bm q) e^{i \bm r \cdot (\bm q + n \bm q_0)} .
\eeq 
Plugging this into Eq.\eqref{eq:linTheta} we get the linear system where the modes $q_{\parallel} \pm n q_0$ are coupled,
\beq 
\label{eq:recurrence}
&& (\lambda + Q_n) c_n + Q_n^+ c_{n+1} + Q_n^- c_{n-1}=0 \\ \nonumber
&& Q_n = i \alpha q_0 (q_\parallel+ n q_0) + D_p q_\perp^2 + D_l(q_\parallel+ n q_0)^2 \\  \nonumber
&& Q_{n}^+= - \frac{\hat{\beta}}{2} (1-i) q_\perp (q_\parallel+ (n+1) q_0) \\ \nonumber
&&  Q_{n}^{-}= -\frac{\hat{\beta}}{2} (1+i) q_\perp (q_\parallel+ (n-1) q_0) \ .
\eeq 
This expression is a tridiagonal recurrence relation of an infinite non-Hermitian matrix that couples adjacent modes. A similar structure is typical of non-linear systems with periodic forcing \cite{risken1989fokker} while, in this case, the periodicity stems from the lack of rotational symmetry in the space of density fields. In Eq.~\eqref{eq:linTheta}, the chiral term appears as an anisotropic diffusion with sinusoidal modulations following the phase of the wave. The Fourier space representation in Eq.~\eqref{eq:recurrence} makes it apparent that the leading order contribution is linear in $q$ meaning that it can compete successfully with the stabilising diffusive terms in Eq.~\eqref{eq:linTheta}.

The real part of the largest eigenvalue $\lambda_{max}$ determines the stability of the patterned solution. Due to the coupling of modes, an analytical computation of $\lambda_{max}$ to obtain a closed expression is not feasible.  We solve it numerically by truncating to mode $N$, or semi-analytically  in terms of continuous fractions (see Methods \ref{sec:travelingwaves}) and summarise the results in Fig.~\ref{fig:LinStabilityanalysis}. The real part of the largest eigenvalue becomes positive for large enough $\beta$, driving a finite wave-number instability that can destroy the pattern. In Methods \ref{sec:travelingwaves}, we compute the wave-number $q_c$ where the system becomes unstable as a function of the parameters $\alpha, \beta$ and the wave-number of the reference pattern $q_0$. The stronger is the chirality, the smaller is $q_c$, asymptotically approaching zero. This instability is largely anisotropic, indeed the most unstable direction is at $\psi=\pi/2$, where $\psi$ is the angle between $\bm q_0$ and the wave-number $\bm q$ of the perturbation (inset of Fig.~\ref{fig:LinStabilityanalysis}(a)). 

Finally, this analysis captures the interplay between linear nonreciprocity and chiral currents in the regime of large $\alpha$. As shown in Fig.~\ref{fig:LinStabilityanalysis}(b), strong nonreciprocal interactions stabilize the moving pattern against the chiral currents. This result is in agreement with the numerical simulations shown in Fig.~\ref{fig:phase_diagram} that, we remind the reader, are performed starting with random initial conditions and with a different equilibrium potential. 
In Methods \ref{sec:travelingwaves}, we present additional simulations of Eq.~\eqref{eq:chiralComplexLG} initialized with banded patterns, which constitute exact solutions of the deterministic equation. To probe their stability, we introduce a small random perturbation by adding weak stochastic noise, following the approach of \cite{saha2020scalar, pisegna2024emergent}. Within the region of the phase diagram identified as unstable by the linear stability analysis (Fig.~\ref{fig:LinStabilityanalysis}(b)), the waves break down, giving rise to either chiral phase separation or disordered dynamics (see movie M4). However, this approach does not allow us to uniquely predict one of these two outcomes, as both are governed by strong non-linear effects.

\subsection{Phase separation with interfacial chiral fluxes}
\label{sec:PhaseSeparation}
For $\beta \gg \alpha$ (see Fig.~\ref{fig:phase_diagram}(g)), we find phase-separation into two bulk domains of constant densities separated by dynamic interfaces. The interfaces carry currents that travel along them, breaking a discrete left-right symmetry. If the final steady state is a round droplet, then the currents travel either clock-wise or counter clock-wise as shown in panels (d-f) of Fig.~\ref{fig:phase_diagram} (movies M2, M3). If the steady state consists of stripes, then the currents can move in opposing directions left or right, see panels (a-b) of Fig.~\ref{fig:PressureBalance} and movie M5. 

As discussed in Section~\ref{Sec:Phenomenology}, the chemical potential $M_a$ consists of a local part $\mu_a$ and a non-local part $\mu_a^c$. Fig.~\ref{fig:PressureBalance}(a) shows the two contributions $\mu_1^c$ and $\mu_1$, and Fig.~\ref{fig:PressureBalance}(b) shows $M_1$ which fluctuates at the interface, but is nearly constant in the bulk. By averaging over sufficient configurations of the fields in the steady state we find that the profiles in the direction perpendicular to the interface resemble classic bulk phase separation with a smoothly varying interface that separates the nearly constant densities, as illustrated in Fig.~\ref{fig:PressureBalance}(c). The profiles of $\mu_{a},\mu_a^c$, and $M_a$ shows that $M_a$ is constant throughout the domain of the simulation, in sharp contrast to the two contributing parts. We emphasise that the concept of equality of chemical potentials holds when applied to the time-average density profiles as seen in Fig.~\ref{fig:PressureBalance}(a-b). As $\alpha$ is increased keeping $\beta$ fixed, this balance is not achieved any longer and the phase separated structure breaks apart.

To explore the similarity between phase separation in $\NRCHs$ and standard bulk phase separation, we now try to find a pressure like quantity in this system. As $M_a$ are constant, we follow the process outlined in \cite{saha2024phase} to obtain a relation that constrains the density profiles in the steady state, see Methods~\ref{sec:phaseComposition} for details.  We multiply $M_1$ with $\phi_1$ and integrate from a point deep within a phase (1) to a point inside the other phase (2) to obtain the relation 
\beq \label{eq:PressureBalance}
\Delta P = \langle \Delta \sigma \rangle,
\eeq 
where $\Delta$ denotes the difference between the two quantities in the two phases. $P$ is associated with the effective free energy of the nonreciprocal dynamics,
\beq \label{eq:NonreciprocalPressure}
&& P = M_2 \phi_2 - M_1 \phi_1 -F_{\rm eff}, \nonumber \\
&& F_{\rm eff} = F_2 - F_1 - \alpha \phi_1 \phi_2,
\eeq 
while $\Delta \sigma$ represents the mean of the chiral contribution
\beq \label{eq:Sigma}
&& \Delta \sigma(x)= \int_1^2 \mbox{d}y  \partial_y \sigma = \int_1^2 \mbox{d} y  \left[ \mu_{2}^c \partial_y \phi_2   - \partial_y \phi_1 \mu_{1}^c \right] .
\eeq 
The balance condition in Eq.~\eqref{eq:PressureBalance} with terms originating from nonreciprocity on the L.H.S. and chirality on the R.H.S. is satisfied whenever we observe bulk-phase separation with fluxes that are restricted to the interface. 

We have verified Eq.~\eqref{eq:PressureBalance} in numerical simulations. $\Delta P$ is calculated using the steady state profile of $\phi_{1,2}$ while $\Delta \sigma$ is calculated by integrating the average profile of $\partial_x \sigma$ between the points (1) and (2) in Fig.~\ref{fig:PressureBalance}(d) and averaging along the interface. For vanishing mean density, i.e. $\bar{\phi}_1 = \bar{\phi}_2 = 0$, we can argue on the grounds of symmetry that bulk phases of $\NRCHs$ are equal and opposite, such that $M_a$ are zero, which further leads to the conclusion that $\Delta P = 0$ (see Methods~\ref{sec:phaseComposition} for details). On changing the average composition, $\Delta P$ is no longer zero but is balanced by the active Laplace pressure that develops at the interface. $\Delta \sigma$ (or $\Delta P$) fluctuates in the steady state, as seen also in the probability distribution function in Fig.~\ref{fig:PressureBalance}(f).

An important difference between conventional phase separation and the one we observe here is that $M_a$ receives contributions from both nonreciprocity and chirality meaning that it is not possible to obtain the compositions in the two phases from a convex-hull construction on the effective free energy connected to the nonreciprocal Cahn-Hilliard model \cite{saha2020scalar, you2020nonreciprocity}, although $\Delta P = 0 $ for vanishing mean density. The reason lies in the non-locality of the chiral contribution to the effective chemical potential.

\subsection{Edge waves}
\noindent
As discussed in Sec.~\ref{sec:PhaseSeparation}, in the chiral phase separated state of Figs.~\ref{fig:phase_diagram}(d, f) the fields $\phi_{1,2}$ coexist within a droplet. The interior and exterior of the droplet are separated by well defined interfaces for both fields, and a complex interaction between them gives rise to chiral edge waves. To study this phenomenon, we  move to the reference frame of the mean interfacial profiles and we inspect the dynamics of the height fields $h_a$ that describe small deviations from the mean. At relevant timescales, we expect the profiles which we assume are in the $y$-direction to remain unchanged, while heights fluctuate in the transverse $x$-axis. Therefore, the density fields are written as 
\beq 
\label{eq:profiles_inter}
&& \phi_1 = f(y-h_1(x,t)), \nonumber \\
&& \phi_2 = g(y-h_2(x,t)),
\eeq 
where $f,g$ are the interfacial profiles that interpolate between the bulk values. In the repulsive case, as in Fig.~\ref{fig:PressureBalance}(c), \ref{fig:phase_diagram}(d,f), $f$ and $g$ are of opposite sign. To proceed analytically we further consider a double-well potential $V_d$ such that,  following \cite{bray2001interface}, we can rewrite the dynamics for $\phi_a$ in terms of $h_a$ using a sharp-interface limit. Introducing the variables $u= y - h_1$ and $u_2 = y - h_2 = u + h_1 - h_2$, we obtain for the first field,
\beq 
\nonumber
 && (- \nabla^2)^{-1}[f'(u) \partial_t h_1 + \beta_1 f'(u)g'(u) \partial_x \Delta h]  = \\
  && -K[(1+|\nabla h_1|^2) f''(u) - f'(u) \nabla^2 h_1]  \\ \nonumber
&& + V_1'(f)+\alpha[ g(u)+g'(u)\Delta h]
\eeq 
where $\Delta h = h_1 - h_2$ and we are expanding the nonreciprocal and chiral terms using $g(y- h_1 + \Delta h) = g(u) + \partial_u g(u) \Delta h$. We project on the interface and integrate along it, such that we express the results in terms of the partial Fourier transform $\tilde G(q_x,y) = \frac{1}{2 |q_x|} \int dy' \ \exp{(- |q_x||y-y'|)} \tilde F(q_x,y')$, where $G$ and $F$ solve $G(x,y)= (-\nabla^2)^{-1}F(x,y)$ in real space.
A detailed calculation of these passages, and analogous steps for the second field, are given in Methods \ref{sec:interfaces}.

The final linear equations for the coupled interfaces fluctuations are,
 \beq 
 \label{eq:interfaces_fluc}
 \nonumber
 && \partial_t \tilde h_1(q_x,t) = (- \bar \beta_1 i q_x + 2 \bar{\alpha} |q_x| )\Delta \tilde h - 2 \bar K |q_x|^3 \tilde h_1  \\ 
 && \partial_t \tilde h_2(q_x,t) = (\bar \beta_2 i q_x + 2 |q_x| \bar \alpha )
 \Delta \tilde h - 2 \bar K |q_x|^3 \tilde h_2 
 \eeq
The barred parameters are proportional to the integrated contributions on the interface, for instance indicating the surface tension $\bar K = K \int du [f'(u)]^2/ \int du \int dv \ \exp(-|q_x||u-v|) f'(u)f'(v)$; the rest are fully reported in the Methods \ref{sec:interfaces}. Note that the chiral coefficients $\bar{\beta}_i$ represent advection of the height fields along the interface while the nonreciprocal coefficient $\bar{\alpha}$ does not.

\begin{figure}[t]
  \centering
	\includegraphics[width=\linewidth]{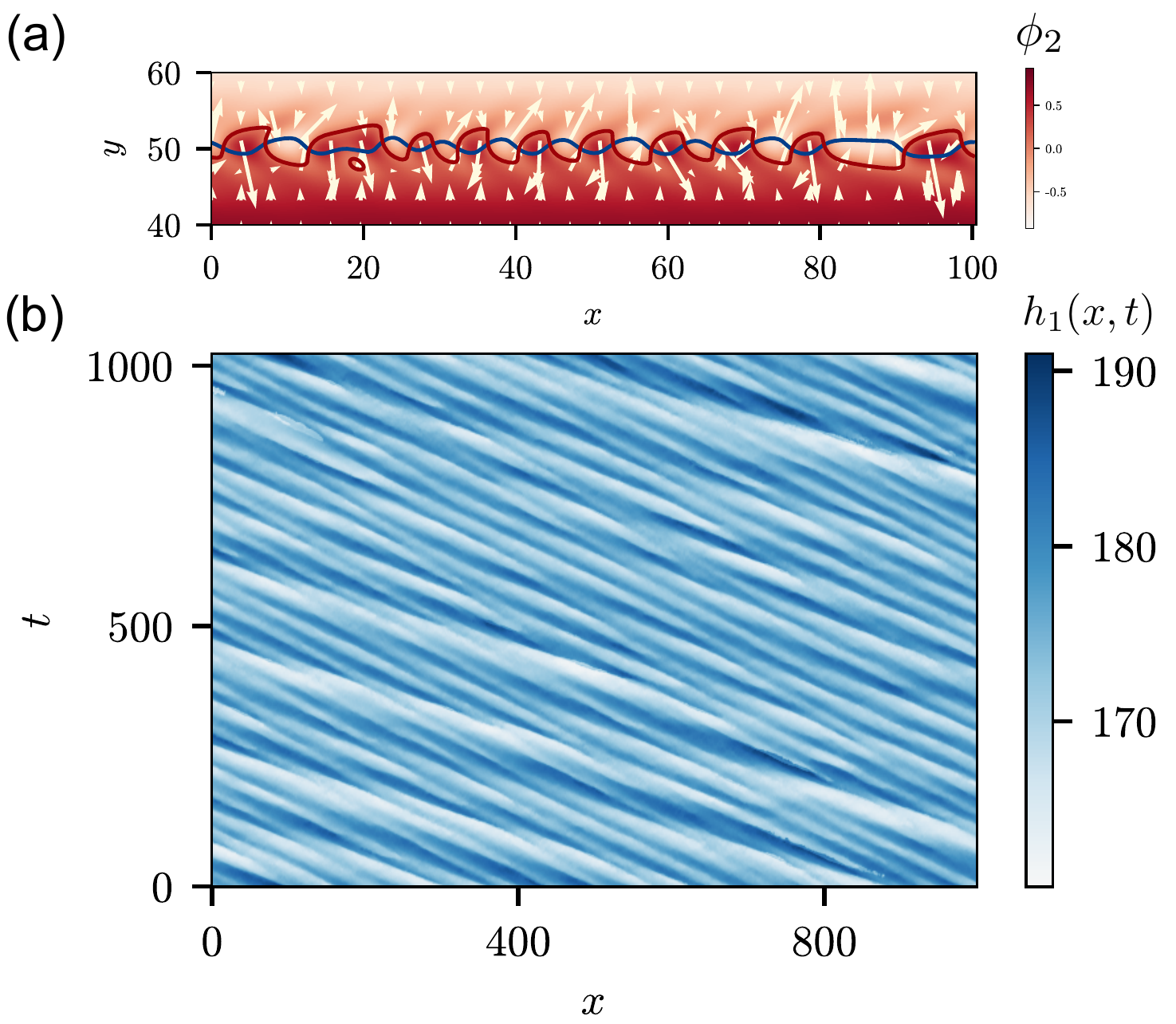}
\caption{\textbf{Interfaces.} (a) Closeup on propagating interfaces in a chiral phase separated state. The red line signals the interface of the field $\phi_2$, the dark blue line of the field $\phi_1$, and the vector field is the polar order parameter $\bm J$. Parameters of numerical simulation: $\alpha=1, \beta=10$, Mexican-hat potential. (b) Kymograph of interface fluctuations of the field $\phi_1$, with double-well potential.
}
\label{fig:interfaces}
\end{figure}

A linear stability analysis of Eq.~\eqref{eq:interfaces_fluc} reveals a completely diffusive mode with eigenvalue $\lambda_1 = - 2 \bar K |q_x|^3$ and eigenvector $\bm v_1 =(1,1)$. The second mode propagates while still diffusing, and is given by
\beq 
&& \lambda_2 = - 2 \bar K |q_x|^3 - i(\bar \beta_1+ \bar \beta_2) q_x \\ \nonumber
&& \bm v_2 = (\rho e^{i \varphi},1) \\ \nonumber
&& \rho^2 =\frac{ 4 \bar \alpha^2 + \bar \beta_1^2}{ 4 \bar \alpha^2 + \bar \beta_2^2} \quad \tan \varphi = \frac{2 \bar \alpha ( \bar \beta_1 + \bar \beta_2) \text{sign} (q_x)}{\bar \beta_1 \bar \beta_2- 4 \bar \alpha^2} \ .
\eeq 
The imaginary part of the eigenvalue describes the propagation of the edge currents with speed $ v_g = (\bar \beta_1+ \bar \beta_2)$ determined by the chiral parameters. A combination of nonreciprocity and chirality induces a phase shift $\varphi$ between the interfacial profiles of the two fields clearly visible in Fig.~\ref{fig:interfaces}(a), breaking the discrete symmetry necessary for propagating waves that chase each other. 

It is interesting to notice that if $\bar{\beta}_1 = - \bar \beta_2$, this calculation seems to predict an absence of propagation at the edge of the phase separating system, even though microscopic particles are chiral. In fact, $v_g$ is computed under the assumption that interface profiles are repulsive (namely $g=-f$, see Methods \ref{sec:interfaces} for explicit expressions of $\bar{\beta}_a$). In the case of opposite chirality, the system spontaneously organizes in phase separating attracting droplets ($g=f$), thus  generating emergent large scale chirality, also when $|\bar \beta_1| \neq |\bar \beta_2|$ (see movies M3a, b). Therefore, mixtures with the same handedness interact repulsively producing profiles shown in Fig.~\ref{fig:phase_diagram}(d,e) where one species is enhanced in the phase were the second is depleted. Those with opposite handedness are effectively attractive with both the species coexisting in the same phase. These predictions match the phenomenology observed in numerical simulations of Eq.~\eqref{eq:chiralComplexLG} with a Mexican-hat potential reported in Fig.~\ref{fig:interfaces}. 

Observing the vector polar field $\bm J$ at the edge of the phase separated state, we can provide an heuristic argument to rationalize the role of nonreciprocity in this dynamics: when the two profiles acquire a phase-shift with modulations of the interfaces, the field $\phi_2$ chases locally $\phi_1$, generating sources of vorticity of $\bm J$ (see Figs.~\ref{fig:phase_diagram}(g) and \ref{fig:interfaces}). This interaction thus induces a local transport of mass, which is redistributed along the $x$-direction through the chiral interaction $\beta$ and the consequent active Laplace pressure. 

Using the same representation as before, we can finally express the non-local chiral chemical potential as,
\begin{equation}
\tilde \mu_a^c(q_x,y) = \frac{\beta_a}{2 |q_x|} \int dy' e^{-|q_x||y-y'|}f'(y') g'(y)i q_x \Delta \tilde h
\end{equation}
and thus obtain,
\beq 
\Delta P(x) = \frac{i (\bar \beta_1+ \bar \beta_2)}{2} \int \frac{d q_x}{2 \pi} e^{i q_x x}\text{sign}(q_x) \Delta \tilde h,
\eeq 
directly connecting the difference of height fluctuations to the variation of pressure along the interface. In line with the conclusions of the previous section~\ref{sec:PhaseSeparation}, this relation confirms that, on average along the $x$-axis, the pressure difference is zero.

\section{Discussion}
In this work, we studied a field-theoretical model describing two interacting density fields of chiral particles with nonreciprocal interactions. We demonstrated that these ingredients are generic in multispecies systems, as they naturally arise in coarse-grained descriptions of chiral chemical colloids or quorum-sensing active particles with nonreciprocal interactions (see Results and Methods \ref{MethodsA}). The resulting mesoscopic theory, the $\NRCHs$ model, is characterized by three essential elements: (i) phase separation that creates inhomogeneities in the density fields; (ii) nonreciprocal couplings that produce local phase shifts between them described by a nonreciprocal current $\bm{J}$ (Fig.~\ref{fig:schematic}(b)); and (iii) chirality, that locally twists $\bm{J}$ (Fig.~\ref{fig:schematic}(c)). The interplay between them generates new active phase separated states summarised in Fig.~\ref{fig:phase_diagram} that we investigated both analytically and numerically. 

Active phase separation has been extensively studied, with much of the work focusing on single-species active systems \cite{tjhung2018cluster, tjhung2011nonequilibrium, fausti2021capillary, solon2018generalized, zwicker2015suppression},  whereas large-scale chiral active matter has drawn interest more recently~\cite{kole2021layered, maitra2025activity}. Generalisation to a two-species mixture marks a new development: chiral contributions involve terms with  two gradients appearing in the dynamics with the same level of importance as nonreciprocity. 
Indeed, they fundamentally modify the nonreciprocal dynamics of the mixture, producing disordered chiral states and chiral phase separated regions with edge currents when chirality is dominant. 

The phenomenology of $\NRCHs$ summarised in Figs.~\ref{fig:phase_diagram}-\ref{fig:coarsening} stands out from the current scenarios of conserved scalar field theories \cite{cates2025active}. Concentrating on the role played by chirality, we show that it emerges as the non-local contribution to the chemical potential, and it is localized in domains where the curl of the nonreciprocal current is large. These domains act as charges of positive and negative signs, thereby preserving the conservation law that characterizes the system. The resulting charges can either be dispersed randomly across the simulation domain, characteristic of the disordered chiral state, or can be arranged linearly along an interface, typical of the phase-separated state as shown in Fig.~\ref{fig:coarsening}(b). 
In this configuration, the alternating signs of these charges effectively screen the long-range nature of chirality, allowing it to counterbalance the role of nonreciprocity in driving phase separation. This balance is quantified through the condition expressed in Eq.~\eqref{eq:PressureBalance}, and its validity is verified in numerical simulations in Fig.~\ref{fig:PressureBalance}. The scenario can be contrasted with the dynamics in Active Model B+, where active currents can be re-expressed as the gradient of a nonlocal chemical potential, which retains its long-ranged nature leading to the selection of a  length-scale and dynamical arrest \cite{tjhung2011nonequilibrium}.  Furthermore, the picture is reminiscent and yet distinct from the semi-classical quantum-hall effect where quantised orbits with a fixed sense of rotation cancel in the bulk producing edge currents~\cite{Cohen2019}, with significance in active matter~\cite{tang2021topology,SriramChiral2025}. In our system, the effect of chirality vanishes in the bulk due to the pairwise cancellation of charges which survive only at the boundary where the chasing dynamics prevents their mutual-annihilation.  

When nonreciprocity dominates over chirality, this cancellation breaks down, leading to a proliferation of charges and the emergence of a chiral disordered state. Unlike other bulk disordered phases, such as bacterial turbulence, where microscopic chirality is lost at large scales, this state retains charges of both chiralities at a well-defined characteristic length scale.

Our theoretical predictions offer a promising route toward experimental verification in mixtures of active agents with intrinsic chirality. Such systems provide a natural playground for testing our ideas, with potential implications ranging from uncovering new principles of self-organization in biological collectives \cite{beppu2021edge} to designing novel functionalities in synthetic chiral metamaterials \cite{ChiralMetamaterial_PhysRevB.79.035407,souslov2017topological}. Beyond these immediate perspectives, our model also opens several exciting avenues for theoretical exploration. A natural step forward is the further analysis of stochastic noise and non-linearities, which would allow us to probe the universality class of the model both at criticality and in the vicinity of traveling-wave states, an extension that is expected to refine, and possibly overturn, current predictions \cite{pisegna2024emergent}. Another important direction is the incorporation of hydrodynamic interactions \cite{pisegna2025non, tucci2025hydrodynamic}, which are known to play a crucial role in the dynamics of both living \cite{tan2022odd} and synthetic chiral systems \cite{chen2025self}. Together, these developments would not only bring the model closer to experimental realizations but also deepen our understanding of how chirality and nonreciprocity conspire to drive emergent collective behavior across a wide class of active matter systems.

\section{METHODS}

\renewcommand{\thefigure}{S\arabic{figure}}
\setcounter{figure}{0}
\subsection{Microscopic models of chiral active mixtures}
\label{MethodsA}

We show that the chiral currents of Eq.~\eqref{eq:ChiralCurrent} arise naturally in a multicomponent system of intrinsically spinning interacting particles when the orientation field is eliminated retaining only the conserved density fields. An \emph{odd} rotational diffusivity naturally leads to the contributions listed in the previous section. We discuss two different systems, a chemically interacting system and a quorum sensing system. Furthermore, using chemical colloids, we demonstrate that chiral currents can arise also from the passive response of non-axisymmetric particles to external gradients.
Unless explicitly stated otherwise, the Stratonovich convention for stochastic calculus is assumed throughout the coarse-graining calculations.

\subsubsection{Chiral bacteria with quorum sensing interactions} \label{sec:ChiralBacteris}
\noindent
Consider a binary system of active particles in two dimensions with quorum sensing interactions. In addition, they spin with angular speed $\omega_a$ about a body-fixed axis that points away from the two dimensional plane to which they are constrained. Their microscopic dynamics can be described as a multi-species generalisation of chiral active Brownian motion \cite{volpe2014simulation},
\beq \label{eq:chiralABP}
&& \dot{\bm r}_a^\gamma = v_a(\{\rho\}) \bm n_a^\gamma + \boldsymbol \zeta_a^\gamma\\ \nonumber
&& \dot{\bm n}_a^\gamma = \bm n_a^\gamma \times (\omega_a \hat{\bm{z}} + \bm \xi_a^\gamma)
\eeq 
The speed of self-propulsion $v_a$ of species $a$ in Eq.~\eqref{eq:chiralABP} is determined by the coarse-grained density fields of the others. The noise terms are assumed white and Gaussian with variances, $\langle \bm \zeta^\gamma (\bm r,t) \cdot \bm \zeta^\nu (\bm r',t') \rangle = 2d D_a \delta_{\gamma \nu} \delta^d(\bm r-\bm r') \delta (t-t')$ and $\langle \bm \xi^\gamma (\bm r,t) \cdot \bm \xi^\nu (\bm r',t') \rangle = 2d D_{ra} \delta_{\gamma \nu} \delta^d(\bm r-\bm r') \delta (t-t')$.  Without chirality, the model in Eq.~\eqref{eq:chiralABP} was introduced and studied in \cite{dinelli2023non,duan2023dynamical}, showing traveling bands, spirals and chaotic chasing phases.  Nonreciprocal interactions are encoded in the $v_a$, and specifically in $\eta_{ab} = \rho_a \partial \ln v_a/\partial \rho_b$ whose sign determines if species $a$ slows down or goes faster in $b$-rich regions. Nonreciprocity emerges when $\eta_{ab} \rho_b\neq \eta_{ba} \rho_a$, see \cite{dinelli2023non,duan2023dynamical} for details. 

Here we are interested in the contribution to the mesoscopic theory given by the chiral terms. Performing a systematic coarse-graining procedure similar to \cite{duan2023dynamical}, and introducing the mesoscopic fields, $\rho_a(\bm r,t) = \langle \sum_\gamma \delta^d(\bm r - \bm r_a^\gamma) \rangle$ and $\bm p_a(\bm r,t)= \langle \sum_\gamma \bm n_a^\gamma  \delta^d(\bm r - \bm r_a^\gamma) \rangle$, we find  
\beq 
&& \partial_t \rho_a = - \bm \nabla \cdot [ v_a(\{\rho\}) \bm p_a - D_a \bm \nabla \rho_a] \\ \nonumber
&& \partial_t \bm p_a = - \mathbb{D}_{ra} \cdot \bm p_a - \frac{1}{2} \bm \nabla [v_a(\{ \rho \}) \rho_a]
\eeq 
with $\mathbb{D}_{ra} = 2 D_{ra} \mathbb I + \omega_a \mathbb A$. The angular speed modifies the rotational diffusivity adding a \emph{odd} contribution to it. We should of course keep in mind that the $\omega_a$ is due to persistent spinning while $D_{ra}$ arises due to noise. If rotational diffusion is fast, we can then work in a quasi-stationary limit such that,
\beq 
\bm p_a = - \frac12 \mathbb D_{ra}^{-1} \cdot \bm \nabla [ v_a \rho_a]
\eeq 
and substituting it into the density's dynamics we get
\beq 
\label{eq:qs_coarse}
\nonumber
&& \partial_t \rho_a = D_a \nabla^2 \rho_a  + \hat{D}_{ra} \bm \nabla \cdot [ v_a \cdot \bm \nabla (\rho_a v_a) ]  \\ \nonumber
&& - \hat{\omega}_a \bm \nabla \cdot [ v_a \mathbb A \cdot \bm \nabla (\rho_a v_a)]
\eeq 
with $\hat{D}_{ra} = D_{ra}/ (2 \det \mathbb D_{ra})$ and $\hat{\omega}_a= \omega_a/ (2 \det \mathbb D_{ra})$. 

A useful model for quorum-sensing interactions expresses the particle speed through a logistic function, $v_a(\{\rho\}) = v_0 L_{aa}(\rho_a) L_{ab}(\rho_b)$, with $L_{ss'}(\rho_s) = 1+ \kappa \tanh (\eta_{ss'}(\rho_s - \rho_0)/\kappa \rho_0)$  \cite{duan2023dynamical, dinelli2022self}. Using this, we  expand Eq.~\eqref{eq:qs_coarse} in fluctuations $\phi_a = \rho_a- \rho_0$ and $\phi_b = \rho_b- \rho_0$ deriving,
\beq 
&& \partial_t \phi_a = \bar{D}_a \nabla^2 \phi_a + v_0^2 \hat{D}_{ra} \eta_{ab}\nabla^2 \phi_b  \\ \nonumber
&& + \hat{\omega}_a v_0^2 \frac{\eta_{ab}}{\rho_0 }\bm \nabla \phi_a \cdot \mathbb A \cdot \bm \nabla \phi_b
\eeq 
with $\bar{D}_a = D_a + v_0^2(1+\eta_{aa})$. We can see how linear nonreciprocity emerges, and how it combines with chirality to produce the multispecies chiral current. 
Of course, we kept only terms relevant for our physical description but additional higher order terms are produced.

\subsubsection{Spinning catalytic colloids} \label{sec:SpinningColloids}
\noindent 
Next, we consider a system composed of two species of chemically active Janus particles which are coated with catalysts in an axisymmetric pattern  about a unit vector $\bm{n}_a^{\gamma}$. Particle $\gamma$ of species $a$ with position $\br_a^\gamma$ and orientation $\bn_a^\gamma$ evolves as
\beq 
\label{eq:chemicalpos}
\dot{\br}_{a}^{\gamma} &=& - \mu_a \bm{\nabla} c + v_a \bn_{a}^{\gamma} + \bm{\zeta}^\gamma , \\ 
\label{eq:oneaxisn}
\dot{\bn}_{a}^{\gamma} &=& \bm{n}_a^{\gamma} \times \left[ \bm{R}^{\gamma}_a + \bm{\xi}^\gamma \right],
\eeq 
where $\bm{R}^{\gamma}_a$ is the angular velocity of the $\gamma$-th particle. The chemical field $c$ is generated by chemical activity on the surface of the particles
\beq
&& \partial_t c - D_c (\nabla^2 - \kappa^2) c  = \sum_{a,\gamma} [ \sigma_a \delta^{d}(\bm r- \bm r_a^\gamma) \\ \nonumber
&& + \bar{\sigma}_a \bm{n}^{\gamma}_a \cdot \bm{\nabla} \delta(\bm r - \bm r_a^{\gamma}) ]
\eeq
and $\sigma_a, \bar \sigma_a$ are the surface activities of the colloids producing or consuming the chemical $c$. $\kappa$ is the inverse of a screening length scale describing spontaneous degradation \cite{tucci2024hydrodynamic}. Interactions between the particles are driven by their chemotactic response  to gradients created autonomously. The colloids move with a velocity determined by the diffusiophoretic mobility $\mu_a$.  

The particle $\gamma$ aligns with the local chemical gradient if $\bm{R}^{\gamma}_a = 0$. By tailoring $\bm{R}^{\gamma}_a$ suitably it is possible to tune a stable fixed point of the orientational dynamics such that $\bm{R}$ vanishes at a given inclination between $\bm{\nabla}c$ and $\bm{n}^{\gamma}_a$.

A realisation of this case is when particles spin at constant frequency $\omega_a$, and chemotactically align with strength $\Omega_a$, such that
\beq 
\bm{R}^{\gamma}_a = -\Omega_a (\bm{n}_{a}^\gamma \times \bm{\nabla}c) - \omega_a \hat{\bm{z}} .
\eeq  
Performing a coarse-graining of this system \cite{tucci2024nonreciprocal, tucci2025hydrodynamic} we get,
\beq\label{eq:cg_densityEq}
\partial_t \rho_a = -\bm{\nabla} \cdot [( v_a \bm{p}_a-\mu_a  \rho_a \bm{\nabla} c  - D_a \bm{\nabla}  \rho_a)]
\eeq 
and 
\beq 
\partial_t \bm p_a = - \mathbb D_{ra} \cdot \bm p_a - v_a \bm \nabla \rho_a + 2 \Omega_a \rho_a \bm \nabla c + \mathcal O(\nabla^2)
\eeq 
where, also in this case, we identify an effective odd rotational diffusivity,
\beq
{\mathbb{D}}_{ra} = 2 D_{ra} \mathbb{I} + \omega_a \mathbb{A}.
\eeq
As shown in \cite{saha2014clusters,tucci2024nonreciprocal}, the orientational dynamics does not show pattern formation in the large-scale limit therefore, at time scales larger than $D_{ra}^{-1}$, it is reasonable to assume that orientations' relaxation can be enslaved to the dynamics of the densities through the relation 
\beq 
\bm p_a =  {\mathbb{D}}_{ra}^{-1} \cdot \left[ 2 \Omega_a \rho_a \bm{\nabla} c - \bm{\nabla} \rho_a v_a \right],
\eeq 
which we can write explicitly into two parts 
\beq\label{eq:Nparts}
&& \bm p_a = \hat{D}_{ra} \mathbb I \cdot  \left[ 2 \Omega_a \rho_a \bm{\nabla} c - \bm{\nabla} \rho_a v_a \right] \\ \nonumber
&& -  \hat{\omega}_{a} \mathbb{A} \cdot \left[  2 \Omega_a \rho_a \bm{\nabla} c - \bm{\nabla} \rho_a v_a  \right]
\eeq 
with $\hat{D}_{ra}= 2 D_{ra}/\det \mathbb{D}_{ra}$, and $\hat{\omega}_a= \omega_a/ \det \mathbb{D}_{ra}$.
Considering length scales $ l \gg 1/\kappa$, we can approximate $c(\bm r,t) \simeq \sum_a (\sigma_a \rho_a)/(D_c \kappa^2)$. Substituting this expression into Eq\eqref{eq:Nparts} and Eq\eqref{eq:cg_densityEq} we find,
\beq 
\label{eq:chemicalrhoa}
&& \partial_t \rho_a = \bar{D}_a \nabla^2 \rho_a  + \sum_b \chi_{ab} \bm \nabla \cdot (\rho_a \bm \nabla \rho_b) \\ \nonumber
&& + \sum_b \beta_{ab} \bm \nabla \cdot \mathbb A \cdot (\rho_a \bm \nabla \rho_b) \\
&& \bar{D}_a = D_a + v_a^2 \hat{D}_{ra}\\
&& \chi_{ab} = \left( \mu_a - 2 v_a \hat{D}_{ra} \right) \frac{\sigma_b}{D_c \kappa^2} \\
&& \beta_{ab} = \frac{2 \hat{\omega}_a \Omega_a v_a \sigma_b}{ D_c \kappa^2}
\eeq 
The last term of Eq\eqref{eq:chemicalrhoa} is the chiral axial current Eq\eqref{eq:ChiralCurrent}. From this derivation we can see how nonreciprocity and effective chirality stem from a phoretic non-equilibrium dynamics. Moreover the same chiral term $\beta_{ab}$ can be apriori nonreciprocal, i.e. $\beta_{ab} \neq \beta_{ba}$.

\subsubsection{Chiral chemotaxis of catalytic colloids} \label{sec:ChiralChemotaxis}
In the examples above, chirality acts as an intrinsic source of activity, manifesting as a constant rotation of the particle’s polarity axis. By contrast, using chemical colloids as an example, we show that chirality can also arise passively as a response to external concentration gradients.

For this, we need to consider a particle $\gamma$ of species $a$ with two axis of polarization $\bm n_{a,1}^\gamma$ and  $\bm n_{a,2}^\gamma$, separated by a fixed angle $\theta_0$. Moreover, we associate to this the two first moments of the Laplace expansion of surface mobility $\mu_{1,1}$ and $\mu_{1,2}$, whose non-null values usually produce terms of alignment of the polarization to $\bm \nabla c$ \cite{saha2019pairing, tucci2025hydrodynamic}.  
This configuration is achievable building a catalytic particle with mobility patches non-axisymmetric on its surface \cite{saha2019pairing, lisicki2018autophoretic}. Using $\bm n_{a,2}^\gamma=(\mathbb I \cos \theta_0 + \mathbb A \sin \theta_0) \cdot \bm n_{a,1}^\gamma $, we derive the dynamics of the orientational vector,
\begin{equation}
\label{eq:twoaxisn}
 \frac{d \bm n_{a,1}^\gamma }{dt} = \bm n_{a,1}^\gamma \times [ \Omega_a \bm \nabla c \times \bm n_{a,1}^\gamma  +  \bar{\Omega}_a \bm \nabla c \times (\mathbb A \cdot \bm n_{a,1}^\gamma) + \bm \xi^\gamma ] 
\end{equation}
with $\Omega_a = 3 (\mu_{1,1} + \mu_{1,2} \cos \theta_0)/ (8R)$ and $\bar \Omega_a = 3 ( \mu_{1,2} \sin \theta_0)/ (8R)$, $R$ is the radius of the particle. Coarse-graining Eq.~\eqref{eq:twoaxisn} with the positions evolution Eq.~\eqref{eq:chemicalpos} we obtain,
\beq 
&& \partial_t \rho_a = - \bm \nabla \cdot [ v_a \bm p_a - \mu_a \rho_a \bm \nabla c - D_a \bm \nabla \rho_a] \\ \nonumber
&& \partial \bm p_a = - 2 D_{ra} \bm p_a - v_a \bm \nabla \rho_a + (\Omega_a \mathbb I - (\bar \Omega_a/3) \mathbb A ) \cdot \bm \nabla c \ \rho_a
\eeq 
Thus, from the latter 
\beq 
\bm p_a = \frac{1}{2 D_{ra}} [- v_a \bm \nabla \rho_a + (\Omega_a \mathbb I - ( \frac{\bar \Omega_a}{3} \mathbb A ) )\cdot \bm \nabla c \rho_a] 
\eeq 
and finally,
\beq 
\nonumber
&& \partial_t \rho_a = (D_a+ v_a^2/ 2 D_{ra}) \nabla^2 \rho_a + (\mu_a - \frac{v_a \Omega_a}{2 D_{ra}}) \bm \nabla \cdot ( \rho_a \bm \nabla c) \\ 
&& + \frac{v_a \bar \Omega_a}{6 D_{ra}} \bm \nabla \cdot ( \mathbb A \cdot \bm \nabla c \rho_a) .
\eeq 
Following the same steps as before, we obtain the chiral currents studied in the main text from the last term of this expression. 

\begin{figure}[t]
    \centering
    \includegraphics[width=\linewidth]{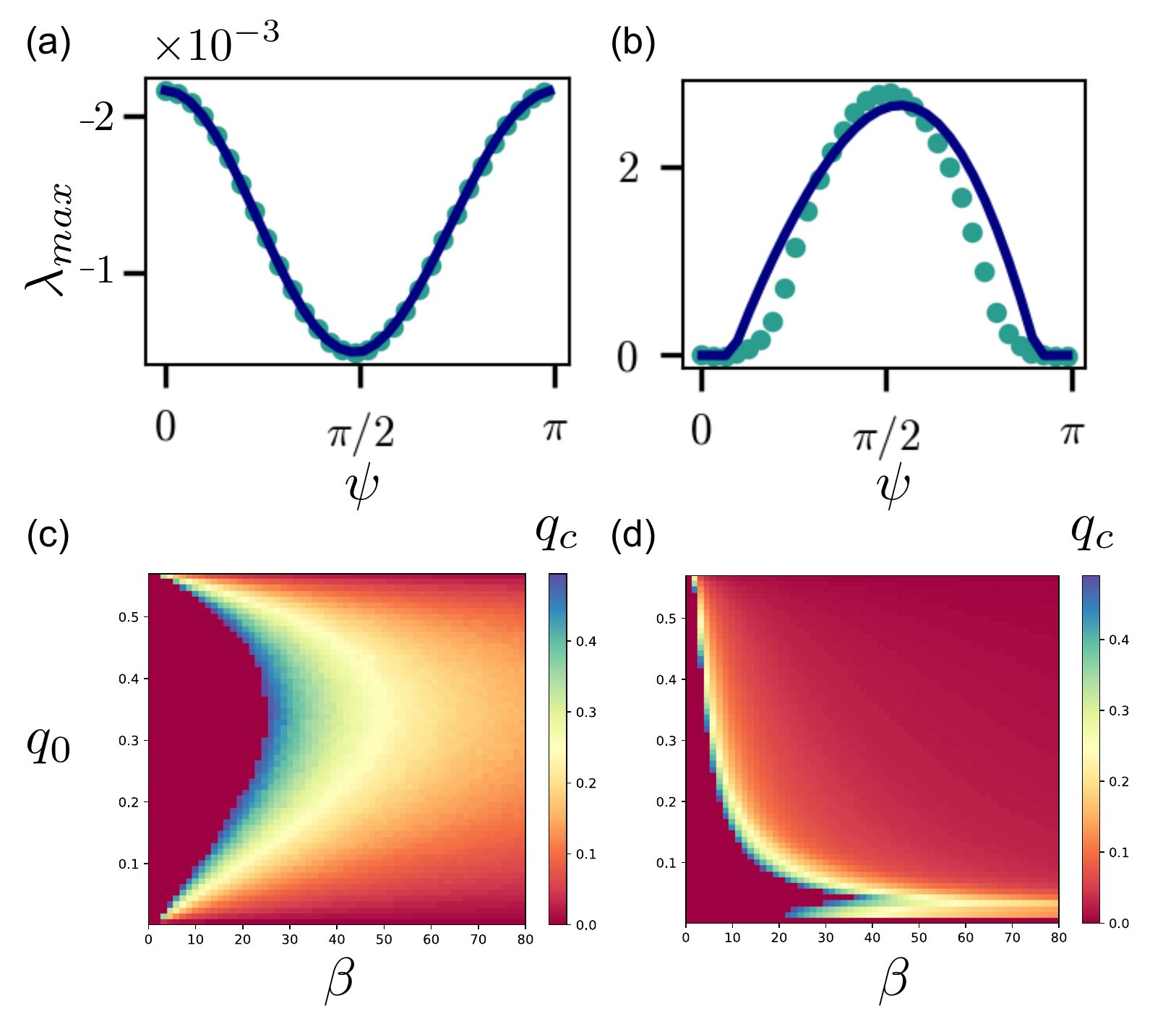}
    \caption{ \textbf{Tridiagonal recurrence relation.} Top: comparison of real largest eigenvalue between direct numerical evaluation of eigenvalue of the matrix (points), and computation with continuous fraction (solid line). Parameters: $N=100,7$ for first method and second method respectively. Panel (a), $
    \beta=1, q_0=0.1, \alpha=1,K=1$, panel (b),$
    \beta=5, q_0=0.5, \alpha=1,K=1$. Bottom: evaluation of $q_c$, wave-number at which the eigenvalue becomes positive as a function of $q_0$ and $\beta$, for $\alpha=0$ (c) and $\alpha=1$ (d). At a given $q_0$, $q_c$ decays to a vanishing value in the limit of very large $\beta$.  }
    \label{fig:recurrence_qc}
\end{figure}

\subsection{Traveling waves as initial condition}
\label{sec:travelingwaves}

In this section, we present the details of the analytical calculations used to linearize the model in Eq. \eqref{eq:chiralComplexLG} around a traveling-wave solution. We then describe an additional semi-analytical method employed to solve the associated tridiagonal recurrence relation Eq. \eqref{eq:recurrence}, which allows us to assess the stability of this solution. Finally, we report numerical simulations of the model, initialized with a striped pattern and subject to weak stochastic noise in the presence of a Mexican-hat potential.

\subsubsection{Linearized equations}

A traveling-wave in polar representation of the density fields reads $\phi_0(\bm x,t)= \rho_0 \exp i \theta_0(\bm x,t)$,
with phase $\theta_0 = \bm q_0 \cdot \bm x - \omega_0 t$.  This is a solution of Eq.\eqref{eq:chiralComplexLG}
with amplitude $\rho_0^2=1-q_0^2$, and frequency $\omega_0 = - \alpha q_0^2$. We study linear fluctuations around this state, namely $\theta= \theta_0+ \delta \theta$ and $\rho=\rho_0+ \delta \rho$. The linear dynamics of these can be computed from Eq.\eqref{eq:chiralComplexLG}, matching $\phi=\phi_0+\delta \phi$ and the corresponding conjugated field,
\beq 
&& \delta \rho  =\frac{ \Re [\delta \phi \ \phi_0^*]} {\rho_0}, \quad  \delta \theta = \frac{ \Im [\delta \phi \ \phi_0^*]} {\rho_0^2}
\eeq 
We obtain,
\beq 
&& \partial_t \delta \theta = 2 i q_0 \alpha \del_\parallel \delta \theta + \frac{2 q_0}{\rho_0}(1+ \rho_0^2) \del_\parallel \delta \rho + K q_0^2 \del_\perp^2 \delta \theta \\ \nonumber
&&+ 5 K q_0^2 \del_\parallel^2 \delta \theta  - \frac{\alpha}{\rho_0}\nabla^2 \delta \rho  - \beta(\sin \theta_0 - \cos \theta_0)  \hat{\bm z} \cdot (\nabla \delta \rho \times \bm q_0) \\ \nonumber \\ \nonumber 
&& \partial_t \delta \rho = - 2 q_0^2 \rho_0^2 \delta \rho + 2 q_0 \alpha \del_\parallel \delta \rho - 2K q_0^3 \rho_0 \del_\parallel \delta \theta + \alpha \rho_0 \nabla^2 \delta \theta \\  &&+
  (5-3 \rho_0^2) \del_\parallel^2 \delta \rho + (1+ \rho_0^2) \del_\perp^2 \delta \rho   \\
\nonumber
&& + \beta \rho_0(\sin \theta_0 + \cos \theta_0)  \hat{\bm z} \cdot (\nabla \delta \rho \times \bm q_0)
\eeq 
where the $\parallel$ and $\perp$ directions are meant with respect to the direction of $\bm q_0$. 
The last equation identifies $\delta \rho$ as a fast mode, given the finite relaxation time scale $\sim 1/2 q_0^2 \rho_0^2$. We can therefore enslave it to the dynamics of $\delta \theta$, substituting $\delta \rho = - K q_0/ \rho_0 \del_\parallel \delta \theta$, and reducing to Eq.~\eqref{eq:linTheta}.

\begin{figure*}[t]
  \centering
	\includegraphics[width= 0.8\linewidth]{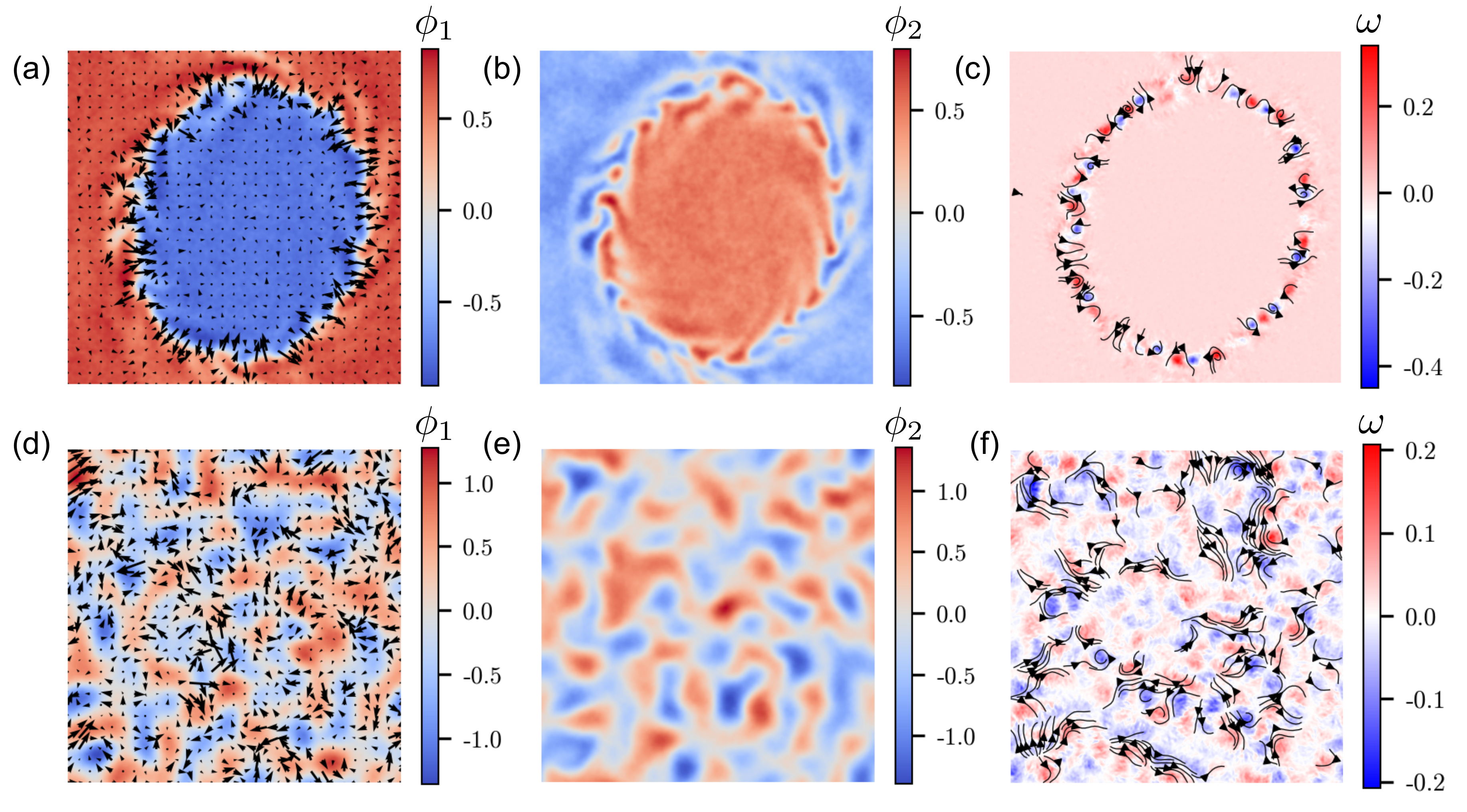}
\caption{\textbf{Simulations with traveling wave as initial condition.} The dynamics is ruled by a Mexican-hat potential in the space $\phi_1,\phi_2$. To activate the instability from this initial condition is necessary to have stochastic noise. A small amplitude is enough as shown by the results of these simulations. Parameters top: $q_0=0.5, \alpha=1, \beta=10, D=0.05 , dt=0.05, T=3\times 10^4$. Parameters bottom: $q_0=0.5, \alpha=5, \beta=10, D=0.05,dt=0.05, T=3\times 10^4$. $\omega = (\bm \nabla \times \bm J)_{\hat{\bm z}}$.}
\label{fig:snapshotTW_IC}
\end{figure*}

\subsubsection{Tridiagonal recurrence relation}
Eq.\eqref{eq:recurrence} can be read as an eigenvalue problem of a tridiagonal recurrence relation \cite{risken1989fokker}. An alternative method to find the eigenvalue $\lambda$ is to use continuous fractions. In our case the index $n$ is not positive definite, hence to solve \eqref{eq:recurrence} we define, 
\beq 
&& r_n^L = \frac{c_{n+1}}{c_n} = - \frac{1}{Q_n^+}\left[ (\lambda + Q_n) + \frac{Q_n^-}{r_{n-1}^L} \right] \\
&& r_n^R = \frac{c_{n-1}}{c_{n}} = - \frac{1}{Q_n^-}\left[ (\lambda + Q_n) + \frac{Q_n^+}{r_{n+1}^R} \right]
\eeq 
These are recursive relations, where the first runs for $n \in [-N,0]$, and the second for the right positive half $ n \in [0,N]$. Here $N$ is our truncation threshold, corresponding to the size of the matrix in the main text. Imposing $c_{N+1}=0$ and $c_{-N-1}=0$ we obtain the boundary conditions $r_{-N}^L$ and $r_{N}^R$, that we use to solve,
\beq 
(\lambda + Q_0) + Q_0^+ r_0^L + Q_0^- r_0^R=0 \ .
\eeq 
The latter is a highly non-linear equation that can be solved numerically in $\lambda$. A comparison between this method and the direct numerical evaluation of the eigenvalues is reported in Fig.~\ref{fig:recurrence_qc}(a,b) for different parameters. When the system is stable the two methods perfectly match (Fig.~\ref{fig:recurrence_qc}(a)), on the other hand when the $\lambda_{max}$ becomes positive the continuous fraction method shows the same features of the other evaluation, but need larger $N$ to reach convergence (Fig.~\ref{fig:recurrence_qc}(b)). 
We show this study as a function of the angle of perturbation $\psi$, but it can be done as well also varying the modulus $q$. 

Within this framework, in Fig.~\ref{fig:LinStabilityanalysis}(a) we see that, after a critical value of $\beta$ the eigenvalue becomes positive crossing the zero at a given $q_c$. In panels (c,d) of Fig.~\ref{fig:recurrence_qc} we report the computation of this quantity as a function of $q_0$ and $\beta$, for two different values of $\alpha$. When $q_c=0$ the system is stable, increasing chirality this acquires non-null values, which decrease monotonically in the limit of $\beta\to \infty$.

\subsection*{Phase composition} 
\label{sec:phaseComposition}

For vanishing average concentration $\bar{\phi}_a = 0$, the chiral phase separated system is invariant under the simultaneous transformation, $\phi_i \to - \phi_a$, and $\beta \to -\beta$. As clear form the expressions of $M_a$, it is actually odd under the transformation meaning that $M_a$ have to vanish for zero mean densities. We will now derive the relations of Sec.\ref{sec:PhaseSeparation}, starting from
\beq
M_1 = \mu_1 + \mu_1^{c}
\eeq
where
\beq 
&& \mu_1 = \frac{\delta F_1}{\delta \phi_1} - K \partial_x^2 \phi_1  + \alpha \phi_2, \nonumber \\
&& \mu_2 = \frac{\delta F_2}{\delta \phi_1}- K \partial_x^2 \phi_2  - \alpha \phi_1.
\eeq 
Multiplying both sides by $\partial_x \phi_a$ and integrating from phase 1 to phase 2, we get
\beq 
&& M_1 \Delta \phi_1 = \Delta F_1 - \left[ \frac{K}{2} |\partial_x \phi_1|^2 \right]_1^2 + \\ \nonumber
&&  \int_1^2 dx  \left( \alpha \phi_2 \partial_x \phi_1 + \mu_1^c \partial_x \phi_1 \right) \nonumber \\
&& M_2 \Delta \phi_2 = \Delta F_2 - \left[ \frac{K}{2} |\partial_x \phi_2|^2 \right]_1^2  + \nonumber \\
&&\int_1^2 dx  \left( -\alpha \phi_1 \partial_x \phi_2 + \mu_2^c \partial_x \phi_2 \right).
\eeq 
We subtract these two equations, obtaining
\beq 
&& \Delta (M_2 \phi_2 - M_1 \phi_1 -F_{\rm eff} ) + \frac{K}{2} \Delta (|\partial_x \phi_1|^2 - |\partial_x \phi_2|^2)  =\nonumber \\
&& \int_1^2 dx (\mu_2^c \partial_x \phi_2 - \mu_1^c \partial_x \phi_1 ) .
\eeq
Deep in the ordered phase, gradients in $\phi_{1,2}$ vanish, and we can define
\beq 
F_{\rm eff} = F_2 -F_1 - \alpha \phi_1 \phi_2
\eeq
or, equivalently, $P=M_2 \phi_2 - M_1 \phi_1 -F_{\rm eff}$. Therefore,
\beq 
\Delta P = \int_1^2 dy \ (\mu_2^c \partial_y \phi_2 - \mu_1^c \partial_y \phi_1 ) \equiv \int_1^2 dy \ \sigma (y)
\eeq 
where we have defined a stress density $\sigma$. 

In the steady state there are currents that travel along the interface meaning that the $M_{a}$ is not constant. The fluctuations in $M_a$ are however restricted to the interface, such that at any instant of time, $x-$averaged quantities resemble profiles that occur in passive phase separation. In steady state, the fields $\phi_{1,2}$ fluctuate about the mean values. The average value is determined from the peaks of the probability distribution of $\phi_{1,2}$.  

\subsection{Interface equations and edge currents}
\label{sec:interfaces}

We report here the details of the calculations to obtain the coupled interface equations of Eq. \eqref{eq:interfaces_fluc}.  We write the passages for the field $\phi_1$, but the same considerations follow for $\phi_2$. Multiplying both side of the densities dynamics for $(-\nabla^2)^{-1}$, we get 
\beq
\nonumber
&& (-\nabla^2)^{-1} [\partial_t \phi_1+\beta_1 ( \partial_x \phi_1 \partial_y \phi_2 -  \partial_y \phi_1 \partial_x \phi_2)] = \nonumber \\
&& + [K \nabla^2 \phi_1 - V_d'(\phi_1) - \alpha \phi_2]  \ . 
\eeq 
In a sharp interface approximation, we use Eq. \eqref{eq:profiles_inter} to compute the derivatives of the fields in space and time, such as  $\partial_t \phi_1 = - f'(u_1) \partial_t h_1$,  $\partial_x \phi_1 = - f'(u_1) \partial_x h_1$, and  $\partial_x \phi_2 = - g'(u_2) \partial_x h_2$,  with $u_2= y - h_2 $,
\beq 
&&  (- \nabla^2)^{-1}[f'(u) \partial_t h_1 + \beta_1 f'(u)g'(u) \partial_x \Delta h]  = \nonumber \\
&&-K[(1+|\nabla h_1|^2) f''(u) - f'(u) \nabla^2 h_1] + V_d'(f) \\ \nonumber
&& +\alpha[ g(u)+g'(u)\Delta h]
\eeq 
where we have expanded in small $\Delta h = h_1-h_2$
\beq 
g'(u_2)=g'(u_1+h_1-h_2) \simeq g'(u_1)+ g''(u_1) \Delta h
\eeq 
 and replaced $u_1=u$. We can now project the equation on the interface, multiplying both sides for $f'(u)$ and integrate over the interface profile:
\beq 
\nonumber
&& \int du f'(u) (- \nabla^2)^{-1}[f'(u) \partial_t h_1 + \beta_1 f'(u)g'(u) \partial_x \Delta h]  \\
&& = \int du f'(u) [ K f'(u) \nabla^2 h_1  +\alpha g'(u)\Delta h ]  
\eeq 
neglecting some terms that are null by symmetry. 

If two dimensional functions $G(x,y)$, and $F(x,y)$ satisfy $G(x,y) = (-\nabla^2)^{-1} F(x,y) $, their partial Fourier transform reads \cite{bray2001interface}, 
\beq 
 \tilde G(k_x,y) = \frac{1}{2 |q_x|} \int_{-\infty}^\infty dy' e^{- |q_x||y-y'|} \tilde F(q_x,y')
\eeq 
We use this to rewrite:
\beq 
&& \frac{1}{2|q_x|}[A_1(q_x) \partial_t \tilde h_1 + B_1(q_x)\beta_1 i q_x (\tilde h_1 - \tilde h_2)]  =\nonumber \\
&&  - K' q_x^2 \tilde h_1 +\alpha' (\tilde h_1 - \tilde h_2)
\eeq
with 
\beq 
&&A_1(q_x)= \int du f'(u) \int dv  e^{-|q_x||u-v|} f'(v) \\
&& B_1(q_x)= \int du f'(u) \int dv  e^{-|q_x||u-v|} f'(v) g'(v) \\
&&  K' = K \int dv [f'(v)]^2 \\
&& \alpha' = \alpha \int dv f'(v) g'(v) 
\eeq 
and finally get,
\begin{equation} \partial_t \tilde h_1 = \frac{1}{A_1(q_x)} [ - i \beta_1 B_1(q_x) q_x \Delta \tilde h - 2 K' |q_x|^3 \tilde h_1 + 2 |q_x| \alpha'   \Delta \tilde h] 
\end{equation}
Because we are interested in length scales much larger than the interfacial width, we can approximate the functions $A_1$ and $B_1$ with their constant values at $q_x=0$. Therefore, 
\beq 
\partial_t \tilde h_1 = - i \bar \beta_1  q_x \Delta \tilde h - 2 \bar K |q_x|^3 \tilde h_1 - 2 |q_x| \bar \alpha \Delta \tilde h 
\eeq 
with $\bar \beta_1= \beta_1 B_1(q_x=0)/A_1(q_x=0)$, $\bar K = K'/A_1(q_x=0)$ and $\bar \alpha = \alpha'/A_1(q_x=0)$. To evaluate these constant factors, we used $g=-f$ as seen in the numerical simulations, assuming perfectly separated systems. For this latter case we can indeed write $\bar \beta_2= \beta_2 B_1(q_x=0)/A_1(q_x=0)$, and recover Eq. \eqref{eq:interfaces_fluc}. This is paramount when writing down the dynamics of $\tilde h_2$, and evaluate the linear stability analysis. 

\subsection{Numerical methods.}\label{Sec:methodsNumerics}

Numerical simulations have been performed using a pseudospectral method with periodic boundary conditions. The algorithm combines the evaluation of linear terms in Fourier space and non-linear terms in real space in order to obtain a stable solution of the non-linear partial differential equations; more details can be found in \cite{saha2020scalar}. We use a backward Euler–Maruyama method to perform the time integration, and the size of the square box is fixed to $L=32 \pi$, with $N=512$ of spatial resolution. For the simulations shown in the main text we initialize the two fields as random configurations of zero mean and variance of $1/12$. 

The analytical analysis expanding the traveling wave solution has been performed using a Mexican-hat potential, instead of a double well. To prove that our analytical results apply to this second case and are independent of the choice of the potential, we run simulations of Eq.(\ref{eq:chiralComplexLG}), with a band pattern as initial condition $\phi(\bm x,t)= \rho_0 \exp (i \bm q_0 \cdot \bm x)$. To observe the instability, we add a small source of conserved noise as $\partial_t \phi \sim \sqrt{2 D} (\bm \nabla \cdot \bm \xi)$ following \cite{pisegna2024emergent} and assuming Ito convention for stochastic calculus. For small enough $D$, as shown in Fig.~\ref{fig:snapshotTW_IC}, we observe the same phenomenology with the same parameters of Fig.~\ref{fig:phase_diagram}.

\begin{figure*}[t]
  \centering
	\includegraphics[width= 0.8\linewidth]{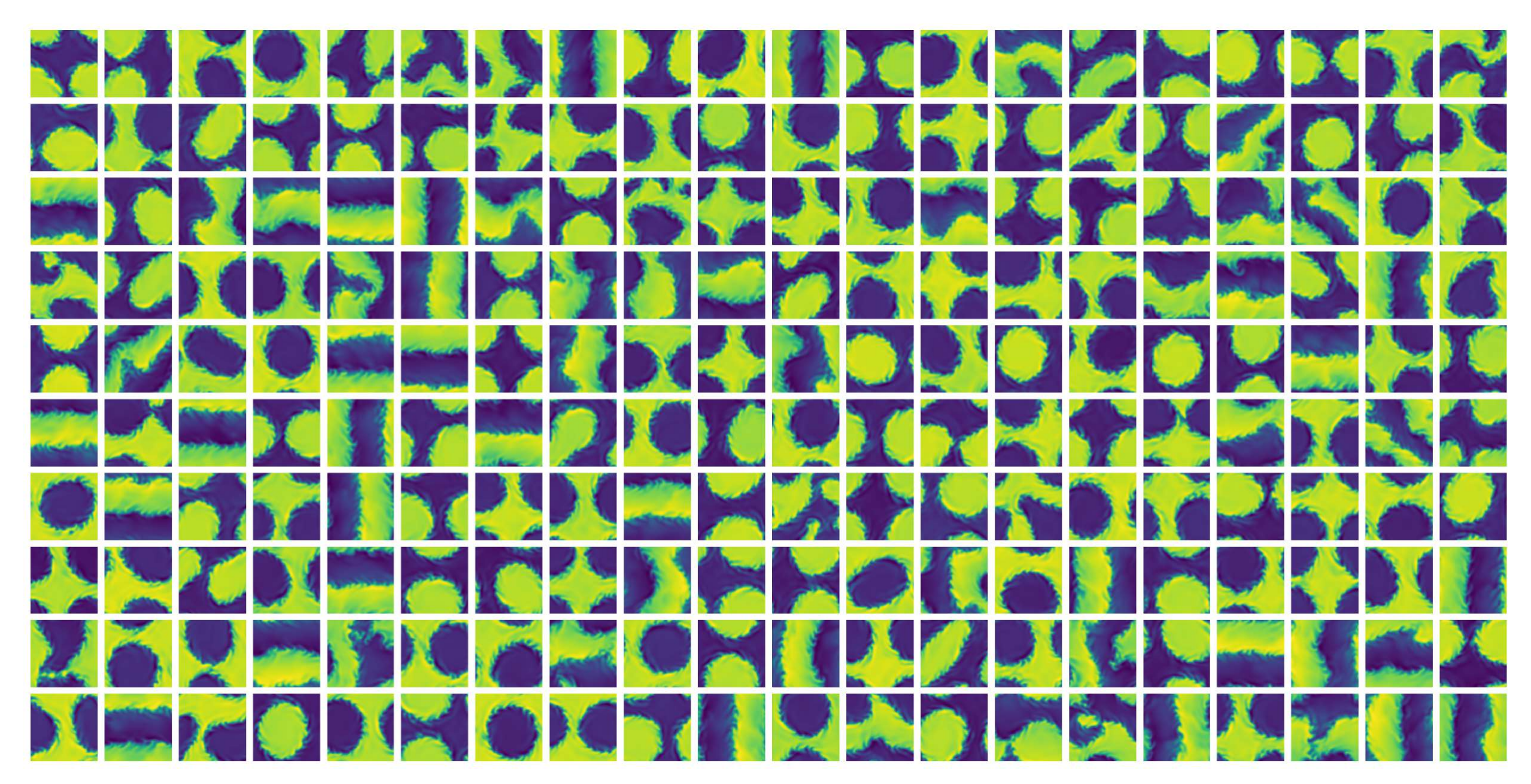}
\caption{\textbf{Varied morphologies of the condensates formed during chiral phase separation.}  A bubble or a stripe is selected through the complex kinetics of the two fields. Approximately $25 \%$ of the ensemble of 200 simulations evolved into a bubble. Parameters of simulations: $\beta=10, \alpha=1, N=512, L=32 \pi$. Mexican-hat potential. }
\label{fig:S3}
\end{figure*}

\subsubsection*{Final configuration of chiral phase separated states.}

We find that the system can converge to configurations of either stripes or a droplet even when all physical parameters are identical. To investigate this, we performed 200 independent simulations that differ only in their initial conditions fixing $\alpha=1, \beta=10$, and the Mexican-hat symmetry for the chemical potential. Specifically, the values of $\phi_1$ and $\phi_2$ at each lattice site were independently drawn from a normal distribution, $\mathcal{N}(0,1/12)$. This is the only source of randomness in the system which then evolves deterministically, following the method adopted to study the phase diagram of Fig.~\ref{fig:phase_diagram}. The steady-state configurations obtained after $6\times10^5$ time steps, reveal that both stripe and bubble morphologies can, with the stripe configuration occurring in approximately $25 \%$ of the realizations, see Fig.~\ref{fig:S3}. Therefore, the system appears to exhibit multi-stability among different spatial patterns. We interpret this phenomenon as a kinetic effect due to the slow coarsening-process happening through annihilation of vorticity's defects. The final morphology depends on the specific, stochastic sequence in which charges of opposite sign happen to meet and annihilate during the early transient, rather than being uniquely selected by the parameters $\alpha, \beta$.

However, we also find that the relative prevalence of the two morphologies is determined by the ratio between $\alpha$ and $\beta$, as well as by the symmetry of the equilibrium potential. These factors interplay, leading to competition between the two topologies, with the kinetics selecting one of them. In addition to those reported in Fig.~\ref{fig:S3}, we ran additional simulations for another parameter set, $\alpha=3$ and $\beta=10$. For this set, we consistently observe a striped pattern. Moreover, we performed another 200 simulations using the parameters of the first analysis, replacing the Mexican-hat potential with the double-well potential, and observed a substantial decrease in the number of striped patterns.

\begin{figure*}[t]
  \centering
	\includegraphics[width= 0.9\linewidth]{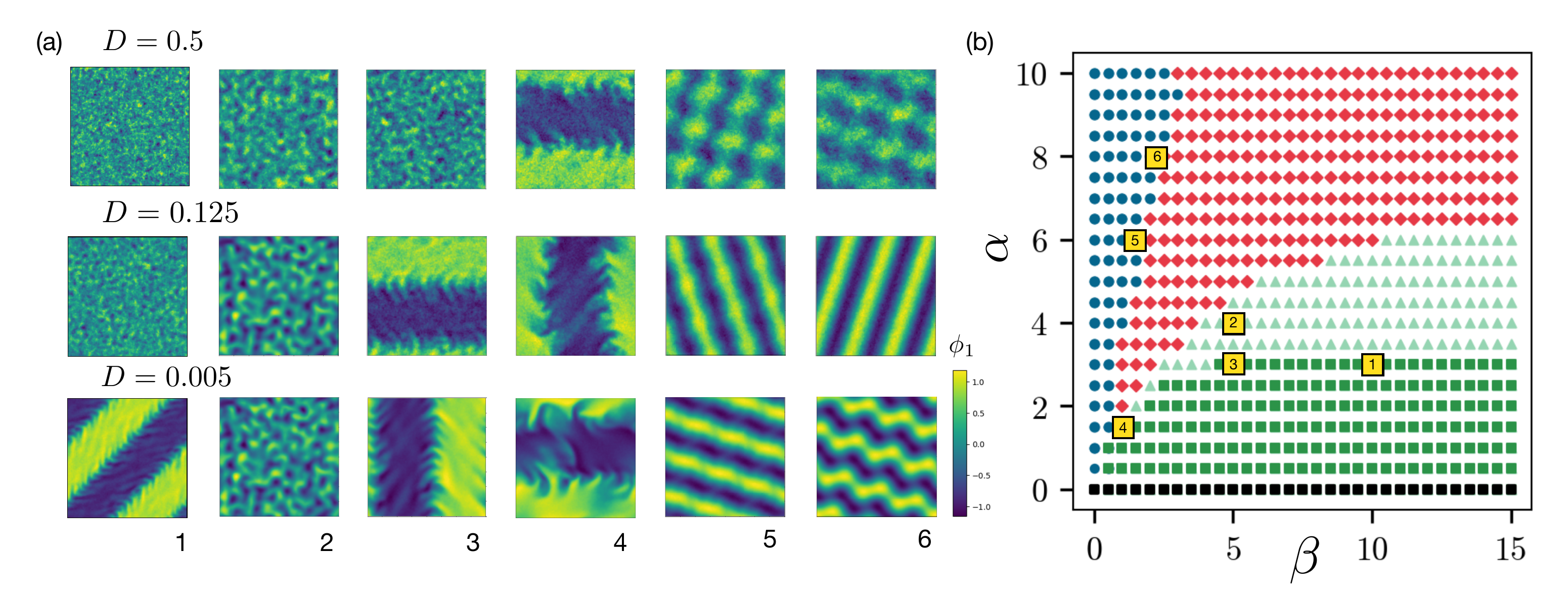}
\caption{\textbf{Stability of the state diagram under fluctuations.} (a) Stochastic numerical simulations for different combinations of $\alpha$ and $\beta$, marked in (b), and different noise amplitudes, indicated in the panels, using a double-well equilibrium potential. The phase-separated pattern with chiral currents persists at low noise strengths, while, with increasing noise, the system settles into a disordered state in cases 1 and 3.
 }
\label{fig:S4}
\end{figure*}

\subsubsection*{Robustness of results to stochastic noise}

Finally, we investigate how the boundaries of the phase diagram in Fig.~\ref{fig:phase_diagram} change upon retaining the stochastic terms with variance $2D$ in Eq.~\eqref{eq:betaterm} the numerical evolution. We expect that, for sufficiently small noise amplitudes, the locations of the phase boundaries remain essentially unchanged, whereas stronger fluctuations should favour the disordered chiral phase over the phase-separated one. To test this hypothesis, we selected representative parameter sets lying close to the main boundaries of the phase diagram (Fig.~\ref{fig:phase_diagram}g). For these sets, we performed additional simulations at different noise amplitudes, $D=0.005$--$0.5$. The results are reported in Fig.~\ref{fig:S4}, which contains snapshots of the configurations at sufficiently long times. We see that, for low noise amplitudes, the patterns persist, while a disordered chaotic state is reached at higher values. Hence, we deduce that the boundaries of the phase diagram do not change drastically under weak noise.

\section*{ACKNOWLEDGMENTS}

We acknowledge insightful discussions with Ananyo Maitra. We are grateful for valuable support and funding by the Max Planck Society. We acknowledge support from the Department of Living Matter Physics, and in particular from Ramin Golestanian.

\section*{AUTHOR CONTRIBUTIONS}


G.P. and S.S. designed the research, conducted the research, analyzed the data, and wrote the paper.

\bibliography{references}

\begin{thebibliography}{73}%
\makeatletter
\providecommand \@ifxundefined [1]{%
 \@ifx{#1\undefined}
}%
\providecommand \@ifnum [1]{%
 \ifnum #1\expandafter \@firstoftwo
 \else \expandafter \@secondoftwo
 \fi
}%
\providecommand \@ifx [1]{%
 \ifx #1\expandafter \@firstoftwo
 \else \expandafter \@secondoftwo
 \fi
}%
\providecommand \natexlab [1]{#1}%
\providecommand \enquote  [1]{``#1''}%
\providecommand \bibnamefont  [1]{#1}%
\providecommand \bibfnamefont [1]{#1}%
\providecommand \citenamefont [1]{#1}%
\providecommand \href@noop [0]{\@secondoftwo}%
\providecommand \href [0]{\begingroup \@sanitize@url \@href}%
\providecommand \@href[1]{\@@startlink{#1}\@@href}%
\providecommand \@@href[1]{\endgroup#1\@@endlink}%
\providecommand \@sanitize@url [0]{\catcode `\\12\catcode `\$12\catcode
  `\&12\catcode `\#12\catcode `\^12\catcode `\_12\catcode `\%12\relax}%
\providecommand \@@startlink[1]{}%
\providecommand \@@endlink[0]{}%
\providecommand \url  [0]{\begingroup\@sanitize@url \@url }%
\providecommand \@url [1]{\endgroup\@href {#1}{\urlprefix }}%
\providecommand \urlprefix  [0]{URL }%
\providecommand \Eprint [0]{\href }%
\providecommand \doibase [0]{https://doi.org/}%
\providecommand \selectlanguage [0]{\@gobble}%
\providecommand \bibinfo  [0]{\@secondoftwo}%
\providecommand \bibfield  [0]{\@secondoftwo}%
\providecommand \translation [1]{[#1]}%
\providecommand \BibitemOpen [0]{}%
\providecommand \bibitemStop [0]{}%
\providecommand \bibitemNoStop [0]{.\EOS\space}%
\providecommand \EOS [0]{\spacefactor3000\relax}%
\providecommand \BibitemShut  [1]{\csname bibitem#1\endcsname}%
\let\auto@bib@innerbib\@empty
\bibitem [{\citenamefont {Harris}\ \emph {et~al.}(1999)\citenamefont {Harris},
  \citenamefont {Kamien},\ and\ \citenamefont
  {Lubensky}}]{Harris_Kamien_Lubensky_RevModPhys.71.1745}%
  \BibitemOpen
  \bibfield  {author} {\bibinfo {author} {\bibfnamefont {A.~B.}\ \bibnamefont
  {Harris}}, \bibinfo {author} {\bibfnamefont {R.~D.}\ \bibnamefont {Kamien}},\
  and\ \bibinfo {author} {\bibfnamefont {T.~C.}\ \bibnamefont {Lubensky}},\
  }\bibfield  {title} {\bibinfo {title} {Molecular chirality and chiral
  parameters},\ }\href {https://doi.org/10.1103/RevModPhys.71.1745} {\bibfield
  {journal} {\bibinfo  {journal} {Rev. Mod. Phys.}\ }\textbf {\bibinfo {volume}
  {71}},\ \bibinfo {pages} {1745} (\bibinfo {year} {1999})}\BibitemShut
  {NoStop}%
\bibitem [{\citenamefont {Tverdislov}(2013)}]{tverdislov2013chirality}%
  \BibitemOpen
  \bibfield  {author} {\bibinfo {author} {\bibfnamefont {V.}~\bibnamefont
  {Tverdislov}},\ }\bibfield  {title} {\bibinfo {title} {Chirality as a primary
  switch of hierarchical levels in molecular biological systems},\ }\href@noop
  {} {\bibfield  {journal} {\bibinfo  {journal} {Biophysics}\ }\textbf
  {\bibinfo {volume} {58}},\ \bibinfo {pages} {128} (\bibinfo {year}
  {2013})}\BibitemShut {NoStop}%
\bibitem [{\citenamefont {Athani}\ \emph {et~al.}(2025)\citenamefont {Athani},
  \citenamefont {Prouse}, \citenamefont {Sarpangala}, \citenamefont {Noerr},
  \citenamefont {Schiano-Lomoriello}, \citenamefont {Kumar}, \citenamefont
  {Memarian}, \citenamefont {Gaillard}, \citenamefont {Blanchoin},
  \citenamefont {Hirst} \emph {et~al.}}]{athani2025gliding}%
  \BibitemOpen
  \bibfield  {author} {\bibinfo {author} {\bibfnamefont {M.~G.}\ \bibnamefont
  {Athani}}, \bibinfo {author} {\bibfnamefont {N.}~\bibnamefont {Prouse}},
  \bibinfo {author} {\bibfnamefont {N.}~\bibnamefont {Sarpangala}}, \bibinfo
  {author} {\bibfnamefont {P.}~\bibnamefont {Noerr}}, \bibinfo {author}
  {\bibfnamefont {G.}~\bibnamefont {Schiano-Lomoriello}}, \bibinfo {author}
  {\bibfnamefont {A.~G.}\ \bibnamefont {Kumar}}, \bibinfo {author}
  {\bibfnamefont {F.~L.}\ \bibnamefont {Memarian}}, \bibinfo {author}
  {\bibfnamefont {J.}~\bibnamefont {Gaillard}}, \bibinfo {author}
  {\bibfnamefont {L.}~\bibnamefont {Blanchoin}}, \bibinfo {author}
  {\bibfnamefont {L.~S.}\ \bibnamefont {Hirst}}, \emph {et~al.},\ }\bibfield
  {title} {\bibinfo {title} {Gliding microtubules exhibit tunable collective
  rotation driven by chiral active forces},\ }\href@noop {} {\bibfield
  {journal} {\bibinfo  {journal} {arXiv preprint arXiv:2507.00245}\ } (\bibinfo
  {year} {2025})}\BibitemShut {NoStop}%
\bibitem [{\citenamefont {Berg}(2004)}]{berg2004coli}%
  \BibitemOpen
  \bibfield  {author} {\bibinfo {author} {\bibfnamefont {H.~C.}\ \bibnamefont
  {Berg}},\ }\href@noop {} {\emph {\bibinfo {title} {E. coli in Motion}}}\
  (\bibinfo  {publisher} {Springer},\ \bibinfo {year} {2004})\BibitemShut
  {NoStop}%
\bibitem [{\citenamefont {Di~Leonardo}\ \emph {et~al.}(2011)\citenamefont
  {Di~Leonardo}, \citenamefont {Dell’Arciprete}, \citenamefont {Angelani},\
  and\ \citenamefont {Iebba}}]{di2011swimming}%
  \BibitemOpen
  \bibfield  {author} {\bibinfo {author} {\bibfnamefont {R.}~\bibnamefont
  {Di~Leonardo}}, \bibinfo {author} {\bibfnamefont {D.}~\bibnamefont
  {Dell’Arciprete}}, \bibinfo {author} {\bibfnamefont {L.}~\bibnamefont
  {Angelani}},\ and\ \bibinfo {author} {\bibfnamefont {V.}~\bibnamefont
  {Iebba}},\ }\bibfield  {title} {\bibinfo {title} {Swimming with an image},\
  }\href@noop {} {\bibfield  {journal} {\bibinfo  {journal} {Physical review
  letters}\ }\textbf {\bibinfo {volume} {106}},\ \bibinfo {pages} {038101}
  (\bibinfo {year} {2011})}\BibitemShut {NoStop}%
\bibitem [{\citenamefont {Inaki}\ \emph {et~al.}(2016)\citenamefont {Inaki},
  \citenamefont {Liu},\ and\ \citenamefont {Matsuno}}]{inaki2016cell}%
  \BibitemOpen
  \bibfield  {author} {\bibinfo {author} {\bibfnamefont {M.}~\bibnamefont
  {Inaki}}, \bibinfo {author} {\bibfnamefont {J.}~\bibnamefont {Liu}},\ and\
  \bibinfo {author} {\bibfnamefont {K.}~\bibnamefont {Matsuno}},\ }\bibfield
  {title} {\bibinfo {title} {Cell chirality: its origin and roles in
  left--right asymmetric development},\ }\href@noop {} {\bibfield  {journal}
  {\bibinfo  {journal} {Philosophical Transactions of the Royal Society B:
  Biological Sciences}\ }\textbf {\bibinfo {volume} {371}},\ \bibinfo {pages}
  {20150403} (\bibinfo {year} {2016})}\BibitemShut {NoStop}%
\bibitem [{\citenamefont {Chen}\ \emph
  {et~al.}(2025{\natexlab{a}})\citenamefont {Chen}, \citenamefont {G{\"o}kmen},
  \citenamefont {Fruchart}, \citenamefont {Krumbein}, \citenamefont
  {Silberzan}, \citenamefont {Yashunsky},\ and\ \citenamefont
  {Vitelli}}]{chen2025chirality}%
  \BibitemOpen
  \bibfield  {author} {\bibinfo {author} {\bibfnamefont {S.}~\bibnamefont
  {Chen}}, \bibinfo {author} {\bibfnamefont {D.~E.}\ \bibnamefont
  {G{\"o}kmen}}, \bibinfo {author} {\bibfnamefont {M.}~\bibnamefont
  {Fruchart}}, \bibinfo {author} {\bibfnamefont {M.}~\bibnamefont {Krumbein}},
  \bibinfo {author} {\bibfnamefont {P.}~\bibnamefont {Silberzan}}, \bibinfo
  {author} {\bibfnamefont {V.}~\bibnamefont {Yashunsky}},\ and\ \bibinfo
  {author} {\bibfnamefont {V.}~\bibnamefont {Vitelli}},\ }\bibfield  {title}
  {\bibinfo {title} {Chirality across scales in tissue dynamics},\ }\href@noop
  {} {\bibfield  {journal} {\bibinfo  {journal} {arXiv preprint
  arXiv:2506.12276}\ } (\bibinfo {year} {2025}{\natexlab{a}})}\BibitemShut
  {NoStop}%
\bibitem [{\citenamefont {Beppu}\ \emph {et~al.}(2021)\citenamefont {Beppu},
  \citenamefont {Izri}, \citenamefont {Sato}, \citenamefont {Yamanishi},
  \citenamefont {Sumino},\ and\ \citenamefont {Maeda}}]{beppu2021edge}%
  \BibitemOpen
  \bibfield  {author} {\bibinfo {author} {\bibfnamefont {K.}~\bibnamefont
  {Beppu}}, \bibinfo {author} {\bibfnamefont {Z.}~\bibnamefont {Izri}},
  \bibinfo {author} {\bibfnamefont {T.}~\bibnamefont {Sato}}, \bibinfo {author}
  {\bibfnamefont {Y.}~\bibnamefont {Yamanishi}}, \bibinfo {author}
  {\bibfnamefont {Y.}~\bibnamefont {Sumino}},\ and\ \bibinfo {author}
  {\bibfnamefont {Y.~T.}\ \bibnamefont {Maeda}},\ }\bibfield  {title} {\bibinfo
  {title} {Edge current and pairing order transition in chiral bacterial
  vortices},\ }\href@noop {} {\bibfield  {journal} {\bibinfo  {journal}
  {Proceedings of the National Academy of Sciences}\ }\textbf {\bibinfo
  {volume} {118}},\ \bibinfo {pages} {e2107461118} (\bibinfo {year}
  {2021})}\BibitemShut {NoStop}%
\bibitem [{\citenamefont {Dunkel}\ \emph {et~al.}(2013)\citenamefont {Dunkel},
  \citenamefont {Heidenreich}, \citenamefont {Drescher}, \citenamefont
  {Wensink}, \citenamefont {B{\"a}r},\ and\ \citenamefont
  {Goldstein}}]{dunkel2013fluid}%
  \BibitemOpen
  \bibfield  {author} {\bibinfo {author} {\bibfnamefont {J.}~\bibnamefont
  {Dunkel}}, \bibinfo {author} {\bibfnamefont {S.}~\bibnamefont {Heidenreich}},
  \bibinfo {author} {\bibfnamefont {K.}~\bibnamefont {Drescher}}, \bibinfo
  {author} {\bibfnamefont {H.~H.}\ \bibnamefont {Wensink}}, \bibinfo {author}
  {\bibfnamefont {M.}~\bibnamefont {B{\"a}r}},\ and\ \bibinfo {author}
  {\bibfnamefont {R.~E.}\ \bibnamefont {Goldstein}},\ }\bibfield  {title}
  {\bibinfo {title} {Fluid dynamics of bacterial turbulence},\ }\href@noop {}
  {\bibfield  {journal} {\bibinfo  {journal} {Physical review letters}\
  }\textbf {\bibinfo {volume} {110}},\ \bibinfo {pages} {228102} (\bibinfo
  {year} {2013})}\BibitemShut {NoStop}%
\bibitem [{\citenamefont {Wolgemuth}(2008)}]{wolgemuth2008collective}%
  \BibitemOpen
  \bibfield  {author} {\bibinfo {author} {\bibfnamefont {C.~W.}\ \bibnamefont
  {Wolgemuth}},\ }\bibfield  {title} {\bibinfo {title} {Collective swimming and
  the dynamics of bacterial turbulence},\ }\href@noop {} {\bibfield  {journal}
  {\bibinfo  {journal} {Biophysical journal}\ }\textbf {\bibinfo {volume}
  {95}},\ \bibinfo {pages} {1564} (\bibinfo {year} {2008})}\BibitemShut
  {NoStop}%
\bibitem [{\citenamefont {Li}\ \emph {et~al.}(2024)\citenamefont {Li},
  \citenamefont {Chat{\'e}}, \citenamefont {Sano}, \citenamefont {Shi},\ and\
  \citenamefont {Zhang}}]{li2024robust}%
  \BibitemOpen
  \bibfield  {author} {\bibinfo {author} {\bibfnamefont {H.}~\bibnamefont
  {Li}}, \bibinfo {author} {\bibfnamefont {H.}~\bibnamefont {Chat{\'e}}},
  \bibinfo {author} {\bibfnamefont {M.}~\bibnamefont {Sano}}, \bibinfo {author}
  {\bibfnamefont {X.-q.}\ \bibnamefont {Shi}},\ and\ \bibinfo {author}
  {\bibfnamefont {H.}~\bibnamefont {Zhang}},\ }\bibfield  {title} {\bibinfo
  {title} {Robust edge flows in swarming bacterial colonies},\ }\href@noop {}
  {\bibfield  {journal} {\bibinfo  {journal} {Physical Review X}\ }\textbf
  {\bibinfo {volume} {14}},\ \bibinfo {pages} {041006} (\bibinfo {year}
  {2024})}\BibitemShut {NoStop}%
\bibitem [{\citenamefont {Digregorio}\ \emph {et~al.}(2025)\citenamefont
  {Digregorio}, \citenamefont {Pagonabarraga},\ and\ \citenamefont
  {Reyes}}]{digregorio2025phaseseparationchiralactive}%
  \BibitemOpen
  \bibfield  {author} {\bibinfo {author} {\bibfnamefont {P.}~\bibnamefont
  {Digregorio}}, \bibinfo {author} {\bibfnamefont {I.}~\bibnamefont
  {Pagonabarraga}},\ and\ \bibinfo {author} {\bibfnamefont {F.~V.}\
  \bibnamefont {Reyes}},\ }\href {https://arxiv.org/abs/2504.08533} {\bibinfo
  {title} {Phase separation in a chiral active fluid of inertial self-spinning
  disks}} (\bibinfo {year} {2025}),\ \Eprint {https://arxiv.org/abs/2504.08533}
  {arXiv:2504.08533 [cond-mat.soft]} \BibitemShut {NoStop}%
\bibitem [{\citenamefont {Caprini}\ and\ \citenamefont {Marini
  Bettolo~Marconi}(2025)}]{Caprini2025}%
  \BibitemOpen
  \bibfield  {author} {\bibinfo {author} {\bibfnamefont {L.}~\bibnamefont
  {Caprini}}\ and\ \bibinfo {author} {\bibfnamefont {U.}~\bibnamefont {Marini
  Bettolo~Marconi}},\ }\bibfield  {title} {\bibinfo {title} {Bubble phase
  induced by odd interactions in chiral systems},\ }\href
  {https://doi.org/10.1063/5.0262594} {\bibfield  {journal} {\bibinfo
  {journal} {The Journal of Chemical Physics}\ }\textbf {\bibinfo {volume}
  {162}},\ \bibinfo {pages} {161101} (\bibinfo {year} {2025})}\BibitemShut
  {NoStop}%
\bibitem [{\citenamefont {Liebchen}\ and\ \citenamefont
  {Levis}(2017)}]{liebchen2017collective}%
  \BibitemOpen
  \bibfield  {author} {\bibinfo {author} {\bibfnamefont {B.}~\bibnamefont
  {Liebchen}}\ and\ \bibinfo {author} {\bibfnamefont {D.}~\bibnamefont
  {Levis}},\ }\bibfield  {title} {\bibinfo {title} {Collective behavior of
  chiral active matter: Pattern formation and enhanced flocking},\ }\href@noop
  {} {\bibfield  {journal} {\bibinfo  {journal} {Physical review letters}\
  }\textbf {\bibinfo {volume} {119}},\ \bibinfo {pages} {058002} (\bibinfo
  {year} {2017})}\BibitemShut {NoStop}%
\bibitem [{\citenamefont {Ses{\'e}-Sansa}\ \emph {et~al.}(2022)\citenamefont
  {Ses{\'e}-Sansa}, \citenamefont {Levis},\ and\ \citenamefont
  {Pagonabarraga}}]{sese2022microscopic}%
  \BibitemOpen
  \bibfield  {author} {\bibinfo {author} {\bibfnamefont {E.}~\bibnamefont
  {Ses{\'e}-Sansa}}, \bibinfo {author} {\bibfnamefont {D.}~\bibnamefont
  {Levis}},\ and\ \bibinfo {author} {\bibfnamefont {I.}~\bibnamefont
  {Pagonabarraga}},\ }\bibfield  {title} {\bibinfo {title} {Microscopic field
  theory for structure formation in systems of self-propelled particles with
  generic torques},\ }\href@noop {} {\bibfield  {journal} {\bibinfo  {journal}
  {The Journal of Chemical Physics}\ }\textbf {\bibinfo {volume} {157}}
  (\bibinfo {year} {2022})}\BibitemShut {NoStop}%
\bibitem [{\citenamefont {Caporusso}\ \emph {et~al.}(2024)\citenamefont
  {Caporusso}, \citenamefont {Gonnella},\ and\ \citenamefont
  {Levis}}]{caporusso2024phase}%
  \BibitemOpen
  \bibfield  {author} {\bibinfo {author} {\bibfnamefont {C.~B.}\ \bibnamefont
  {Caporusso}}, \bibinfo {author} {\bibfnamefont {G.}~\bibnamefont
  {Gonnella}},\ and\ \bibinfo {author} {\bibfnamefont {D.}~\bibnamefont
  {Levis}},\ }\bibfield  {title} {\bibinfo {title} {Phase coexistence and edge
  currents in the chiral lennard-jones fluid},\ }\href@noop {} {\bibfield
  {journal} {\bibinfo  {journal} {Physical Review Letters}\ }\textbf {\bibinfo
  {volume} {132}},\ \bibinfo {pages} {168201} (\bibinfo {year}
  {2024})}\BibitemShut {NoStop}%
\bibitem [{\citenamefont {Soni}\ \emph {et~al.}(2019)\citenamefont {Soni},
  \citenamefont {Bililign}, \citenamefont {Magkiriadou}, \citenamefont
  {Sacanna}, \citenamefont {Bartolo}, \citenamefont {Shelley},\ and\
  \citenamefont {Irvine}}]{soni2019odd}%
  \BibitemOpen
  \bibfield  {author} {\bibinfo {author} {\bibfnamefont {V.}~\bibnamefont
  {Soni}}, \bibinfo {author} {\bibfnamefont {E.~S.}\ \bibnamefont {Bililign}},
  \bibinfo {author} {\bibfnamefont {S.}~\bibnamefont {Magkiriadou}}, \bibinfo
  {author} {\bibfnamefont {S.}~\bibnamefont {Sacanna}}, \bibinfo {author}
  {\bibfnamefont {D.}~\bibnamefont {Bartolo}}, \bibinfo {author} {\bibfnamefont
  {M.~J.}\ \bibnamefont {Shelley}},\ and\ \bibinfo {author} {\bibfnamefont
  {W.~T.}\ \bibnamefont {Irvine}},\ }\bibfield  {title} {\bibinfo {title} {The
  odd free surface flows of a colloidal chiral fluid},\ }\href@noop {}
  {\bibfield  {journal} {\bibinfo  {journal} {Nature physics}\ }\textbf
  {\bibinfo {volume} {15}},\ \bibinfo {pages} {1188} (\bibinfo {year}
  {2019})}\BibitemShut {NoStop}%
\bibitem [{\citenamefont {Chen}\ \emph
  {et~al.}(2025{\natexlab{b}})\citenamefont {Chen}, \citenamefont {Weady},
  \citenamefont {Atis}, \citenamefont {Matsuzawa}, \citenamefont {Shelley},\
  and\ \citenamefont {Irvine}}]{chen2025self}%
  \BibitemOpen
  \bibfield  {author} {\bibinfo {author} {\bibfnamefont {P.}~\bibnamefont
  {Chen}}, \bibinfo {author} {\bibfnamefont {S.}~\bibnamefont {Weady}},
  \bibinfo {author} {\bibfnamefont {S.}~\bibnamefont {Atis}}, \bibinfo {author}
  {\bibfnamefont {T.}~\bibnamefont {Matsuzawa}}, \bibinfo {author}
  {\bibfnamefont {M.~J.}\ \bibnamefont {Shelley}},\ and\ \bibinfo {author}
  {\bibfnamefont {W.~T.}\ \bibnamefont {Irvine}},\ }\bibfield  {title}
  {\bibinfo {title} {Self-propulsion, flocking and chiral active phases from
  particles spinning at intermediate reynolds numbers},\ }\href@noop {}
  {\bibfield  {journal} {\bibinfo  {journal} {Nature Physics}\ }\textbf
  {\bibinfo {volume} {21}},\ \bibinfo {pages} {146} (\bibinfo {year}
  {2025}{\natexlab{b}})}\BibitemShut {NoStop}%
\bibitem [{\citenamefont {Levis}\ and\ \citenamefont
  {Liebchen}(2019)}]{ChiralMixtures_PhysRevE.100.012406}%
  \BibitemOpen
  \bibfield  {author} {\bibinfo {author} {\bibfnamefont {D.}~\bibnamefont
  {Levis}}\ and\ \bibinfo {author} {\bibfnamefont {B.}~\bibnamefont
  {Liebchen}},\ }\bibfield  {title} {\bibinfo {title} {Simultaneous phase
  separation and pattern formation in chiral active mixtures},\ }\href
  {https://doi.org/10.1103/PhysRevE.100.012406} {\bibfield  {journal} {\bibinfo
   {journal} {Phys. Rev. E}\ }\textbf {\bibinfo {volume} {100}},\ \bibinfo
  {pages} {012406} (\bibinfo {year} {2019})}\BibitemShut {NoStop}%
\bibitem [{\citenamefont {Levis}\ \emph {et~al.}(2019)\citenamefont {Levis},
  \citenamefont {Pagonabarraga},\ and\ \citenamefont
  {Liebchen}}]{Levis_Pagonabarraga_Liebchen_PhysRevResearch.1.023026}%
  \BibitemOpen
  \bibfield  {author} {\bibinfo {author} {\bibfnamefont {D.}~\bibnamefont
  {Levis}}, \bibinfo {author} {\bibfnamefont {I.}~\bibnamefont
  {Pagonabarraga}},\ and\ \bibinfo {author} {\bibfnamefont {B.}~\bibnamefont
  {Liebchen}},\ }\bibfield  {title} {\bibinfo {title} {Activity induced
  synchronization: Mutual flocking and chiral self-sorting},\ }\href
  {https://doi.org/10.1103/PhysRevResearch.1.023026} {\bibfield  {journal}
  {\bibinfo  {journal} {Phys. Rev. Res.}\ }\textbf {\bibinfo {volume} {1}},\
  \bibinfo {pages} {023026} (\bibinfo {year} {2019})}\BibitemShut {NoStop}%
\bibitem [{\citenamefont {Nguyen}\ \emph {et~al.}(2014)\citenamefont {Nguyen},
  \citenamefont {Klotsa}, \citenamefont {Engel},\ and\ \citenamefont
  {Glotzer}}]{nguyen2014emergent}%
  \BibitemOpen
  \bibfield  {author} {\bibinfo {author} {\bibfnamefont {N.~H.}\ \bibnamefont
  {Nguyen}}, \bibinfo {author} {\bibfnamefont {D.}~\bibnamefont {Klotsa}},
  \bibinfo {author} {\bibfnamefont {M.}~\bibnamefont {Engel}},\ and\ \bibinfo
  {author} {\bibfnamefont {S.~C.}\ \bibnamefont {Glotzer}},\ }\bibfield
  {title} {\bibinfo {title} {Emergent collective phenomena in a mixture of hard
  shapes through active rotation},\ }\href@noop {} {\bibfield  {journal}
  {\bibinfo  {journal} {Physical review letters}\ }\textbf {\bibinfo {volume}
  {112}},\ \bibinfo {pages} {075701} (\bibinfo {year} {2014})}\BibitemShut
  {NoStop}%
\bibitem [{\citenamefont {Yeo}\ \emph {et~al.}(2015)\citenamefont {Yeo},
  \citenamefont {Lushi},\ and\ \citenamefont {Vlahovska}}]{yeo2015collective}%
  \BibitemOpen
  \bibfield  {author} {\bibinfo {author} {\bibfnamefont {K.}~\bibnamefont
  {Yeo}}, \bibinfo {author} {\bibfnamefont {E.}~\bibnamefont {Lushi}},\ and\
  \bibinfo {author} {\bibfnamefont {P.~M.}\ \bibnamefont {Vlahovska}},\
  }\bibfield  {title} {\bibinfo {title} {Collective dynamics in a binary
  mixture of hydrodynamically coupled microrotors},\ }\href@noop {} {\bibfield
  {journal} {\bibinfo  {journal} {Physical review letters}\ }\textbf {\bibinfo
  {volume} {114}},\ \bibinfo {pages} {188301} (\bibinfo {year}
  {2015})}\BibitemShut {NoStop}%
\bibitem [{\citenamefont {Soto}\ and\ \citenamefont
  {Golestanian}(2014)}]{soto2014self}%
  \BibitemOpen
  \bibfield  {author} {\bibinfo {author} {\bibfnamefont {R.}~\bibnamefont
  {Soto}}\ and\ \bibinfo {author} {\bibfnamefont {R.}~\bibnamefont
  {Golestanian}},\ }\bibfield  {title} {\bibinfo {title} {Self-assembly of
  catalytically active colloidal molecules: Tailoring activity through surface
  chemistry},\ }\href {https://doi.org/10.1103/PhysRevLett.112.068301}
  {\bibfield  {journal} {\bibinfo  {journal} {Phys. Rev. Lett.}\ }\textbf
  {\bibinfo {volume} {112}},\ \bibinfo {pages} {068301} (\bibinfo {year}
  {2014})}\BibitemShut {NoStop}%
\bibitem [{\citenamefont {Agudo-Canalejo}\ and\ \citenamefont
  {Golestanian}(2019)}]{agudo2019active}%
  \BibitemOpen
  \bibfield  {author} {\bibinfo {author} {\bibfnamefont {J.}~\bibnamefont
  {Agudo-Canalejo}}\ and\ \bibinfo {author} {\bibfnamefont {R.}~\bibnamefont
  {Golestanian}},\ }\bibfield  {title} {\bibinfo {title} {Active phase
  separation in mixtures of chemically interacting particles},\ }\href
  {https://doi.org/10.1103/PhysRevLett.123.018101} {\bibfield  {journal}
  {\bibinfo  {journal} {Phys. Rev. Lett.}\ }\textbf {\bibinfo {volume} {123}},\
  \bibinfo {pages} {018101} (\bibinfo {year} {2019})}\BibitemShut {NoStop}%
\bibitem [{\citenamefont {Ouazan-Reboul}\ \emph {et~al.}(2023)\citenamefont
  {Ouazan-Reboul}, \citenamefont {Agudo-Canalejo},\ and\ \citenamefont
  {Golestanian}}]{ouazan2023self}%
  \BibitemOpen
  \bibfield  {author} {\bibinfo {author} {\bibfnamefont {V.}~\bibnamefont
  {Ouazan-Reboul}}, \bibinfo {author} {\bibfnamefont {J.}~\bibnamefont
  {Agudo-Canalejo}},\ and\ \bibinfo {author} {\bibfnamefont {R.}~\bibnamefont
  {Golestanian}},\ }\bibfield  {title} {\bibinfo {title} {Self-organization of
  primitive metabolic cycles due to non-reciprocal interactions},\ }\href@noop
  {} {\bibfield  {journal} {\bibinfo  {journal} {Nature Communications}\
  }\textbf {\bibinfo {volume} {14}},\ \bibinfo {pages} {4496} (\bibinfo {year}
  {2023})}\BibitemShut {NoStop}%
\bibitem [{\citenamefont {Saha}\ \emph {et~al.}(2020)\citenamefont {Saha},
  \citenamefont {Agudo-Canalejo},\ and\ \citenamefont
  {Golestanian}}]{saha2020scalar}%
  \BibitemOpen
  \bibfield  {author} {\bibinfo {author} {\bibfnamefont {S.}~\bibnamefont
  {Saha}}, \bibinfo {author} {\bibfnamefont {J.}~\bibnamefont
  {Agudo-Canalejo}},\ and\ \bibinfo {author} {\bibfnamefont {R.}~\bibnamefont
  {Golestanian}},\ }\bibfield  {title} {\bibinfo {title} {Scalar active
  mixtures: The nonreciprocal cahn-hilliard model},\ }\href@noop {} {\bibfield
  {journal} {\bibinfo  {journal} {Physical Review X}\ }\textbf {\bibinfo
  {volume} {10}},\ \bibinfo {pages} {041009} (\bibinfo {year}
  {2020})}\BibitemShut {NoStop}%
\bibitem [{\citenamefont {You}\ \emph {et~al.}(2020)\citenamefont {You},
  \citenamefont {Baskaran},\ and\ \citenamefont
  {Marchetti}}]{you2020nonreciprocity}%
  \BibitemOpen
  \bibfield  {author} {\bibinfo {author} {\bibfnamefont {Z.}~\bibnamefont
  {You}}, \bibinfo {author} {\bibfnamefont {A.}~\bibnamefont {Baskaran}},\ and\
  \bibinfo {author} {\bibfnamefont {M.~C.}\ \bibnamefont {Marchetti}},\
  }\bibfield  {title} {\bibinfo {title} {Nonreciprocity as a generic route to
  traveling states},\ }\href {https://doi.org/10.1073/pnas.2010318117}
  {\bibfield  {journal} {\bibinfo  {journal} {Proceedings of the National
  Academy of Sciences}\ }\textbf {\bibinfo {volume} {117}},\ \bibinfo {pages}
  {19767–19772} (\bibinfo {year} {2020})}\BibitemShut {NoStop}%
\bibitem [{\citenamefont {Pisegna}\ \emph {et~al.}(2024)\citenamefont
  {Pisegna}, \citenamefont {Saha},\ and\ \citenamefont
  {Golestanian}}]{pisegna2024emergent}%
  \BibitemOpen
  \bibfield  {author} {\bibinfo {author} {\bibfnamefont {G.}~\bibnamefont
  {Pisegna}}, \bibinfo {author} {\bibfnamefont {S.}~\bibnamefont {Saha}},\ and\
  \bibinfo {author} {\bibfnamefont {R.}~\bibnamefont {Golestanian}},\
  }\bibfield  {title} {\bibinfo {title} {Emergent polar order in nonpolar
  mixtures with nonreciprocal interactions},\ }\href@noop {} {\bibfield
  {journal} {\bibinfo  {journal} {Proceedings of the National Academy of
  Sciences}\ }\textbf {\bibinfo {volume} {121}},\ \bibinfo {pages}
  {e2407705121} (\bibinfo {year} {2024})}\BibitemShut {NoStop}%
\bibitem [{\citenamefont {Rana}\ and\ \citenamefont
  {Golestanian}(2024{\natexlab{a}})}]{rana2023defect}%
  \BibitemOpen
  \bibfield  {author} {\bibinfo {author} {\bibfnamefont {N.}~\bibnamefont
  {Rana}}\ and\ \bibinfo {author} {\bibfnamefont {R.}~\bibnamefont
  {Golestanian}},\ }\bibfield  {title} {\bibinfo {title} {Defect solutions of
  the nonreciprocal cahn-hilliard model: Spirals and targets},\ }\href
  {https://doi.org/10.1103/PhysRevLett.133.078301} {\bibfield  {journal}
  {\bibinfo  {journal} {Phys. Rev. Lett.}\ }\textbf {\bibinfo {volume} {133}},\
  \bibinfo {pages} {078301} (\bibinfo {year} {2024}{\natexlab{a}})}\BibitemShut
  {NoStop}%
\bibitem [{\citenamefont {Kole}\ \emph {et~al.}(2021)\citenamefont {Kole},
  \citenamefont {Alexander}, \citenamefont {Ramaswamy},\ and\ \citenamefont
  {Maitra}}]{kole2021layered}%
  \BibitemOpen
  \bibfield  {author} {\bibinfo {author} {\bibfnamefont {S.}~\bibnamefont
  {Kole}}, \bibinfo {author} {\bibfnamefont {G.~P.}\ \bibnamefont {Alexander}},
  \bibinfo {author} {\bibfnamefont {S.}~\bibnamefont {Ramaswamy}},\ and\
  \bibinfo {author} {\bibfnamefont {A.}~\bibnamefont {Maitra}},\ }\bibfield
  {title} {\bibinfo {title} {Layered chiral active matter: Beyond odd
  elasticity},\ }\href@noop {} {\bibfield  {journal} {\bibinfo  {journal}
  {Physical Review Letters}\ }\textbf {\bibinfo {volume} {126}},\ \bibinfo
  {pages} {248001} (\bibinfo {year} {2021})}\BibitemShut {NoStop}%
\bibitem [{\citenamefont {Maitra}\ \emph {et~al.}(2020)\citenamefont {Maitra},
  \citenamefont {Lenz},\ and\ \citenamefont {Voituriez}}]{maitra2020chiral}%
  \BibitemOpen
  \bibfield  {author} {\bibinfo {author} {\bibfnamefont {A.}~\bibnamefont
  {Maitra}}, \bibinfo {author} {\bibfnamefont {M.}~\bibnamefont {Lenz}},\ and\
  \bibinfo {author} {\bibfnamefont {R.}~\bibnamefont {Voituriez}},\ }\bibfield
  {title} {\bibinfo {title} {Chiral active hexatics: Giant number fluctuations,
  waves, and destruction of order},\ }\href@noop {} {\bibfield  {journal}
  {\bibinfo  {journal} {Physical Review Letters}\ }\textbf {\bibinfo {volume}
  {125}},\ \bibinfo {pages} {238005} (\bibinfo {year} {2020})}\BibitemShut
  {NoStop}%
\bibitem [{\citenamefont {Maitra}(2025{\natexlab{a}})}]{Maitra_annurev}%
  \BibitemOpen
  \bibfield  {author} {\bibinfo {author} {\bibfnamefont {A.}~\bibnamefont
  {Maitra}},\ }\bibfield  {title} {\bibinfo {title} {Activity unmasks chirality
  in liquid-crystalline matter},\ }\href
  {https://doi.org/https://doi.org/10.1146/annurev-conmatphys-032922-101439}
  {\bibfield  {journal} {\bibinfo  {journal} {Annual Review of Condensed Matter
  Physics}\ }\textbf {\bibinfo {volume} {16}},\ \bibinfo {pages} {275}
  (\bibinfo {year} {2025}{\natexlab{a}})}\BibitemShut {NoStop}%
\bibitem [{\citenamefont {Hohenberg}\ and\ \citenamefont
  {Halperin}(1977)}]{hohenberg1977theory}%
  \BibitemOpen
  \bibfield  {author} {\bibinfo {author} {\bibfnamefont {P.~C.}\ \bibnamefont
  {Hohenberg}}\ and\ \bibinfo {author} {\bibfnamefont {B.~I.}\ \bibnamefont
  {Halperin}},\ }\bibfield  {title} {\bibinfo {title} {Theory of dynamic
  critical phenomena},\ }\href@noop {} {\bibfield  {journal} {\bibinfo
  {journal} {Reviews of Modern Physics}\ }\textbf {\bibinfo {volume} {49}},\
  \bibinfo {pages} {435} (\bibinfo {year} {1977})}\BibitemShut {NoStop}%
\bibitem [{Note1()}]{Note1}%
  \BibitemOpen
  \bibinfo {note} {Three different choices lead to the same term on the R.H.S.
  of Eq.~\protect \eqref {eq:ChiralCon}. They are $\protect \mathcal {G}=c_1,
  \protect \, \protect \mathcal {G}' = c_2 \phi _1 \phi _2$, $\protect \mathcal
  {G}=c_1 \phi _1, \protect \, \protect \mathcal {G}' = c_2 \phi _2$, and
  $\protect \mathcal {G}=c_1 \phi _2, \protect \, \protect \mathcal {G}' =c_2
  \phi _1$, with the redefinition $\beta \equiv \beta c_1 c_2$}\BibitemShut
  {NoStop}%
\bibitem [{\citenamefont {Bureković}\ \emph {et~al.}(2026)\citenamefont
  {Bureković}, \citenamefont {Kole}, \citenamefont {Maitra},\ and\
  \citenamefont {Nardini}}]{bureković2026bulkmicrophaseseparationchiral}%
  \BibitemOpen
  \bibfield  {author} {\bibinfo {author} {\bibfnamefont {S.}~\bibnamefont
  {Bureković}}, \bibinfo {author} {\bibfnamefont {S.~J.}\ \bibnamefont
  {Kole}}, \bibinfo {author} {\bibfnamefont {A.}~\bibnamefont {Maitra}},\ and\
  \bibinfo {author} {\bibfnamefont {C.}~\bibnamefont {Nardini}},\ }\href
  {https://arxiv.org/abs/2607.12045} {\bibinfo {title} {Bulk and microphase
  separation in chiral active systems}} (\bibinfo {year} {2026}),\ \Eprint
  {https://arxiv.org/abs/2607.12045} {arXiv:2607.12045 [cond-mat.stat-mech]}
  \BibitemShut {NoStop}%
\bibitem [{\citenamefont {Cahn}\ and\ \citenamefont
  {Hilliard}(1958)}]{cahn1958free}%
  \BibitemOpen
  \bibfield  {author} {\bibinfo {author} {\bibfnamefont {J.~W.}\ \bibnamefont
  {Cahn}}\ and\ \bibinfo {author} {\bibfnamefont {J.~E.}\ \bibnamefont
  {Hilliard}},\ }\bibfield  {title} {\bibinfo {title} {Free energy of a
  nonuniform system. i. interfacial free energy},\ }\href@noop {} {\bibfield
  {journal} {\bibinfo  {journal} {The Journal of chemical physics}\ }\textbf
  {\bibinfo {volume} {28}},\ \bibinfo {pages} {258} (\bibinfo {year}
  {1958})}\BibitemShut {NoStop}%
\bibitem [{\citenamefont {Frohoff-H{\"u}lsmann}\ \emph
  {et~al.}(2021)\citenamefont {Frohoff-H{\"u}lsmann}, \citenamefont {Wrembel},\
  and\ \citenamefont {Thiele}}]{frohoff2021suppression}%
  \BibitemOpen
  \bibfield  {author} {\bibinfo {author} {\bibfnamefont {T.}~\bibnamefont
  {Frohoff-H{\"u}lsmann}}, \bibinfo {author} {\bibfnamefont {J.}~\bibnamefont
  {Wrembel}},\ and\ \bibinfo {author} {\bibfnamefont {U.}~\bibnamefont
  {Thiele}},\ }\bibfield  {title} {\bibinfo {title} {Suppression of coarsening
  and emergence of oscillatory behavior in a cahn-hilliard model with
  nonvariational coupling},\ }\href@noop {} {\bibfield  {journal} {\bibinfo
  {journal} {Physical Review E}\ }\textbf {\bibinfo {volume} {103}},\ \bibinfo
  {pages} {042602} (\bibinfo {year} {2021})}\BibitemShut {NoStop}%
\bibitem [{\citenamefont {Johnsrud}\ and\ \citenamefont
  {Golestanian}(2025)}]{johnsrud2025state}%
  \BibitemOpen
  \bibfield  {author} {\bibinfo {author} {\bibfnamefont {M.~K.}\ \bibnamefont
  {Johnsrud}}\ and\ \bibinfo {author} {\bibfnamefont {R.}~\bibnamefont
  {Golestanian}},\ }\bibfield  {title} {\bibinfo {title} {State diagram of the
  non-reciprocal cahn-hilliard model and the effects of symmetry},\ }\href@noop
  {} {\bibfield  {journal} {\bibinfo  {journal} {arXiv preprint
  arXiv:2503.07579}\ } (\bibinfo {year} {2025})}\BibitemShut {NoStop}%
\bibitem [{\citenamefont {Cates}\ and\ \citenamefont
  {Tailleur}(2015)}]{cates2015motility}%
  \BibitemOpen
  \bibfield  {author} {\bibinfo {author} {\bibfnamefont {M.~E.}\ \bibnamefont
  {Cates}}\ and\ \bibinfo {author} {\bibfnamefont {J.}~\bibnamefont
  {Tailleur}},\ }\bibfield  {title} {\bibinfo {title} {Motility-induced phase
  separation},\ }\href@noop {} {\bibfield  {journal} {\bibinfo  {journal}
  {Annu. Rev. Condens. Matter Phys.}\ }\textbf {\bibinfo {volume} {6}},\
  \bibinfo {pages} {219} (\bibinfo {year} {2015})}\BibitemShut {NoStop}%
\bibitem [{\citenamefont {Dinelli}\ \emph {et~al.}(2022)\citenamefont
  {Dinelli}, \citenamefont {O'Byrne}, \citenamefont {Curatolo}, \citenamefont
  {Zhao}, \citenamefont {Sollich},\ and\ \citenamefont
  {Tailleur}}]{dinelli2022self}%
  \BibitemOpen
  \bibfield  {author} {\bibinfo {author} {\bibfnamefont {A.}~\bibnamefont
  {Dinelli}}, \bibinfo {author} {\bibfnamefont {J.}~\bibnamefont {O'Byrne}},
  \bibinfo {author} {\bibfnamefont {A.}~\bibnamefont {Curatolo}}, \bibinfo
  {author} {\bibfnamefont {Y.}~\bibnamefont {Zhao}}, \bibinfo {author}
  {\bibfnamefont {P.}~\bibnamefont {Sollich}},\ and\ \bibinfo {author}
  {\bibfnamefont {J.}~\bibnamefont {Tailleur}},\ }\bibfield  {title} {\bibinfo
  {title} {Self-organization of bacterial mixtures in the presence of
  quorum-sensing interactions},\ }\href@noop {} {\bibfield  {journal} {\bibinfo
   {journal} {arXiv preprint arXiv:2203.07757}\ } (\bibinfo {year}
  {2022})}\BibitemShut {NoStop}%
\bibitem [{\citenamefont {Duan}\ \emph {et~al.}(2023)\citenamefont {Duan},
  \citenamefont {Agudo-Canalejo}, \citenamefont {Golestanian},\ and\
  \citenamefont {Mahault}}]{duan2023dynamical}%
  \BibitemOpen
  \bibfield  {author} {\bibinfo {author} {\bibfnamefont {Y.}~\bibnamefont
  {Duan}}, \bibinfo {author} {\bibfnamefont {J.}~\bibnamefont
  {Agudo-Canalejo}}, \bibinfo {author} {\bibfnamefont {R.}~\bibnamefont
  {Golestanian}},\ and\ \bibinfo {author} {\bibfnamefont {B.}~\bibnamefont
  {Mahault}},\ }\bibfield  {title} {\bibinfo {title} {Dynamical pattern
  formation without self-attraction in quorum-sensing active matter: The
  interplay between nonreciprocity and motility},\ }\href@noop {} {\bibfield
  {journal} {\bibinfo  {journal} {Physical Review Letters}\ }\textbf {\bibinfo
  {volume} {131}},\ \bibinfo {pages} {148301} (\bibinfo {year}
  {2023})}\BibitemShut {NoStop}%
\bibitem [{\citenamefont {Saha}\ \emph {et~al.}(2019)\citenamefont {Saha},
  \citenamefont {Ramaswamy},\ and\ \citenamefont
  {Golestanian}}]{saha2019pairing}%
  \BibitemOpen
  \bibfield  {author} {\bibinfo {author} {\bibfnamefont {S.}~\bibnamefont
  {Saha}}, \bibinfo {author} {\bibfnamefont {S.}~\bibnamefont {Ramaswamy}},\
  and\ \bibinfo {author} {\bibfnamefont {R.}~\bibnamefont {Golestanian}},\
  }\bibfield  {title} {\bibinfo {title} {Pairing, waltzing and scattering of
  chemotactic active colloids},\ }\href@noop {} {\bibfield  {journal} {\bibinfo
   {journal} {New Journal of Physics}\ }\textbf {\bibinfo {volume} {21}},\
  \bibinfo {pages} {063006} (\bibinfo {year} {2019})}\BibitemShut {NoStop}%
\bibitem [{\citenamefont {Boffi}\ and\ \citenamefont
  {Vanden-Eijnden}(2024)}]{boffi2024deep}%
  \BibitemOpen
  \bibfield  {author} {\bibinfo {author} {\bibfnamefont {N.~M.}\ \bibnamefont
  {Boffi}}\ and\ \bibinfo {author} {\bibfnamefont {E.}~\bibnamefont
  {Vanden-Eijnden}},\ }\bibfield  {title} {\bibinfo {title} {Deep learning
  probability flows and entropy production rates in active matter},\
  }\href@noop {} {\bibfield  {journal} {\bibinfo  {journal} {Proceedings of the
  National Academy of Sciences}\ }\textbf {\bibinfo {volume} {121}},\ \bibinfo
  {pages} {e2318106121} (\bibinfo {year} {2024})}\BibitemShut {NoStop}%
\bibitem [{\citenamefont {Supekar}\ \emph {et~al.}(2023)\citenamefont
  {Supekar}, \citenamefont {Song}, \citenamefont {Hastewell}, \citenamefont
  {Choi}, \citenamefont {Mietke},\ and\ \citenamefont
  {Dunkel}}]{supekar2023learning}%
  \BibitemOpen
  \bibfield  {author} {\bibinfo {author} {\bibfnamefont {R.}~\bibnamefont
  {Supekar}}, \bibinfo {author} {\bibfnamefont {B.}~\bibnamefont {Song}},
  \bibinfo {author} {\bibfnamefont {A.}~\bibnamefont {Hastewell}}, \bibinfo
  {author} {\bibfnamefont {G.~P.}\ \bibnamefont {Choi}}, \bibinfo {author}
  {\bibfnamefont {A.}~\bibnamefont {Mietke}},\ and\ \bibinfo {author}
  {\bibfnamefont {J.}~\bibnamefont {Dunkel}},\ }\bibfield  {title} {\bibinfo
  {title} {Learning hydrodynamic equations for active matter from particle
  simulations and experiments},\ }\href@noop {} {\bibfield  {journal} {\bibinfo
   {journal} {Proceedings of the National Academy of Sciences}\ }\textbf
  {\bibinfo {volume} {120}},\ \bibinfo {pages} {e2206994120} (\bibinfo {year}
  {2023})}\BibitemShut {NoStop}%
\bibitem [{\citenamefont {Pisegna}\ \emph {et~al.}(2025)\citenamefont
  {Pisegna}, \citenamefont {Rana}, \citenamefont {Golestanian},\ and\
  \citenamefont {Saha}}]{pisegna2025non}%
  \BibitemOpen
  \bibfield  {author} {\bibinfo {author} {\bibfnamefont {G.}~\bibnamefont
  {Pisegna}}, \bibinfo {author} {\bibfnamefont {N.}~\bibnamefont {Rana}},
  \bibinfo {author} {\bibfnamefont {R.}~\bibnamefont {Golestanian}},\ and\
  \bibinfo {author} {\bibfnamefont {S.}~\bibnamefont {Saha}},\ }\bibfield
  {title} {\bibinfo {title} {Nonreciprocal mixtures in suspension: The role of
  hydrodynamic interactions},\ }\href@noop {} {\bibfield  {journal} {\bibinfo
  {journal} {Physical Review Letters}\ }\textbf {\bibinfo {volume} {135}},\
  \bibinfo {pages} {108301} (\bibinfo {year} {2025})}\BibitemShut {NoStop}%
\bibitem [{\citenamefont {Bray}(2003)}]{bray2003coarsening}%
  \BibitemOpen
  \bibfield  {author} {\bibinfo {author} {\bibfnamefont {A.}~\bibnamefont
  {Bray}},\ }\bibfield  {title} {\bibinfo {title} {Coarsening dynamics of
  phase-separating systems},\ }\href@noop {} {\bibfield  {journal} {\bibinfo
  {journal} {Philosophical Transactions of the Royal Society of London. Series
  A: Mathematical, Physical and Engineering Sciences}\ }\textbf {\bibinfo
  {volume} {361}},\ \bibinfo {pages} {781} (\bibinfo {year}
  {2003})}\BibitemShut {NoStop}%
\bibitem [{\citenamefont {Harrison}\ \emph {et~al.}(2000)\citenamefont
  {Harrison}, \citenamefont {Adamson}, \citenamefont {Cheng}, \citenamefont
  {Sebastian}, \citenamefont {Sethuraman}, \citenamefont {Huse}, \citenamefont
  {Register},\ and\ \citenamefont {Chaikin}}]{harrison2000mechanisms}%
  \BibitemOpen
  \bibfield  {author} {\bibinfo {author} {\bibfnamefont {C.}~\bibnamefont
  {Harrison}}, \bibinfo {author} {\bibfnamefont {D.~H.}\ \bibnamefont
  {Adamson}}, \bibinfo {author} {\bibfnamefont {Z.}~\bibnamefont {Cheng}},
  \bibinfo {author} {\bibfnamefont {J.~M.}\ \bibnamefont {Sebastian}}, \bibinfo
  {author} {\bibfnamefont {S.}~\bibnamefont {Sethuraman}}, \bibinfo {author}
  {\bibfnamefont {D.~A.}\ \bibnamefont {Huse}}, \bibinfo {author}
  {\bibfnamefont {R.~A.}\ \bibnamefont {Register}},\ and\ \bibinfo {author}
  {\bibfnamefont {P.}~\bibnamefont {Chaikin}},\ }\bibfield  {title} {\bibinfo
  {title} {Mechanisms of ordering in striped patterns},\ }\href@noop {}
  {\bibfield  {journal} {\bibinfo  {journal} {Science}\ }\textbf {\bibinfo
  {volume} {290}},\ \bibinfo {pages} {1558} (\bibinfo {year}
  {2000})}\BibitemShut {NoStop}%
\bibitem [{\citenamefont {Aranson}\ and\ \citenamefont
  {Kramer}(2002)}]{aranson2002world}%
  \BibitemOpen
  \bibfield  {author} {\bibinfo {author} {\bibfnamefont {I.~S.}\ \bibnamefont
  {Aranson}}\ and\ \bibinfo {author} {\bibfnamefont {L.}~\bibnamefont
  {Kramer}},\ }\bibfield  {title} {\bibinfo {title} {The world of the complex
  ginzburg-landau equation},\ }\href {https://doi.org/10.1103/RevModPhys.74.99}
  {\bibfield  {journal} {\bibinfo  {journal} {Rev. Mod. Phys.}\ }\textbf
  {\bibinfo {volume} {74}},\ \bibinfo {pages} {99} (\bibinfo {year}
  {2002})}\BibitemShut {NoStop}%
\bibitem [{\citenamefont {Rana}\ and\ \citenamefont
  {Golestanian}(2024{\natexlab{b}})}]{rana2024defect}%
  \BibitemOpen
  \bibfield  {author} {\bibinfo {author} {\bibfnamefont {N.}~\bibnamefont
  {Rana}}\ and\ \bibinfo {author} {\bibfnamefont {R.}~\bibnamefont
  {Golestanian}},\ }\bibfield  {title} {\bibinfo {title} {Defect solutions of
  the nonreciprocal cahn-hilliard model: Spirals and targets},\ }\href@noop {}
  {\bibfield  {journal} {\bibinfo  {journal} {Physical Review Letters}\
  }\textbf {\bibinfo {volume} {133}},\ \bibinfo {pages} {078301} (\bibinfo
  {year} {2024}{\natexlab{b}})}\BibitemShut {NoStop}%
\bibitem [{\citenamefont {Cross}\ and\ \citenamefont
  {Greenside}(2009)}]{cross2009pattern}%
  \BibitemOpen
  \bibfield  {author} {\bibinfo {author} {\bibfnamefont {M.}~\bibnamefont
  {Cross}}\ and\ \bibinfo {author} {\bibfnamefont {H.}~\bibnamefont
  {Greenside}},\ }\href@noop {} {\emph {\bibinfo {title} {Pattern formation and
  dynamics in nonequilibrium systems}}}\ (\bibinfo  {publisher} {Cambridge
  University Press},\ \bibinfo {year} {2009})\BibitemShut {NoStop}%
\bibitem [{\citenamefont {Risken}(1989)}]{risken1989fokker}%
  \BibitemOpen
  \bibfield  {author} {\bibinfo {author} {\bibfnamefont {H.}~\bibnamefont
  {Risken}},\ }\bibfield  {title} {\bibinfo {title} {Fokker-planck equation},\
  }in\ \href@noop {} {\emph {\bibinfo {booktitle} {The Fokker-Planck equation:
  methods of solution and applications}}}\ (\bibinfo  {publisher} {Springer},\
  \bibinfo {year} {1989})\ pp.\ \bibinfo {pages} {63--95}\BibitemShut {NoStop}%
\bibitem [{\citenamefont {Saha}(2024)}]{saha2024phase}%
  \BibitemOpen
  \bibfield  {author} {\bibinfo {author} {\bibfnamefont {S.}~\bibnamefont
  {Saha}},\ }\href@noop {} {\bibinfo {title} {Phase coexistence in the
  non-reciprocal cahn-hilliard model}} (\bibinfo {year} {2024}),\ \Eprint
  {https://arxiv.org/abs/2402.10057} {arXiv:2402.10057 [cond-mat.soft]}
  \BibitemShut {NoStop}%
\bibitem [{\citenamefont {Bray}\ \emph {et~al.}(2001)\citenamefont {Bray},
  \citenamefont {Cavagna},\ and\ \citenamefont {Travasso}}]{bray2001interface}%
  \BibitemOpen
  \bibfield  {author} {\bibinfo {author} {\bibfnamefont {A.~J.}\ \bibnamefont
  {Bray}}, \bibinfo {author} {\bibfnamefont {A.}~\bibnamefont {Cavagna}},\ and\
  \bibinfo {author} {\bibfnamefont {R.~D.}\ \bibnamefont {Travasso}},\
  }\bibfield  {title} {\bibinfo {title} {Interface fluctuations, burgers
  equations, and coarsening under shear},\ }\href@noop {} {\bibfield  {journal}
  {\bibinfo  {journal} {Physical Review E}\ }\textbf {\bibinfo {volume} {65}},\
  \bibinfo {pages} {016104} (\bibinfo {year} {2001})}\BibitemShut {NoStop}%
\bibitem [{\citenamefont {Tjhung}\ \emph {et~al.}(2018)\citenamefont {Tjhung},
  \citenamefont {Nardini},\ and\ \citenamefont {Cates}}]{tjhung2018cluster}%
  \BibitemOpen
  \bibfield  {author} {\bibinfo {author} {\bibfnamefont {E.}~\bibnamefont
  {Tjhung}}, \bibinfo {author} {\bibfnamefont {C.}~\bibnamefont {Nardini}},\
  and\ \bibinfo {author} {\bibfnamefont {M.~E.}\ \bibnamefont {Cates}},\
  }\bibfield  {title} {\bibinfo {title} {Cluster phases and bubbly phase
  separation in active fluids: Reversal of the ostwald process},\ }\href@noop
  {} {\bibfield  {journal} {\bibinfo  {journal} {Physical Review X}\ }\textbf
  {\bibinfo {volume} {8}},\ \bibinfo {pages} {031080} (\bibinfo {year}
  {2018})}\BibitemShut {NoStop}%
\bibitem [{\citenamefont {Tjhung}\ \emph {et~al.}(2011)\citenamefont {Tjhung},
  \citenamefont {Cates},\ and\ \citenamefont
  {Marenduzzo}}]{tjhung2011nonequilibrium}%
  \BibitemOpen
  \bibfield  {author} {\bibinfo {author} {\bibfnamefont {E.}~\bibnamefont
  {Tjhung}}, \bibinfo {author} {\bibfnamefont {M.~E.}\ \bibnamefont {Cates}},\
  and\ \bibinfo {author} {\bibfnamefont {D.}~\bibnamefont {Marenduzzo}},\
  }\bibfield  {title} {\bibinfo {title} {Nonequilibrium steady states in polar
  active fluids},\ }\href@noop {} {\bibfield  {journal} {\bibinfo  {journal}
  {Soft Matter}\ }\textbf {\bibinfo {volume} {7}},\ \bibinfo {pages} {7453}
  (\bibinfo {year} {2011})}\BibitemShut {NoStop}%
\bibitem [{fau(2021)}]{fausti2021capillary}%
  \BibitemOpen
  \bibfield  {title} {\bibinfo {title} {Capillary interfacial tension in active
  phase separation},\ }\href@noop {} {\bibfield  {journal} {\bibinfo  {journal}
  {Physical review letters}\ }\textbf {\bibinfo {volume} {127}},\ \bibinfo
  {pages} {068001} (\bibinfo {year} {2021})}\BibitemShut {NoStop}%
\bibitem [{\citenamefont {Solon}\ \emph {et~al.}(2018)\citenamefont {Solon},
  \citenamefont {Stenhammar}, \citenamefont {Cates}, \citenamefont {Kafri},\
  and\ \citenamefont {Tailleur}}]{solon2018generalized}%
  \BibitemOpen
  \bibfield  {author} {\bibinfo {author} {\bibfnamefont {A.~P.}\ \bibnamefont
  {Solon}}, \bibinfo {author} {\bibfnamefont {J.}~\bibnamefont {Stenhammar}},
  \bibinfo {author} {\bibfnamefont {M.~E.}\ \bibnamefont {Cates}}, \bibinfo
  {author} {\bibfnamefont {Y.}~\bibnamefont {Kafri}},\ and\ \bibinfo {author}
  {\bibfnamefont {J.}~\bibnamefont {Tailleur}},\ }\bibfield  {title} {\bibinfo
  {title} {Generalized thermodynamics of phase equilibria in scalar active
  matter},\ }\href@noop {} {\bibfield  {journal} {\bibinfo  {journal} {Physical
  Review E}\ }\textbf {\bibinfo {volume} {97}},\ \bibinfo {pages} {020602}
  (\bibinfo {year} {2018})}\BibitemShut {NoStop}%
\bibitem [{\citenamefont {Zwicker}\ \emph {et~al.}(2015)\citenamefont
  {Zwicker}, \citenamefont {Hyman},\ and\ \citenamefont
  {J{\"u}licher}}]{zwicker2015suppression}%
  \BibitemOpen
  \bibfield  {author} {\bibinfo {author} {\bibfnamefont {D.}~\bibnamefont
  {Zwicker}}, \bibinfo {author} {\bibfnamefont {A.~A.}\ \bibnamefont {Hyman}},\
  and\ \bibinfo {author} {\bibfnamefont {F.}~\bibnamefont {J{\"u}licher}},\
  }\bibfield  {title} {\bibinfo {title} {Suppression of ostwald ripening in
  active emulsions},\ }\href@noop {} {\bibfield  {journal} {\bibinfo  {journal}
  {Physical Review E}\ }\textbf {\bibinfo {volume} {92}},\ \bibinfo {pages}
  {012317} (\bibinfo {year} {2015})}\BibitemShut {NoStop}%
\bibitem [{\citenamefont {Maitra}(2025{\natexlab{b}})}]{maitra2025activity}%
  \BibitemOpen
  \bibfield  {author} {\bibinfo {author} {\bibfnamefont {A.}~\bibnamefont
  {Maitra}},\ }\bibfield  {title} {\bibinfo {title} {Activity unmasks chirality
  in liquid-crystalline matter},\ }\href@noop {} {\bibfield  {journal}
  {\bibinfo  {journal} {Annual Review of Condensed Matter Physics}\ }\textbf
  {\bibinfo {volume} {16}} (\bibinfo {year} {2025}{\natexlab{b}})}\BibitemShut
  {NoStop}%
\bibitem [{\citenamefont {Cates}\ and\ \citenamefont
  {Nardini}(2025)}]{cates2025active}%
  \BibitemOpen
  \bibfield  {author} {\bibinfo {author} {\bibfnamefont {M.~E.}\ \bibnamefont
  {Cates}}\ and\ \bibinfo {author} {\bibfnamefont {C.}~\bibnamefont
  {Nardini}},\ }\bibfield  {title} {\bibinfo {title} {Active phase separation:
  new phenomenology from non-equilibrium physics},\ }\href@noop {} {\bibfield
  {journal} {\bibinfo  {journal} {Reports on Progress in Physics}\ }\textbf
  {\bibinfo {volume} {88}},\ \bibinfo {pages} {056601} (\bibinfo {year}
  {2025})}\BibitemShut {NoStop}%
\bibitem [{\citenamefont {Cohen}\ \emph {et~al.}(2019)\citenamefont {Cohen},
  \citenamefont {Larocque}, \citenamefont {Bouchard}, \citenamefont
  {Nejadsattari}, \citenamefont {Gefen},\ and\ \citenamefont
  {Karimi}}]{Cohen2019}%
  \BibitemOpen
  \bibfield  {author} {\bibinfo {author} {\bibfnamefont {E.}~\bibnamefont
  {Cohen}}, \bibinfo {author} {\bibfnamefont {H.}~\bibnamefont {Larocque}},
  \bibinfo {author} {\bibfnamefont {F.}~\bibnamefont {Bouchard}}, \bibinfo
  {author} {\bibfnamefont {F.}~\bibnamefont {Nejadsattari}}, \bibinfo {author}
  {\bibfnamefont {Y.}~\bibnamefont {Gefen}},\ and\ \bibinfo {author}
  {\bibfnamefont {E.}~\bibnamefont {Karimi}},\ }\bibfield  {title} {\bibinfo
  {title} {Geometric phase from aharonov--bohm to pancharatnam--berry
  and beyond},\ }\href {https://doi.org/10.1038/s42254-019-0071-1} {\bibfield
  {journal} {\bibinfo  {journal} {Nature Reviews Physics}\ }\textbf {\bibinfo
  {volume} {1}},\ \bibinfo {pages} {437} (\bibinfo {year} {2019})}\BibitemShut
  {NoStop}%
\bibitem [{\citenamefont {Tang}\ \emph {et~al.}(2021)\citenamefont {Tang},
  \citenamefont {Agudo-Canalejo},\ and\ \citenamefont
  {Golestanian}}]{tang2021topology}%
  \BibitemOpen
  \bibfield  {author} {\bibinfo {author} {\bibfnamefont {E.}~\bibnamefont
  {Tang}}, \bibinfo {author} {\bibfnamefont {J.}~\bibnamefont
  {Agudo-Canalejo}},\ and\ \bibinfo {author} {\bibfnamefont {R.}~\bibnamefont
  {Golestanian}},\ }\bibfield  {title} {\bibinfo {title} {Topology protects
  chiral edge currents in stochastic systems},\ }\href@noop {} {\bibfield
  {journal} {\bibinfo  {journal} {Physical Review X}\ }\textbf {\bibinfo
  {volume} {11}},\ \bibinfo {pages} {031015} (\bibinfo {year}
  {2021})}\BibitemShut {NoStop}%
\bibitem [{\citenamefont {Kant}\ \emph {et~al.}(2025)\citenamefont {Kant},
  \citenamefont {Maitra}, \citenamefont {Sood},\ and\ \citenamefont
  {Ramaswamy}}]{SriramChiral2025}%
  \BibitemOpen
  \bibfield  {author} {\bibinfo {author} {\bibfnamefont {R.}~\bibnamefont
  {Kant}}, \bibinfo {author} {\bibfnamefont {A.}~\bibnamefont {Maitra}},
  \bibinfo {author} {\bibfnamefont {A.~K.}\ \bibnamefont {Sood}},\ and\
  \bibinfo {author} {\bibfnamefont {S.}~\bibnamefont {Ramaswamy}},\ }\href
  {https://arxiv.org/abs/2509.00729} {\bibinfo {title} {Edge states, pairing,
  and sorting of motile chiral particles}} (\bibinfo {year} {2025}),\ \Eprint
  {https://arxiv.org/abs/2509.00729} {arXiv:2509.00729 [cond-mat.soft]}
  \BibitemShut {NoStop}%
\bibitem [{\citenamefont {Plum}\ \emph {et~al.}(2009)\citenamefont {Plum},
  \citenamefont {Zhou}, \citenamefont {Dong}, \citenamefont {Fedotov},
  \citenamefont {Koschny}, \citenamefont {Soukoulis},\ and\ \citenamefont
  {Zheludev}}]{ChiralMetamaterial_PhysRevB.79.035407}%
  \BibitemOpen
  \bibfield  {author} {\bibinfo {author} {\bibfnamefont {E.}~\bibnamefont
  {Plum}}, \bibinfo {author} {\bibfnamefont {J.}~\bibnamefont {Zhou}}, \bibinfo
  {author} {\bibfnamefont {J.}~\bibnamefont {Dong}}, \bibinfo {author}
  {\bibfnamefont {V.~A.}\ \bibnamefont {Fedotov}}, \bibinfo {author}
  {\bibfnamefont {T.}~\bibnamefont {Koschny}}, \bibinfo {author} {\bibfnamefont
  {C.~M.}\ \bibnamefont {Soukoulis}},\ and\ \bibinfo {author} {\bibfnamefont
  {N.~I.}\ \bibnamefont {Zheludev}},\ }\bibfield  {title} {\bibinfo {title}
  {Metamaterial with negative index due to chirality},\ }\href
  {https://doi.org/10.1103/PhysRevB.79.035407} {\bibfield  {journal} {\bibinfo
  {journal} {Phys. Rev. B}\ }\textbf {\bibinfo {volume} {79}},\ \bibinfo
  {pages} {035407} (\bibinfo {year} {2009})}\BibitemShut {NoStop}%
\bibitem [{\citenamefont {Souslov}\ \emph {et~al.}(2017)\citenamefont
  {Souslov}, \citenamefont {Van~Zuiden}, \citenamefont {Bartolo},\ and\
  \citenamefont {Vitelli}}]{souslov2017topological}%
  \BibitemOpen
  \bibfield  {author} {\bibinfo {author} {\bibfnamefont {A.}~\bibnamefont
  {Souslov}}, \bibinfo {author} {\bibfnamefont {B.~C.}\ \bibnamefont
  {Van~Zuiden}}, \bibinfo {author} {\bibfnamefont {D.}~\bibnamefont
  {Bartolo}},\ and\ \bibinfo {author} {\bibfnamefont {V.}~\bibnamefont
  {Vitelli}},\ }\bibfield  {title} {\bibinfo {title} {Topological sound in
  active-liquid metamaterials},\ }\href@noop {} {\bibfield  {journal} {\bibinfo
   {journal} {Nature Physics}\ }\textbf {\bibinfo {volume} {13}},\ \bibinfo
  {pages} {1091} (\bibinfo {year} {2017})}\BibitemShut {NoStop}%
\bibitem [{\citenamefont {Tucci}\ \emph {et~al.}(2025)\citenamefont {Tucci},
  \citenamefont {Pisegna}, \citenamefont {Golestanian},\ and\ \citenamefont
  {Saha}}]{tucci2025hydrodynamic}%
  \BibitemOpen
  \bibfield  {author} {\bibinfo {author} {\bibfnamefont {G.}~\bibnamefont
  {Tucci}}, \bibinfo {author} {\bibfnamefont {G.}~\bibnamefont {Pisegna}},
  \bibinfo {author} {\bibfnamefont {R.}~\bibnamefont {Golestanian}},\ and\
  \bibinfo {author} {\bibfnamefont {S.}~\bibnamefont {Saha}},\ }\bibfield
  {title} {\bibinfo {title} {Hydrodynamic stresses in a multi-species
  suspension of active janus colloids},\ }\href@noop {} {\bibfield  {journal}
  {\bibinfo  {journal} {arXiv preprint arXiv:2502.07744}\ } (\bibinfo {year}
  {2025})}\BibitemShut {NoStop}%
\bibitem [{\citenamefont {Tan}\ \emph {et~al.}(2022)\citenamefont {Tan},
  \citenamefont {Mietke}, \citenamefont {Li}, \citenamefont {Chen},
  \citenamefont {Higinbotham}, \citenamefont {Foster}, \citenamefont {Gokhale},
  \citenamefont {Dunkel},\ and\ \citenamefont {Fakhri}}]{tan2022odd}%
  \BibitemOpen
  \bibfield  {author} {\bibinfo {author} {\bibfnamefont {T.~H.}\ \bibnamefont
  {Tan}}, \bibinfo {author} {\bibfnamefont {A.}~\bibnamefont {Mietke}},
  \bibinfo {author} {\bibfnamefont {J.}~\bibnamefont {Li}}, \bibinfo {author}
  {\bibfnamefont {Y.}~\bibnamefont {Chen}}, \bibinfo {author} {\bibfnamefont
  {H.}~\bibnamefont {Higinbotham}}, \bibinfo {author} {\bibfnamefont {P.~J.}\
  \bibnamefont {Foster}}, \bibinfo {author} {\bibfnamefont {S.}~\bibnamefont
  {Gokhale}}, \bibinfo {author} {\bibfnamefont {J.}~\bibnamefont {Dunkel}},\
  and\ \bibinfo {author} {\bibfnamefont {N.}~\bibnamefont {Fakhri}},\
  }\bibfield  {title} {\bibinfo {title} {Odd dynamics of living chiral
  crystals},\ }\href@noop {} {\bibfield  {journal} {\bibinfo  {journal}
  {Nature}\ }\textbf {\bibinfo {volume} {607}},\ \bibinfo {pages} {287}
  (\bibinfo {year} {2022})}\BibitemShut {NoStop}%
\bibitem [{\citenamefont {Volpe}\ \emph {et~al.}(2014)\citenamefont {Volpe},
  \citenamefont {Gigan},\ and\ \citenamefont {Volpe}}]{volpe2014simulation}%
  \BibitemOpen
  \bibfield  {author} {\bibinfo {author} {\bibfnamefont {G.}~\bibnamefont
  {Volpe}}, \bibinfo {author} {\bibfnamefont {S.}~\bibnamefont {Gigan}},\ and\
  \bibinfo {author} {\bibfnamefont {G.}~\bibnamefont {Volpe}},\ }\bibfield
  {title} {\bibinfo {title} {Simulation of the active brownian motion of a
  microswimmer},\ }\href@noop {} {\bibfield  {journal} {\bibinfo  {journal}
  {American journal of physics}\ }\textbf {\bibinfo {volume} {82}},\ \bibinfo
  {pages} {659} (\bibinfo {year} {2014})}\BibitemShut {NoStop}%
\bibitem [{\citenamefont {Dinelli}\ \emph {et~al.}(2023)\citenamefont
  {Dinelli}, \citenamefont {O’Byrne}, \citenamefont {Curatolo}, \citenamefont
  {Zhao}, \citenamefont {Sollich},\ and\ \citenamefont
  {Tailleur}}]{dinelli2023non}%
  \BibitemOpen
  \bibfield  {author} {\bibinfo {author} {\bibfnamefont {A.}~\bibnamefont
  {Dinelli}}, \bibinfo {author} {\bibfnamefont {J.}~\bibnamefont {O’Byrne}},
  \bibinfo {author} {\bibfnamefont {A.}~\bibnamefont {Curatolo}}, \bibinfo
  {author} {\bibfnamefont {Y.}~\bibnamefont {Zhao}}, \bibinfo {author}
  {\bibfnamefont {P.}~\bibnamefont {Sollich}},\ and\ \bibinfo {author}
  {\bibfnamefont {J.}~\bibnamefont {Tailleur}},\ }\bibfield  {title} {\bibinfo
  {title} {Non-reciprocity across scales in active mixtures},\ }\href@noop {}
  {\bibfield  {journal} {\bibinfo  {journal} {Nature Communications}\ }\textbf
  {\bibinfo {volume} {14}},\ \bibinfo {pages} {7035} (\bibinfo {year}
  {2023})}\BibitemShut {NoStop}%
\bibitem [{\citenamefont {Tucci}\ \emph
  {et~al.}(2024{\natexlab{a}})\citenamefont {Tucci}, \citenamefont {Pisegna},
  \citenamefont {Golestanian},\ and\ \citenamefont
  {Saha}}]{tucci2024hydrodynamic}%
  \BibitemOpen
  \bibfield  {author} {\bibinfo {author} {\bibfnamefont {G.}~\bibnamefont
  {Tucci}}, \bibinfo {author} {\bibfnamefont {G.}~\bibnamefont {Pisegna}},
  \bibinfo {author} {\bibfnamefont {R.}~\bibnamefont {Golestanian}},\ and\
  \bibinfo {author} {\bibfnamefont {S.}~\bibnamefont {Saha}},\ }\bibfield
  {title} {\bibinfo {title} {Hydrodynamic stress tensor in a multi-species
  suspension of janus colloids},\ }\href@noop {} {\bibfield  {journal}
  {\bibinfo  {journal} {in preparation}\ } (\bibinfo {year}
  {2024}{\natexlab{a}})}\BibitemShut {NoStop}%
\bibitem [{\citenamefont {Tucci}\ \emph
  {et~al.}(2024{\natexlab{b}})\citenamefont {Tucci}, \citenamefont
  {Golestanian},\ and\ \citenamefont {Saha}}]{tucci2024nonreciprocal}%
  \BibitemOpen
  \bibfield  {author} {\bibinfo {author} {\bibfnamefont {G.}~\bibnamefont
  {Tucci}}, \bibinfo {author} {\bibfnamefont {R.}~\bibnamefont {Golestanian}},\
  and\ \bibinfo {author} {\bibfnamefont {S.}~\bibnamefont {Saha}},\ }\bibfield
  {title} {\bibinfo {title} {Nonreciprocal collective dynamics in a mixture of
  phoretic janus colloids},\ }\href@noop {} {\bibfield  {journal} {\bibinfo
  {journal} {New Journal of Physics}\ }\textbf {\bibinfo {volume} {26}},\
  \bibinfo {pages} {073006} (\bibinfo {year} {2024}{\natexlab{b}})}\BibitemShut
  {NoStop}%
\bibitem [{\citenamefont {Saha}\ \emph {et~al.}(2014)\citenamefont {Saha},
  \citenamefont {Golestanian},\ and\ \citenamefont
  {Ramaswamy}}]{saha2014clusters}%
  \BibitemOpen
  \bibfield  {author} {\bibinfo {author} {\bibfnamefont {S.}~\bibnamefont
  {Saha}}, \bibinfo {author} {\bibfnamefont {R.}~\bibnamefont {Golestanian}},\
  and\ \bibinfo {author} {\bibfnamefont {S.}~\bibnamefont {Ramaswamy}},\
  }\bibfield  {title} {\bibinfo {title} {Clusters, asters, and collective
  oscillations in chemotactic colloids},\ }\href@noop {} {\bibfield  {journal}
  {\bibinfo  {journal} {Physical Review E}\ }\textbf {\bibinfo {volume} {89}},\
  \bibinfo {pages} {062316} (\bibinfo {year} {2014})}\BibitemShut {NoStop}%
\bibitem [{\citenamefont {Lisicki}\ \emph {et~al.}(2018)\citenamefont
  {Lisicki}, \citenamefont {Reigh},\ and\ \citenamefont
  {Lauga}}]{lisicki2018autophoretic}%
  \BibitemOpen
  \bibfield  {author} {\bibinfo {author} {\bibfnamefont {M.}~\bibnamefont
  {Lisicki}}, \bibinfo {author} {\bibfnamefont {S.~Y.}\ \bibnamefont {Reigh}},\
  and\ \bibinfo {author} {\bibfnamefont {E.}~\bibnamefont {Lauga}},\ }\bibfield
   {title} {\bibinfo {title} {Autophoretic motion in three dimensions},\
  }\href@noop {} {\bibfield  {journal} {\bibinfo  {journal} {Soft Matter}\
  }\textbf {\bibinfo {volume} {14}},\ \bibinfo {pages} {3304} (\bibinfo {year}
  {2018})}\BibitemShut {NoStop}%
\end{thebibliography}%

\bigskip


\end{document}